%% file: paper.tex
\pdfoutput=1
\documentclass[aps,prd,amsmath,floats,floatfix, twocolumn,
superscriptaddress,nofootinbib,showpacs,longbibliography]{revtex4-1}

\usepackage[T1]{fontenc}
\usepackage[utf8]{inputenc}
\usepackage{lmodern}

\usepackage{verbatim}

\usepackage[dvipsnames, usenames]{xcolor}
\definecolor{linkcolor}{rgb}{0.0,0.3,0.5}
\usepackage[hypertexnames=false, unicode, colorlinks=true, linkcolor=linkcolor,
citecolor=linkcolor, filecolor=linkcolor,urlcolor=linkcolor,
pdfusetitle]{hyperref}

\usepackage[all]{hypcap}
\usepackage{graphicx}
\usepackage[caption=false]{subfig}
\usepackage{xspace}
\usepackage{amssymb}
\usepackage[normalem]{ulem} %
\usepackage{bm} %
\usepackage{float}
\usepackage{microtype}
\usepackage{textcase}

\usepackage[english]{babel}
\usepackage{blindtext}
\usepackage{adjustbox}

\graphicspath{%
  {figs/}%
}

\DeclareMathAlphabet{\mathpzc}{OT1}{pzc}{m}{it}

\newif\ifshowcomments

\showcommentsfalse

\newcommand{\red}{\textcolor{red}}

\ifshowcomments
    \newcommand{\vv}[1]{\textcolor{WildStrawberry}{VV: #1}}
    \newcommand{\scott}[1]{\textcolor{blue}{#1}}
    \newcommand{\ar}[1]{\textcolor{OliveGreen}{[AR: #1]}}
    \newcommand{\lcs}[1]{\textcolor{ForestGreen}{[LCS: #1]}}
    \newcommand{\KM}[1]{\textcolor{orange}{#1}}
    \newcommand{\KMcom}[1]{\textcolor{orange}{[KM: #1]}}
    \newcommand{\TODO}[1]{\red{TODO: #1}}
    
\else
    \newcommand{\vv}[1]{}
    \newcommand{\scott}[1]{}
    \newcommand{\ar}[1]{}
    \newcommand{\lcs}[1]{}
    \newcommand{\KM}[1]{}
    \newcommand{\KMcom}[1]{}
    \newcommand{\TODO}[1]{}
    
\fi

\newcommand{\h}{\mathpzc{h}}

\newcommand*{\Scale}[2][4]{\scalebox{#1}{\ensuremath{#2}}}

\newcommand{\insp}{\text{I}}
\newcommand{\ring}{\text{R}}
\newcommand{\I}[1]{~^{\insp}#1}
\newcommand{\R}[1]{~^{\ring}#1}
\newcommand{\IR}[1]{~^{\insp,\ring}#1}
\newcommand{\RE}{\mathrm{Re}}
\newcommand{\IM}{\mathrm{Im}}

\newcommand{\tdb}[1]{t_{#1}}
\newcommand{\TD}{\mathcal{T}}
\newcommand{\TDataset}{\mathcal{D}}
\newcommand{\NDomains}{\mathcal{N}}
\newcommand{\NDataset}{N}
\newcommand{\DMask}[1]{#1{\tau^{\text{d}}}}
\newcommand{\BMask}[1]{#1{\tau^{\text{b}}}}

\newcommand{\BMasks}{\IR{\tau^{\text{b}}}}
\newcommand{\oldsur}{\texttt{NRSur7dq4}\xspace}
\newcommand{\newsur}{\texttt{NRSur7dq4v2}\xspace}

\usepackage{orcidlink}

\input{macros/sxsid_macro.tex}
\input{macros/sxsid_extrap_macro.tex}
\begin{document}

\title{\texttt{NRSur7dq4v2}: A multi-domain precessing surrogate model with improved accuracy}

\newcommand{\AEI}{\affiliation{Max Planck Institute for Gravitational Physics
(Albert Einstein Institute), D-14476 Potsdam, Germany}}
\newcommand{\UMassD}{\affiliation{Department of Mathematics,
    Center for Scientific Computing and Data Science Research,
    University of Massachusetts, Dartmouth, MA 02747, USA}}
\newcommand{\Cornell}{\affiliation{Cornell Center for Astrophysics
    and Planetary Science, Cornell University, Ithaca, New York 14853, USA}}
\newcommand\CornellPhys{\affiliation{Department of Physics, Cornell
    University, Ithaca, New York 14853, USA}}
\newcommand\Caltech{\affiliation{TAPIR 350-17, California Institute of
    Technology, 1200 E California Boulevard, Pasadena, CA 91125, USA}}
\newcommand\Olemiss{\affiliation{Department of Physics and Astronomy,
    The University of Mississippi, University, MS 38677, USA}}

\author{Abhishek Ravishankar~\orcidlink{0009-0006-6519-8996}}
\email{aravishankar@umassd.edu}
\UMassD
\author{Vijay Varma~\orcidlink{0000-0002-9994-1761}}
\UMassD
\author{Scott E. Field~\orcidlink{0000-0002-6037-3277}}
\UMassD
\author{\protect\linebreak Keefe Mitman~\orcidlink{0000-0003-0276-3856}}
\Cornell
\author{Nils Deppe~\orcidlink{0000-0003-4557-4115}}
\Cornell
\author{Leo C. Stein~\orcidlink{0000-0001-7559-9597}}
\Olemiss
\author{Michael Boyle~\orcidlink{0000-0002-5075-5116}}
\Cornell
\author{Mark A. Scheel~\orcidlink{0000-0001-6656-9134}}
\Caltech

\hypersetup{pdfauthor={Ravishankar et al.}}

\date{\today}

\begin{abstract}
Numerical relativity simulations provide the most accurate waveforms for binary black hole coalescences, but are prohibitively expensive for direct use in gravitational-wave data analysis. Surrogate models overcome this cost, and \texttt{NRSur7dq4} is commonly used in parameter estimation for this reason. However, there is evidence in the literature that \texttt{NRSur7dq4}'s accuracy could be improved in the merger-ringdown portion of the waveform, which is particularly important for the analysis of high-mass binary black hole events. Motivated by these observations, we construct \texttt{NRSur7dq4v2}, a multi-domain extension of \texttt{NRSur7dq4} in which overlapping temporal subdomains allow tighter, independent error control over the inspiral and merger-ringdown portions of the waveform before they are smoothly combined. To assess the new model's performance in the merger-ringdown regime, we infer the mass and spin of the remnant black hole from quasi-normal fits to the ringdown portion of the surrogate waveform. We find that \texttt{NRSur7dq4v2} produces significantly improved remnant mass and spin estimates, with gains of roughly factors of $3$ to $10$ over \texttt{NRSur7dq4}. Compared to \texttt{NRSur7dq4}, \texttt{NRSur7dq4v2} also includes modes up to $\ell=5$, includes more accurate modeling of certain subdominant modes, and introduces a runtime model-complexity feature that gives users direct control over the tradeoff between evaluation cost and accuracy. Finally, alongside the model development, we have optimized the \texttt{gwsurrogate} package to achieve a fourfold speedup for precessing surrogates. The resulting precessing surrogates called through \texttt{gwsurrogate} are now slightly faster than their \texttt{LALSimulation} counterparts.
\end{abstract}

\maketitle

\section{Introduction}
\label{sec:introduction}
With the LIGO~\cite{TheLIGOScientific:2014jea}, Virgo~\cite{TheVirgo:2014hva} and KAGRA~\cite{KAGRA:2020tym} gravitational wave detectors having completed four operating runs~\cite{Abbott:2016blz,Abbott:2016nmj,LIGOScientific:2018mvr,LIGOScientific:2020ibl,LIGOScientific:2021usb,KAGRA:2021vkt,LIGOScientific:2025slb,LIGOScientific:2025snk},
we have detected over 200 gravitational wave signals from compact binary
coalescences so far, with next-generation detectors poised to detect many more
events~\cite{LIGOScientific:2016wof,
Punturo:2010zz,Maggiore:2019uih,Reitze:2019iox,LISA:2017pwj, Flaminio:2020lqk,
Fritschel:2023,TianQin:2015yph,TaijiScientific:2021qgx,Kawamura:2006up,Ajith:2024mie}.
An integral part of the gravitational wave data analysis pipeline is the
estimation of source parameters from the detected signals, which relies on the
availability of fast and accurate inspiral-merger-ringdown (IMR) waveform models
for the gravitational waves emitted by these systems, which are predominantly
binary black holes (BBHs).

The most accurate waveforms for binary black hole coalescences are provided by numerical relativity (NR) simulations, which solve the full Einstein equations numerically. However, the computational cost of these simulations is high, making the resulting waveforms typically impractical for direct use in many gravitational-wave data analysis workflows that require repeated waveform evaluations. This has resulted in the development of a host of semi-analytical IMR waveform models, such as the effective-one-body (EOB) family of models~\cite{Buonanno:1998gg, Buonanno:2000ef,
Buonanno:2005xu,Damour:2000we, Damour:2001tu, Buonanno:2014aza, Damour:2008yg,Taracchini:2013rva,Pan:2013rra,Ossokine:2020kjp,Ramos-Buades:2023ehm,Nagar:2018zoe,Albanesi:2025txj}, and the phenomenological family of models~\cite{Khan:2015jqa, Husa:2015iqa, Hannam:2013oca, Pratten:2020ceb,
Pratten:2020fqn, Ajith:2007qp, Ajith:2007kx, Ajith:2009bn, Santamaria:2010yb,
London:2017bcn, Khan:2018fmp, Khan:2019kot, Dietrich:2018nrt, Dietrich:2019nrt,
Thompson:2020nei, Garcia-Quiros:2020qpx, Garcia-Quiros:2020qlt}. These models are calibrated to NR simulations to improve their accuracy, and are fast enough for direct use in parameter estimation. However, as the sensitivity of gravitational wave detectors improves, there is a need for even more accurate waveform models.

NR surrogate models
~\cite{Blackman:2015pia,Blackman:2017dfb,Blackman:2017pcm,Varma:2018mmi,Varma:2019csw,Islam:2021mha,GramaxoFreitas:2024bpk,Nee:2025nmh}
are a data-driven alternate, directly trained on NR simulations, that provide an
efficient means of bridging the gap between the accuracy of NR simulations and
the speed required for parameter estimation. These reduced-order models take
advantage of the lower intrinsic dimensionality in its training data and work by
projecting the training data down onto a lower dimensional subspace, and
recasting the problem as one of regression, resulting in a model that can be
evaluated much more quickly than the original NR simulations. One of the most
widely used models is \oldsur~\cite{Varma:2019csw}, trained on the
7-dimensional parameter space of precessing quasi-circular binary black holes
with mass ratios up to $4$ and dimensionless spin magnitudes up to $0.8$.  This model has been used in a
variety of studies including catalog-wide
analyses~\cite{LIGOScientific:2020ibl,Islam:2023zzj,LIGOScientific:2025slb},
targeted studies of key systems such as GW190521 and GW231123\_135430 (intermediate-mass
black holes), GW200129\_065458 (first observation of orbital
precession)~\cite{LIGOScientific:2020ufj,LIGOScientific:2020iuh,Islam:2020reh,Hannam:2021pit,Islam:2023zzj},
and the extremely loud GW250114\_082203 event used to test Hawking's area
theorem~\cite{LIGOScientific:2025rid,LIGOScientific:2025wao}.

Despite the model's success in accurately modeling precessing BBH systems, there is evidence in the literature~\cite{Finch:2021iip,Siegel:2025xgb} that \oldsur may have room for improved accuracy in the merger-ringdown portion of the waveform. This is consistent with a known limitation of the surrogate construction. Namely, the modeling and validation procedures most directly provide a global error bound over the full waveform, which does not preclude larger errors that are localized in time. The merger-ringdown regime is both more structurally complex than the inspiral and much shorter in duration, making it inherently more challenging to model accurately. This issue is becoming increasingly consequential as more high-mass BBH events are detected, such as GW231123\_135430~\cite{LIGOScientific:2025rsn}, for which the detector in-band signal is dominated by merger-ringdown, and as exceptionally high-SNR events like GW250114\_082203 enable precision tests of strong-field gravity, including tests of Hawking's area theorem~\cite{LIGOScientific:2025rid,LIGOScientific:2025wao}. To analyze these high-leverage observations, it is important that surrogate models be as accurate as possible in this regime. Recent work~\cite{Rink:2024swg,Nee:2025nmh} has demonstrated the promise of multi-domain surrogate modeling for aligned-spin BBH waveforms. Related multi-domain reduced-order modeling efforts for aligned-spin SEOB waveform families have also been developed~\cite{Purrer:2015tud,Cotesta:2020qhw,Pompili:2023tna}.

Here we extend some of these idea in two key ways. First, we generalize the multi-domain surrogate approach to fully precessing systems. Second, whereas Ref.~\cite{Rink:2024swg} employed non-overlapping subdomains without explicitly accounting for model smoothness across domain boundaries, we introduce overlapping subdomains and use smoothness diagnostics to guide their placement and structure. We then apply our generalized methodology to \oldsur, constructing a domain-decomposed extension, \newsur. We show that \newsur achieves improved accuracy relative to \oldsur, with the largest gains in the merger-ringdown regime, and that it produces smooth, well-behaved waveforms across the domain interface. In particular, these improvements largely resolve the merger-ringdown accuracy issues reported in Ref.~\cite{Finch:2021iip}.

The structure of this paper is as follows. In Sec.~\ref{sec:methods}, we summarize the general surrogate modeling process for gravitational waveforms, and present our multi-domain surrogate modeling methodology along with a new smoothness diagnostic. In Sec.~\ref{sec:constructingv2}, we present the construction of \newsur using this methodology. In Sec.~\ref{sec:results}, we compare the accuracy of \newsur against \oldsur, and demonstrate the improved accuracy of \newsur in the merger-ringdown portion of the waveform. In Sec.~\ref{sec:discussion}, we highlight a newly introduced user-level basis truncation feature introduced for \newsur. We conclude in Sec.~\ref{sec:conclusion} with a summary and discussion of future work.

\section{Methodology}
\label{sec:methods}

We begin with a brief summary of the general surrogate modeling process for gravitational waveforms in Sec.~\ref{sec:surrogate_modeling_process}, used to model a time series obtained from numerical relativity data. This is followed by a description of our multi-domain surrogate modeling methodology in Sec.~\ref{sec:multi_domain_method}, which augments the general surrogate modeling process, allowing it to model the time series independently over subdomains of time. Finally, as the transformations in the multi-domain decomposition can be discontinuous, there is a need to ensure that the final waveform is smooth. Hence, in Sec.~\ref{sec:diagnostics} we present a new diagnostic to quantify the smoothness of multi-domain surrogate waveforms.

\subsection{The general surrogate modeling process}
\label{sec:surrogate_modeling_process}
The gravitational waveform produced by a fiducial model is denoted by
$\h(t,\iota,\varphi_0)$, where $t$ is time, and $\iota$ and $\varphi_0$
are the polar and azimuthal angles of the source on the sky.
The gravitational waveform consists of the
two fundamental polarizations, $\h_+(t,\iota,\varphi_0)$ and
$\h_\times(t,\iota,\varphi_0)$, as given by
$\h(t,\iota,\varphi_0)=\h_+(t,\iota,\varphi_0)-i\h_\times(t,\iota,\varphi_0)$.
This complexified waveform is then decomposed as
\begin{equation}
    \h(t,\iota,\varphi_0) = \sum_{\ell=2}^{\infty} \sum_{m=-\ell}^{\ell} \h_{\ell m}(t) ~_{-2}Y_{\ell m}(\iota,\varphi_0),
\end{equation}
where $\h_{\ell m}(t)$ and $~_{-2}Y_{\ell m}(\iota,\varphi_0)$ are
the spin-weighted spherical harmonic mode and spin-weighted spherical harmonic,
of degree $\ell$, order $m$, and spin-weight $-2$. These harmonic modes are the primary targets for the surrogate modeling process. Henceforth, we refer to the data to be modeled, which may signify a spherical harmonic mode or a specific transformation thereof, as data pieces\footnote{Data pieces will refer to the final quantities to be modeled, as well as the quantities obtained through intermediate transformations such as $\h^{\pm}_{\ell m}$ in Sec.~\ref{sec:surrogate_construction}.}, denoted by $X(t;\bm{\lambda})$ for the rest of this subsection. For the surrogate model constructed in this work, the details of the specific transformations used are documented in Secs.~\ref{sec:waveform_decomposition} and~\ref{sec:surrogate_construction}. Here, $\bm{\lambda}$ consists of the mass ratio $q$ and the 6 components of the dimensionless spin vectors $\bm{\chi}_1$ and $\bm{\chi}_2$ of the two black holes.

The surrogate modeling process begins with the construction of a dataset
$\TDataset=\{X(t;\bm{\lambda}_i)\}_{i=1}^{\NDataset}$ of data pieces constructed from a fiducial waveform model
evaluated at a set of $\NDataset$ 
parameter values $\bm{\lambda}_i$. 
This set of
parameters is chosen to cover the region of parameter space to be modeled, and
time $t$ belongs to a discrete time grid of $L$ points. The surrogate modeling
of waveforms at parameter values not belonging to this set proceeds in the
following steps.
\subsubsection{Reduced basis construction}
\label{sec: RB_Construction}
The first step involves using the dataset $\TDataset$ to pick an $n$-dimensional subspace such that $n<L$, $n<\NDataset$, that approximates the full dataset up to a specified tolerance. This is done by constructing an orthonormal basis $\{e_i(t)\}_{i=1}^n$ from $\TDataset$ that spans the aforementioned subspace. 

The reduced basis is constructed using singular value decomposition (SVD).\footnote{A greedy algorithm can also be used for reduced basis construction, but we focus on SVD due to its use in \oldsur. For a discussion on both methods, refer to Appendix B of Ref.~\cite{Blackman:2017dfb}} Given an $L\times N$ matrix $A$ whose columns are the elements of the dataset $\TDataset$, the singular value decomposition of $A$ is given by $A=U\Sigma V^T$, where $U$ and $V$ are orthonormal matrices and $\Sigma$ is a diagonal matrix of singular values, chosen to be in decreasing order. The reduced basis is then constructed from the first $n$ columns of $U$, where $n$ is chosen such that the error term,
\begin{equation}
    \max_{1\le j\le \NDataset}\|A_j - \hat{A}_{n,j}\|,
    \label{eq:SVD_err_term}
\end{equation}
is below a specified basis tolerance, where $A_j$ and $\hat{A}_{n,j}$ are the $j$-th columns of $A$ and its projection onto the reduced basis, respectively.

Having obtained a lower-dimensional space in which to approximate data pieces in the dataset, the problem is restructured as one of interpolation in time using the Empirical Interpolation method~\cite{Maday:2009,chaturantabut2010nonlinear} (EIM). The output of the EIM is a set of times $\{T_i\}_{i=1}^n$ referred to as empirical interpolation (EI) nodes, and the empirical interpolant,
\begin{equation}
    \mathcal{I}_n[X](t;\bm{\lambda}_k) = \sum_{j=1}^{n}  X(T_j;\bm{\lambda}_k)B_j(t).
    \label{eq:emp_interp}
\end{equation}
Here, $B_j(t)$ is constructed from a change-of-basis of the original reduced basis obtained from the SVD, and is independent of the parameter values $\bm{\lambda}$. 
The empirical interpolant is constructed such that for any datapiece element in the dataset $\TDataset$, the representation~\eqref{eq:emp_interp} interpolates over the waveform's values at the EI nodes. The EI nodes are in turn chosen greedily from the time grid such that the interpolation problem is as well conditioned as possible.

\subsubsection{Parametric fitting at empirical interpolation nodes}
As the basis elements $B_j(t)$ are independent of the parameter values $\bm{\lambda}$, the only dependence on the parameter values is through the data piece values at the empirical nodes. Hence, in order to construct a surrogate model over parameter space, we need only model the coefficients $X(T_j;\bm{\lambda})$ as a function of the parameters $\bm{\lambda}$. 

The precise forms of these fits can be varied, such as tensor products of monomials in the individual parameters~\cite{Blackman:2015pia,Blackman:2017dfb,Blackman:2017pcm,Varma:2019csw}, post-Newtonian theory inspired forms~\cite{Field:2013cfa}, fit using forward-stepwise least squares fits, Gaussian processes fit using Gaussian process regression~\cite{Varma:2018mmi,MaganaZertuche:2024ajz,Rink:2024swg} or artificial neural networks~\cite{GramaxoFreitas:2024bpk,Chua:2020stf,Thomas:2022rmc,Thomas:2025rje}. The resultant parametric fits at the empirical interpolation nodes will be denoted by $X_{\text{fit}}(T_j,\bm{\lambda})$:
\begin{equation}
    X_{\text{fit}}(T_j,\bm{\lambda}) \approx X(T_j;\bm{\lambda}),\qquad j\in\{1,\ldots,n\}.
\end{equation}

\subsubsection{Surrogate evaluation}
Finally, during evaluation, the parametric fits are evaluated at the required parameter value $\bm{\lambda}$, resulting in a surrogate model
\begin{equation}
    X_S(t;\bm{\lambda}) \equiv \sum_{j=1}^{n} X_{\text{fit}}(T_j;\bm{\lambda})B_j(t),
\end{equation}
for each data piece. 
As the reduced basis $\{B_j(t)\}_{j=1}^n$ does not depend on parameter values,
it can be precomputed, resulting in the parametric fit evaluation and
interpolation onto a target time grid being the only
steps to be performed during evaluation. 
The evaluated data pieces are then combined by 
reversing the decompositions and frame transformations. 
This yields inertial-frame waveform modes at the requested parameter value.

\subsection{Multi-domain surrogate models}
\label{sec:multi_domain_method}

\begin{figure*}[thb]
    \begin{adjustbox}{trim=0mm 0mm 0mm 0mm, clip}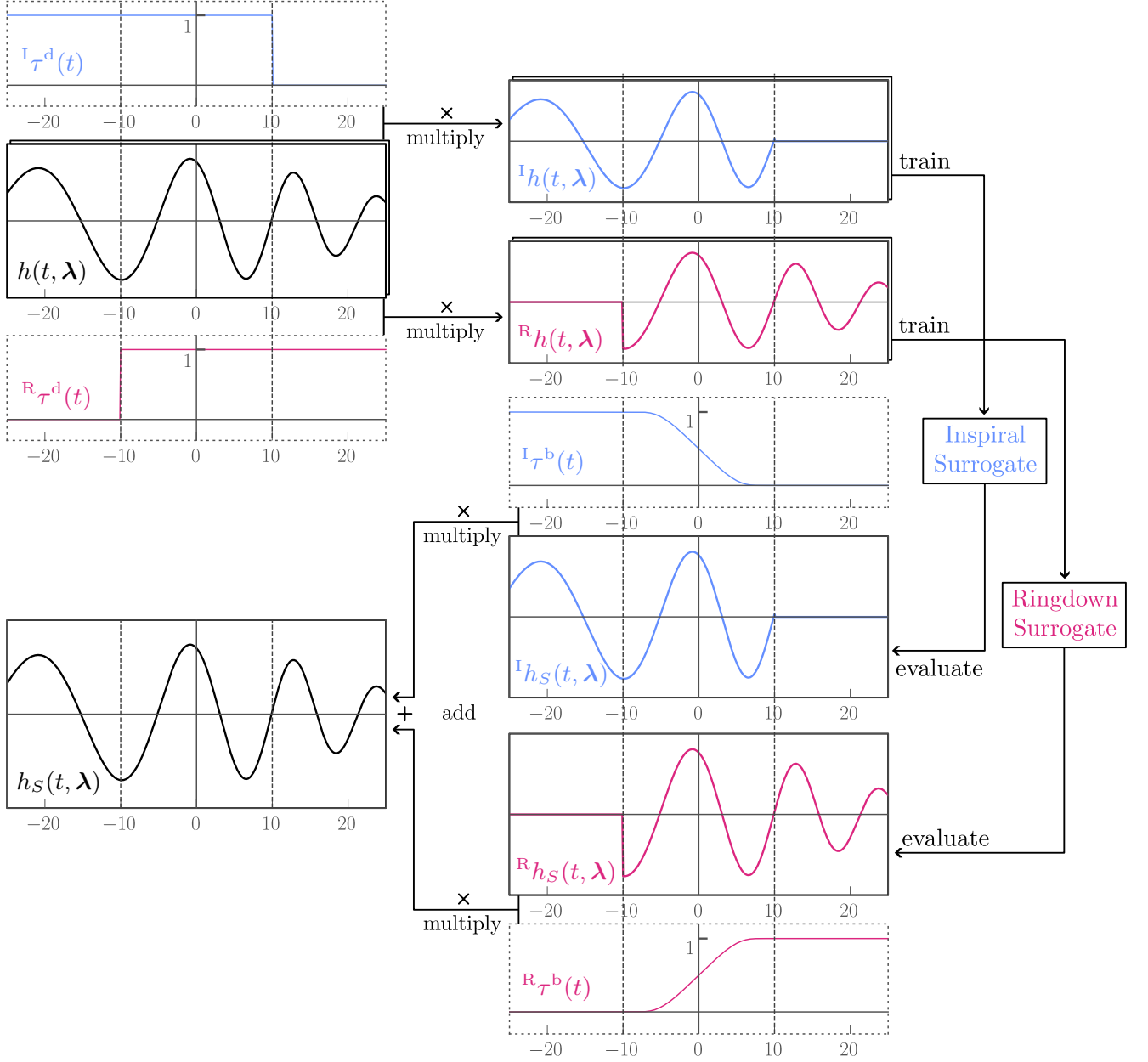
    \end{adjustbox}
    \caption{
    Schematic representing the methodology used to construct multi-domain surrogate models. The horizontal-axis of each plot represents time in units of total mass $M$, truncated to a neighbourhood of the overlap region $[t_{\ring,0},t_{\insp,1}]=[-10M,10M]$. Beginning with a dataset of waveforms $h(t,\bm{\lambda})$ (top left), the waveforms are decomposed into inspiral and ringdown data pieces $\IR{h}(t,\bm{\lambda})$ (top right) as given by Eq.~\eqref{eq:inspringpieces} using decomposition masks $\DMask{\IR}(t)$. These inspiral and ringdown datasets are then modeled independently using the surrogate modeling process resulting in inspiral and ringdown surrogates. Evaluation of these surrogates results in inspiral and ringdown surrogate waveforms $\IR{h_S}(t,\bm{\lambda})$ (bottom right). Finally, these inspiral and ringdown surrogate waveforms are combined as in Eq.~\eqref{eq:inspringblend} using blending masks $\BMask{\IR}(t)$ to obtain the final multi-domain surrogate waveform $h_S(t,\bm{\lambda})$ (bottom left).
    }
    \label{fig:DomainDecompSchematic}
\end{figure*}

In many applications, a gravitational waveform exhibits qualitatively different features across the various phases of a binary black hole coalescence, such as the inspiral and the ringdown. Because the modeling challenges are unique to each phase, surrogate errors are often inherently localized to these specific sub-intervals in the time domain. One intuitive approach to mitigate these localized errors is to modify the global cost function during reduced basis construction to upweight the problematic regions during training. However, this is still restrictive as it forces a single model to simultaneously accommodate different waveform features.

This tension between globally representing a solution and resolving localized features parallels challenges encountered in the numerical solution of partial differential equations (PDEs), where domain decomposition is routinely used to construct separate approximations in different regions. In such methods, separate approximations are constructed on individual, often overlapping, subdomains and then combined to form a smooth global solution.
Drawing inspiration from PDE solving, we treat the complete waveform as a composition of distinct, overlapping temporal domains and to model each separately. This temporal decomposition, in turn, motivates us to independently tailor the reduced basis representation to the dynamics of each regime before blending them into a global model.
While we focus here on a two-domain decomposition and use the labels `inspiral' and `ringdown' for concreteness, the methodology is not tied to this particular choice; a more general description of our multi-domain approach is presented in Appendix \ref{app:general_domain_decomposition}.

We again consider the dataset $\TDataset=\{h(t;\bm{\lambda}_i)\}_{i=1}^{\NDataset}$ of data pieces\footnote{Specifically, we model the real and imaginary parts of linear combinations of the spherical harmonic modes in the coorbital frame as described in Secs.~\ref{sec:waveform_decomposition} and~\ref{sec:surrogate_construction}.} of the fiducial waveform model evaluated at a set of $\NDataset$ parameter values $\bm{\lambda}_i$ over a common time interval $[t_{\text{start}},t_{\text{end}}]$. We would like to independently model the inspiral and ringdown regions of the waveform, separated by merger time $t_m$, while ensuring that the final surrogate waveform is a smooth function of time. To do so, we first define two overlapping subdomains,
\begin{align}
    \text{Inspiral }\mathcal{T}_{\insp} &=[t_{\insp,0},t_{\insp,1}]=[t_{\text{start}},t_m+\Delta t/2],\\
    \text{Ringdown }\mathcal{T}_{\ring} &=[t_{\ring,0},t_{\ring,1}]=[t_m-\Delta t/2,t_{\text{end}}],
\end{align}
with an overlap region $[t_{\ring,0},t_{\insp,1}]$ of width $\Delta t$ centered on $t_m$.

Each waveform in the training dataset is then decomposed into inspiral and ringdown data pieces\,$\IR{h}(t;\bm{\lambda})$:
\begin{align}
    \IR{h}(t;\bm{\lambda})&=\DMask{\IR}(t)\, h(t;\bm{\lambda}),\label{eq:inspringpieces}
\end{align}
where
\begin{align}
\DMask{\I}(t)&=1-\Theta\left(t-(t_m+ \frac{\Delta t}{2})\right),\\
\DMask{\R}(t)&=\Theta\left(t-(t_m-\frac{\Delta t}{2})\right),
\end{align}
and $\Theta$ is the Heaviside step function. We refer to the functions
$\DMask{\I}$ and $\DMask{\R}$ as the \emph{decomposition masks} (with the
superscript $d$ denoting \emph{decomposition}).
The inspiral and ringdown data pieces are supported on the inspiral and ringdown subdomains $\mathcal{T}_{\insp}$ and $\mathcal{T}_{\ring}$ respectively. The independence of these data pieces allows for greater freedom in the modeling choices for each subdomain, and individual surrogate models $\IR{h_S}(t;\bm{\lambda})$ are constructed for each of these data pieces using the surrogate modeling process summarized in Sec.~\ref{sec:surrogate_modeling_process}.

During surrogate evaluation, the inspiral and ringdown surrogate waveforms are combined to obtain a full-domain surrogate waveform $h_S(t;\bm{\lambda})$,
\begin{equation}
    h_S(t;\bm{\lambda}) = \BMask{\I}(t)\I{h_S}(t;\bm{\lambda}) + \BMask{\R}(t)\R{h_S}(t;\bm{\lambda}).\label{eq:inspringblend}
\end{equation}
The functions $\BMask{\I}$ and $\BMask{\R}$,
\begin{align}
    \BMask{\I}(t)&=\begin{cases}
    1 & \text{if } t\leq t_{\ring,0}\\
    \frac{1}{\exp(-z(t))+1} & \text{if } t_{\ring,0} < t < t_{\insp,1}\\
    0 & \text{if } t \geq t_{\insp,1}
\end{cases}\\
\BMask{\R}(t)&=\begin{cases}
    0 & \text{if } t\leq t_{\ring,0}\\
    \frac{1}{\exp(z(t))+1}
    & \text{if } t_{\ring,0} < t < t_{\insp,1}\\
    1 & \text{if } t \geq t_{\insp,1}
\end{cases}
\end{align}
are referred to as the \emph{blending masks} (with the superscript $b$ denoting \emph{blending}), and 
\begin{equation}
    z(t)=\frac{t_{\insp,1}-t_{\ring,0}}{t-t_{\ring,0}}-\frac{t_{\insp,1}-t_{\ring,0}}{t_{\insp,1}-t}.
\end{equation}
We note here that $\BMask{\I}$ and $\BMask{\R}$ are domain-restricted Planck window functions~\cite{McKechan:2010kp}. Like the decomposition masks, the blending masks $\BMask{\I}$ and $\BMask{\R}$ are supported on the inspiral and ringdown subdomains $\mathcal{T}_{\insp}$ and $\mathcal{T}_{\ring}$, but unlike the decomposition masks, they are smooth functions. Since the blending masks are independent of the parameter values $\bm{\lambda}$, they can be precomputed, and the resultant surrogate waveform $h_S(t;\bm{\lambda})$ will be smooth, as we go on to quantify in Sec.~\ref{sec:diagnostics}. This complete multi-domain procedure is schematically represented in Fig.~\ref{fig:DomainDecompSchematic}.

\begin{figure*}[thb]
\includegraphics[width=1\textwidth]{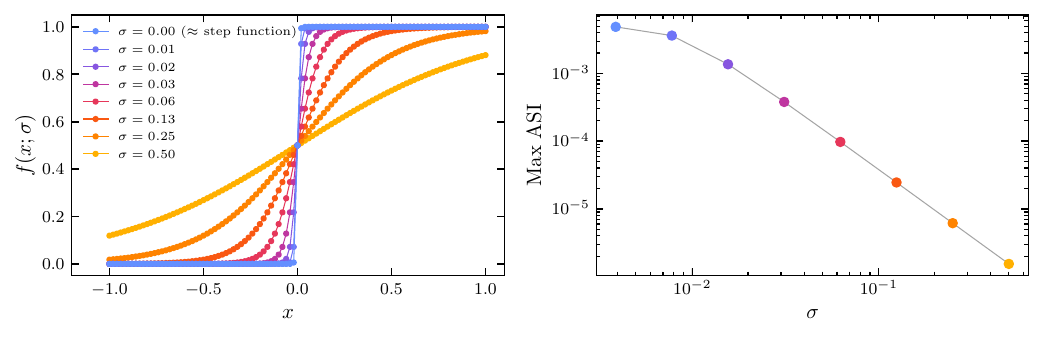}
\caption{
Test of the maximum areal smoothness indicator for logistic functions $f(x;\sigma)$ parametrized by $\sigma$ as defined in Eq.~\eqref{eq:test_function}. \textbf{Left:} Plots of test functions $f(x;\sigma)$ on $x\in[-1,1]$. \textbf{Right:} Maximum areal smoothness indicator over the domain of the functions, as a function of $\sigma$, showing the expected decline in ASI with increasing $\sigma$. This toy problem provides intuition for the behavior of the smoothness indicator we will later apply to our surrogate model.
}
\label{fig:ASI_Example}
\end{figure*}

It is important to mention that this choice of decomposition and blending masks is not unique, and other choices are possible. In order for the final surrogate waveform $h_S(t;\bm{\lambda})$ to be a smooth and accurate approximation of the fiducial waveform $h(t;\bm{\lambda})$, the decomposition and blending masks must be supported on their corresponding subdomains, the trained subdomain surrogates must be sufficiently accurate, and the products $\BMask{\I}\cdot\DMask{\I}$ and $\BMask{\R}\cdot\DMask{\R}$ must form a smooth partition of unity, i.e. 
\begin{equation}
    \BMask{\I}\cdot\DMask{\I} + \BMask{\R}\cdot\DMask{\R} \equiv 1.
\end{equation}

\subsection{Multi-domain diagnostic: Smoothness indicator}
\label{sec:diagnostics}

Since the domains are modeled independently with only a small overlap region, we need to ensure that the resulting multi-domain surrogate waveform is smooth. To quantify the smoothness of data on a grid, we take inspiration from weighted essentially non-oscillatory (WENO) schemes~\cite{JIANG1996202}, where the presence of a shock in a time series is detected by evaluating quantities known as smoothness measurements. 

As the smoothness measurements for WENO schemes are generally defined for a time series on a uniform grid and their extensions to non-uniform grids are unwieldy, we define the \emph{Areal Smoothness Indicator} $\text{ASI}_{j}$ for a time series $\{h(t_{i})\}_{i=1}^{L}$ at time-point $t_{j}$ as follows:
\begin{equation}
    \text{ASI}\left(t_{j}\right)\equiv\text{ASI}_{j}\equiv
    \frac{1}{2}\begin{vmatrix}
    t_{j-1} & h\left(t_{j-1}\right) & 1\\
    t_{j} & h\left(t_{j}\right) & 1\\
    t_{j+1} & h\left(t_{j+1}\right) & 1
    \end{vmatrix}.
\end{equation}
This quantity, representing the area of the triangle formed by the three consecutive points $\left(t_{j-1}, h(t_{j-1})\right)$, $\left(t_{j}, h(t_{j})\right)$ and $\left(t_{j+1}, h(t_{j+1})\right)$, should vanish for a sufficiently smooth function on a sufficiently dense grid. We use the ASI at specific points of interest on the grid, or the maximum ASI over an interval of interest of the grid, to monitor the smoothness of surrogate waveforms.

As a simple example to demonstrate the utility of this quantity, we consider a set of parametrized test functions $f(x;\sigma)$ defined to be zero-centered logistic functions of inverse-growth rate $\sigma$,
\begin{equation}
    f(x;\sigma) = \frac{1}{1+e^{-x/\sigma}},
    \label{eq:test_function}
\end{equation}
for $x\in[-1,1]$ and $\sigma\in\{2^{-n}\mid n=1,2,\ldots,8\}$. Each function,
parameterized by a given value of $\sigma$, is plotted in the left panel of
Fig.~\ref{fig:ASI_Example}.
For each of these functions, we compute the maximum ASI over the interval
$[-1,1]$ on a uniform grid of 101 points, and plot the maximum ASI as a
function of $\sigma$ in the right panel of Fig.~\ref{fig:ASI_Example}. As
expected, as $\sigma$ increases, the sharpness of the functions decreases, and
this is reflected by a decreasing maximum ASI.

\section{\texorpdfstring{\NoCaseChange{\newsur}}{\newsur}}
\label{sec:constructingv2}
We now present the results of the domain decomposition methodology described in Sec.~\ref{sec:methods}: the new domain-decomposed surrogate \newsur. Unless specified otherwise, all details of this model's construction are consistent with those of \oldsur presented in Ref.~\cite{Varma:2019csw}.
\subsection{Dataset}
We use the 1528 NR simulations used to construct the surrogate model presented
in Ref.~\cite{Varma:2019csw}, as well as 4 additional precessing NR simulations
and 1 additional aligned-spin simulation, resulting in a total of 1533 NR
simulations. This dataset corresponds to simulations in the SXS public catalog~\cite{Scheel:2025jct} with identifiers \SXSIDs. While the waveforms in this catalog include memory, we remove the memory contribution from the waveforms in our dataset to maintain consistency with the memory-free waveforms used in \oldsur. This is achieved by subtracting the energy flux contribution to the electric part of the strain from the total strain as described by Eq.~(17b) of Ref.~\cite{Mitman:2020bjf}.

\subsection{Waveform decomposition}
\label{sec:waveform_decomposition}
A precessing waveform's spherical harmonic modes $\h_{\ell m}(t;\bm{\lambda})$ are simplified
into a set of data pieces starting with the transformation to the coprecessing
frame, wherein the orbital angular momentum is always aligned along the
$z$-axis. An important deviation from \oldsur is the handling of this
coprecessing frame in the ringdown. 
Instead of extending the coprecessing frame quaternions into the ringdown, we
taper the quaternion from its value at the peak of the total waveform
amplitude\footnote{The total waveform amplitude is defined in Eq.~(5) of
Ref.~\cite{Varma:2019csw}.} at $t=0$ to a constant value at $t=20M$ using a
smooth transition function. 
This ensures that the coprecessing frame is well-behaved in the late ringdown. Despite this deviation in the ringdown, we will continue to refer to this frame as the coprecessing frame for simplicity.\\

Next, this coprecessing
waveform $\h^{\text{copr}}_{\ell m}(t;\bm{\lambda})$ is transformed into the
coorbital frame in which the black holes are always on the $x$-axis, with the
heavier black hole on the positive $x$-axis.\footnote{As in \cite{Varma:2019csw}, the coorbital frame is defined using the waveform itself as opposed to the BH trajectories. Nevertheless, we will continue to refer to this frame as the coorbital
frame for simplicity.} 
These coordinate
transformations require the modeling of the dynamics of the coprecessing frame quaternions $\hat{Q}(t)$, 
and the orbital phase $\phi(t)$. These quantities,
along with the individual coorbital frame black hole spins
$\bm{\chi}^{\text{coorb}}_{1,2}$ are modeled by a dynamics surrogate as
described in Sec. IV B of Ref.~\cite{Varma:2019csw}. 

Another important deviation
from \oldsur is the computation of the orbital phase $\phi(t)$. While
\oldsur uses the phases of the $(2,2)$ and $(2,-2)$ modes to compute
$\phi(t)$, \newsur uses the angular velocity function $\vec{\omega}(t)$ defined in
Ref.~\cite{Boyle:2013nka} to compute $\phi(t)$ as
\begin{equation}
    \phi(t) = \int_{-4300M}^t \|\vec{\omega}(t')\| dt'.
\end{equation}
The result is a smoother and more robustly defined coorbital
frame.\footnote{While Ref.~\cite{Boyle:2013nka} uses $\vec{\omega}$ to define a corotating frame directly, we use $\vec{\omega}$ solely to compute the orbital phase, and then use this orbital phase to define the coorbital frame from the coprecessing frame.}

\subsection{Waveform surrogate}
\label{sec:surrogate_construction}
Our new surrogate mainly differs from \oldsur in the modeling of the coorbital frame waveform $\h^{\text{coorb}}_{\ell m}(t;\bm{\lambda})$. To further simplify the data to be modeled, linear mode combinations of the form
\begin{equation}
    \h^{\pm}_{\ell m}=\frac{\h^{\text{coorb}}_{\ell, m}\pm\h^{\text{coorb}}_{\ell, -m}~^*}{2},
\end{equation}
are used for $m\neq 0$. The surrogate \oldsur models the real and imaginary parts of $\h^{\text{coorb}}_{\ell, 0}$ and $\h^{\pm}_{\ell m}$ as its data pieces.

Since we seek to obtain higher accuracy in the surrogate ringdown waveforms of \newsur, we further decompose $\h^{\text{coorb}}_{\ell, 0}$ and $\h^{\pm}_{\ell m}$ onto an inspiral subdomain and a ringdown subdomain as described in \ref{sec:multi_domain_method}. In particular, we choose the following inspiral and ringdown subdomains:
\begin{align}
    \text{Inspiral Subdomain }\mathcal{T}_{\insp} &=[-4300M,10M],\\
    \text{Ringdown Subdomain }\mathcal{T}_{\ring} &=[ -10M,100M].
\end{align}
The overlap region is centered on the peak of the total waveform amplitude as a natural choice for the transition between the inspiral and ringdown. The width of the overlap region is chosen to be $20M$ by varying the width and monitoring the resultant surrogate accuracy and smoothness, as quantified by the diagnostics described in Sec.~\ref{sec:diagnostics}. The choice of $20M$ is found to be large enough to ensure smoothness of the surrogate waveforms, while being small enough to ensure that the inspiral and ringdown subdomains are sufficiently distinct.

Each of the data pieces $\h^{\pm}_{\ell m}$ and $\h^{\text{coorb}}_{\ell, 0}$ is partitioned into inspiral and ringdown data pieces $\IR{\h^{\pm}_{\ell m}}$ and $\IR{\h^{\text{coorb}}_{\ell, 0}}$ as detailed in Sec.~\ref{sec:multi_domain_method}. An independent surrogate model is constructed for the real and imaginary parts of each of these inspiral and ringdown data pieces, and the number of basis elements for each is chosen using a cross-validation methodology described in Appendix \ref{app:cv_basis_size_testing}. During surrogate evaluation, these subdomain data pieces are combined using the blending masks $\BMasks$ to obtain full-domain data pieces $\h^{\pm}_{\ell m,S}$ and $\h^{\text{coorb}}_{\ell, 0,S}$. 
Finally, the full coorbital frame surrogate waveform $\h^{\text{coorb}}_{\ell m,S}$ is reconstructed from these mode linear combinations, and transformed back to the coprecessing and inertial frames using the dynamics surrogate as described in Sec. IV B of Ref.~\cite{Varma:2019csw}.

\section{Results}
\label{sec:results}
We evaluate the performance of \newsur utilizing a multitude of diagnostics. Comparisons between the waveforms from NR simulations and the surrogate model \newsur include the use of the areal smoothness indicator to verify the smoothness of the resultant multi-domain surrogate waveforms, and various mismatches to quantify the difference between these waveforms. These are performed through 20-fold cross-validation studies as in Ref.~\cite{Varma:2019csw}, where the training set is divided up into 20 ``folds'', each of size $\sim76$. For each fold, a surrogate is trained on the remaining $\sim1457$ out-of-fold waveforms and tested on the fold, with the testing results aggregated across folds. For the purpose of these cross-validation studies, we also construct a version of \oldsur that is trained on the same 1533 NR simulations as \newsur with the same design choices except for domain decomposition, and we refer to this version as simply \oldsur for these studies. This allows for a more direct test of the effects of domain decomposition on the surrogate waveform. Beyond these cross-validation-studies, we perform surrogate evaluation timing comparisons with \oldsur, and remnant error testing to quantify the accuracy of the ringdown waveform.
\begin{figure*}[bht]
\includegraphics[width=\textwidth, clip=true, trim=0mm 0mm 0mm 0mm]{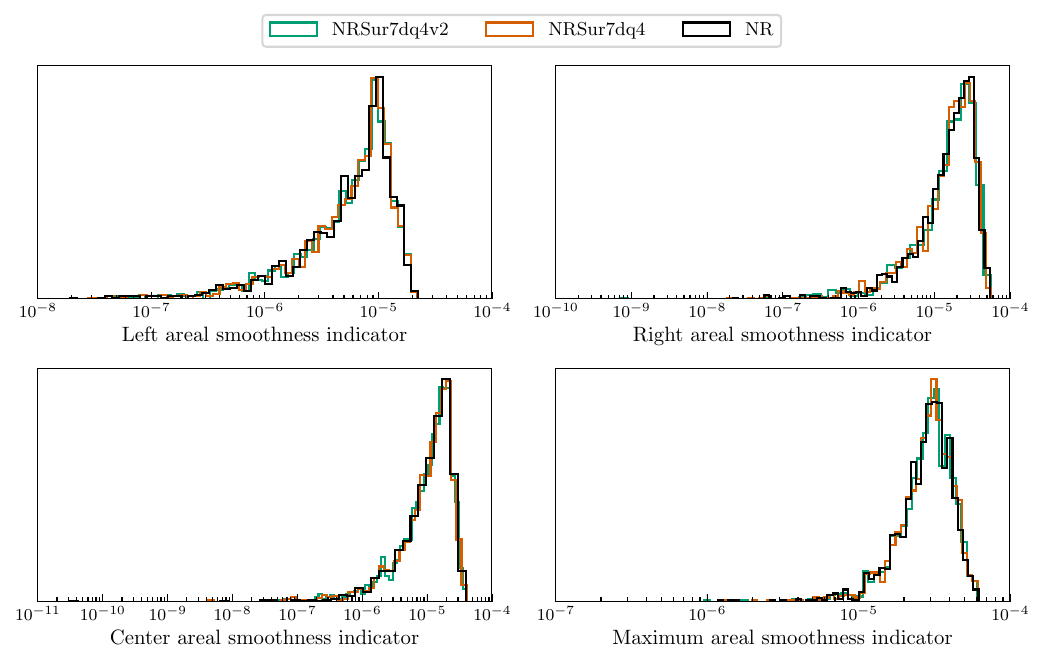}
\caption{
Distribution of areal smoothness indicators of $\RE\left(\h_{22}\right)$ over the cross-validation dataset for \newsur (green), \oldsur (orange) and fiducial NR waveform (black) at the following regions of interest: \textbf{Top Left:} Left boundary of the overlap region, as defined in Eq.~\eqref{eq:left_boundary}. \textbf{Top Right:} Right boundary of the overlap region, as defined in Eq.~\eqref{eq:right_boundary}. \textbf{Bottom Left:} Center of the overlap region, as defined in Eq.~\eqref{eq:center_overlap}. \textbf{Bottom Right:} Maximum areal smoothness indicator over the overlap region, as defined in Eq.~\eqref{eq:max_asi}. The smoothness indicators for $\RE\left(\h_{22}\right)$ are shown as they are representative of the smoothness indicators for all modes modeled by \newsur. The close agreement among the three distributions provides evidence that the multi-domain construction does not introduce sharp or non-smooth features.
}
\label{fig:ASI_Sur}
\end{figure*}
\subsection{Cross-validation comparison: Smoothness}
 We use the areal smoothness indicator introduced in Sec.~\ref{sec:diagnostics} to quantify smoothness, restricting its use to comparing the smoothness of two or more waveforms on the same grid, since its value is sensitive to grid spacing. During cross-validation testing, we compute the areal smoothness indicator for the real and imaginary parts of each waveform mode of \newsur, \oldsur, and the fiducial NR simulation at time-points of interest. We choose these points as follows:
\begin{align}
    &\text{Left-Boundary of Overlap: } &&t=-\frac{\Delta t}{2}\label{eq:left_boundary}\\
    &\text{Right-Boundary of Overlap: } &&t=+\frac{\Delta t}{2}\label{eq:right_boundary}\\
    &\text{Center of Overlap: } &&t=0\label{eq:center_overlap} 
\end{align}
The boundary points are monitored to ensure that the blending is smooth, while the center point is monitored to ensure a well-behaved merger. Furthermore, as a measure of worst-case non-smoothness, we also calculate the maximum areal smoothness indicator,
\begin{equation}
    \text{Max ASI} = \max_{t\in[-\Delta t, \Delta t]} \text{ASI}(t)\label{eq:max_asi} \,,
\end{equation}
over a region encompassing the overlap region.

The distributions of these smoothness indicator values over the aggregate of 20 folds are then plotted as histograms for both surrogate waveforms and the fiducial NR waveform. An example of the resultant histogram is shown in Figure.~\ref{fig:ASI_Sur}. The matching profiles indicate that the multi-domain surrogate \newsur is comparable in smoothness to both \oldsur and the fiducial NR waveform at all points of interest. This observation holds true for all harmonic modes modeled by \newsur, up to $\ell=5$. Having established the smoothness of \newsur, we now move on to quantifying its accuracy.

\begin{figure*}
    \centering
    
    \includegraphics[width=0.485\textwidth]{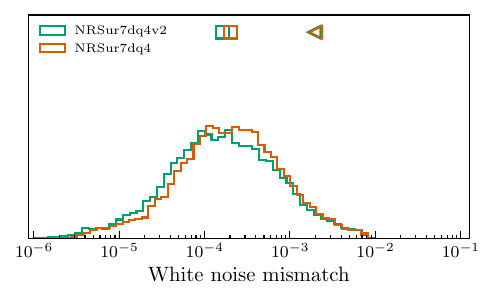}
    \hfill
    \includegraphics[width=0.485\textwidth]{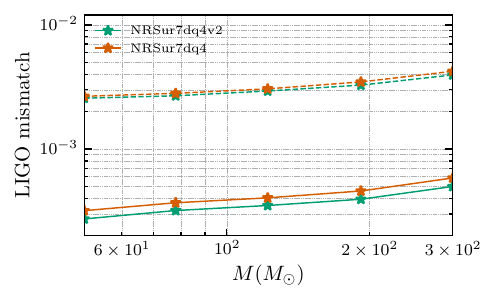}
    
    \caption{Distribution of noise-weighted mismatches over the cross-validation dataset for \newsur (green) and \oldsur (orange) against NR simulations. Mismatches are calculated using all modes with $\ell\leq 5$ for mass ratio $q\in[1,4]$ and spin magnitudes $\chi_{1,2}\leq 0.8$. Each cross-validation waveform is used to evaluate mismatches at $27$ sky locations, and the mismatches are optimized over shifts in time, phase and polarization angle. \textbf{Left:} Mismatches calculated with a white noise curve, with the median and 95\textsuperscript{th} percentile mismatches of each distribution indicated by the square and triangle markers, respectively. \textbf{Right:} Mismatches calculated with the Advanced LIGO design sensitivity noise curve, as a function of total mass. The median (95\textsuperscript{th} percentile) mismatches at each total mass are indicated by the solid (dashed) lines.
    }

    \label{fig:MM_Full_Pre}
\end{figure*}
\begin{figure*}
    \centering

    \includegraphics[width=0.485\textwidth]{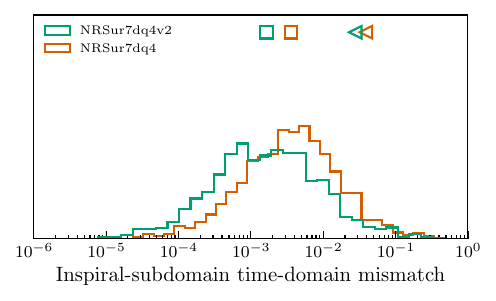}
    \hfill
    \includegraphics[width=0.485\textwidth]{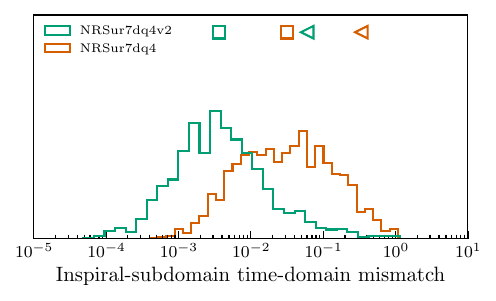}
    \caption{Distribution of time-domain mismatches of higher modes in the inspiral subdomain $t\in[-4300M,10M]$ over the cross-validation dataset for \newsur (green) and \oldsur (orange) against NR simulations. The median and 95\textsuperscript{th} percentile mismatches of each distribution are indicated by the square and triangle markers, respectively. \textbf{Left:} Mismatches for the $(3,2)$ mode. \textbf{Right:} Mismatches for the $(4,1)$ mode. \newsur matches or outperforms \oldsur across the entire dataset. The $(3,2)$ and $(4,1)$ modes are featured to illustrate the most significant reductions in mismatch.}
    \label{fig:MM_Time_Domain_Full_Pre}
\end{figure*}
\subsection{Cross-validation comparison: Mismatches}
\label{sec:mismatches}

To quantify the difference between the surrogate waveforms and the fiducial NR waveforms, we we rely on the mismatch $\mathcal{MM}$. We define the general mismatch between two waveforms $\h_1$ and $\h_2$ as
\begin{equation}
    \mathcal{MM} = 1-\mathcal{O} = 1 - \frac{\langle \h_1, \h_2 \rangle}{\sqrt{\langle \h_1, \h_1 \rangle \langle \h_2, \h_2 \rangle}},
    \label{eq:mismatch}
\end{equation}
where overlap $\mathcal{O}$ is defined in terms of an inner product $\langle\cdot,\cdot\rangle$. By changing the definition of this inner product, we can adapt the mismatch to evaluate different aspects of the waveform error. We first compute noise-weighted mismatches, as they represent the most physically appropriate measure of waveform distinguishability for a true detector. We then compute time-domain mismatches across the full time domain, as well as the isolated pre- and post-merger subdomains. This time-domain approach is specifically adopted to measure accuracy on localized time intervals, where Fourier transforms over subdomains are not meaningful.
\subsubsection{Noise-weighted mismatches}
\label{sec:Noisemismatch}
Noise-weighted mismatches are defined using the noise-weighted inner product
\begin{equation}
    \langle \h_1, \h_2 \rangle = 4\RE\int_{f_{\text{min}}}^{f_{\text{max}}} \frac{\tilde{\h}_1(f)\tilde{\h}_2^*(f)}{S_n(f)} df,
\end{equation}
where $\tilde{\h}(f)$ is the Fourier transform of strain $\h(t)$, $^{*}$ denotes the complex conjugate, $\RE$ denotes the real part, and $S_n(f)$ is the one-sided power spectral density of the detector. The waveforms are tapered using a Planck window of width of $500M$ at the start of the waveform, and a Planck window of width $20M$ at the end of the waveform. The waveform is then padded to the next power of 2 in length with zeros, and we further pad to higher powers of 2 for larger total masses to ensure that the frequency resolution is sufficiently high. The Fourier transforms are computed with $f_{\text{min}}$ set to twice the orbital angular velocity at $t=-3800M$, the end of the tapering window at the start of the waveform, and $f_{\text{max}}$ set to 
$10$ times the orbital angular velocity at the peak of the waveform amplitude, where $10$ is chosen to account for all modes up to $m=5$ with a factor of $2$ as a safety margin. The mismatches are 
optimized over shifts in time, orbital phase, and polarization angle, and each waveform is evaluated at 
$27$ locations uniformly distributed in the sky in the source frame. 

The power spectral densities used are those for white noise, and the Advanced LIGO design noise sensitivity
curve~\cite{aLIGODesignNoiseCurve}, and the resultant mismatches are summarized
in Fig.~\ref{fig:MM_Full_Pre}. The white noise mismatch
distributions of \newsur and \oldsur over the aggregated
out-of-fold datasets are shown in the left panel, where it is clear that
the mismatches produced by the two surrogate models are comparable, with
\newsur mismatches having a marginally lower median value as
indicated by square markers. In the case of the LIGO noise curve, where mismatch
will be a function of total mass, we evaluate the mismatches at total masses in
the range $[50M_{\odot}, 300M_{\odot}]$ uniformly in log-scale. The median and
95\textsuperscript{th} percentile of the mismatch distributions over the
aggregated out-of-fold datasets are then plotted as a function of total mass for
\oldsur and \newsur in the right panel of Fig.~\ref{fig:MM_Full_Pre}. Again, it
is clear that the mismatches produced by the two surrogate models are comparable, with
\newsur consistently having marginally lower
median and 95\textsuperscript{th} percentile values. 

\subsubsection{Time-domain mismatches}
In addition to the noise-weighted mismatches, we also compute simple time-domain mismatches between the surrogate and NR waveforms over the full time domain and inspiral and ringdown subdomains for each mode. This serves the purpose of testing the accuracy of the surrogate model over the specific subdomains of interest. These mismatches are defined in terms of an inner product in the time-domain,
\begin{equation}
    \langle \h_1, \h_2 \rangle = \int_{t_{\text{start}}}^{t_{\text{end}}} \h_1(t)\h_2^*(t) dt,
\end{equation}
and the time domains of interest are the full time domain $t\in[-4300M,100M]$,
the inspiral subdomain $t\in[-4300M,10M]$ and the ringdown subdomain
$t\in[-10M,100M]$. It should be noted here that unlike the noise-weighted mismatches, there is no optimization performed in the computation of the time-domain mismatches. Fig.~\ref{fig:MM_Time_Domain_Full_Pre} shows the time-domain mismatch distributions for the
$(3,2)$ and $(4,1)$ modes. In every mode, \newsur is at least as accurate as \oldsur, and it frequently provides superior results in higher order harmonics. We highlight the $(3,2)$ and $(4,1)$ modes here because they exhibit the most significant reductions in mismatch, with improvements reaching up to an order of magnitude.

\subsection{Ringdown accuracy: Remnant error}
\label{subsec:rem_err}
While mismatches quantify disagreement between the surrogate and fiducial waveforms, they do not sufficiently capture the physical information contained in the waveform, especially the ringdown. Following Ref.~\cite{Finch:2021iip}, we use \emph{remnant error} (introduced in Ref.~\cite{Giesler:2019uxc}) as a measure of the accuracy of the ringdown waveform of the surrogate model, and we repeat its definition here for completeness.

Given a harmonic mode $\h_{\ell m}(t)$, we can write out a quasi-normal mode (QNM) expansion of the form:
\begin{align}
    \h^{Q}_{\ell m}(t) &= \sum_{\ell'=\ell_{\text{min}}}^{\ell_{\text{max}}}\sum_{n=0}^{n_{\text{max}}}\left[\mathcal{A}^{+}_{\ell' m n} e^{-i\omega^{+}_{\ell' m n}\left(t-t_{0}\right)}C_{\ell\ell'm}(a\omega^{+}_{\ell'mn})\right.\nonumber\\
    &\quad\left.+\mathcal{A}^{-}_{\ell' m n} e^{-i\omega^{-}_{\ell' m n}\left(t-t_{0}\right)}
  C_{\ell\ell'm}(a\omega^{-}_{\ell'mn})\right]\,.
  \label{eq:qnm_expansion}
\end{align}
where $\ell_{\text{min}}=\max\left(|m|,2\right)$. The functions $C_{\ell\ell'm}(c)$ are known as the spherical-spheroidal expansion coefficients and are functions of the oblateness-parameter $c=a\omega$. The complex numbers $\omega^{+}_{\ell' m n}$ and $\omega^{-}_{\ell' m n}$ are the prograde and retrograde mode frequencies of QNMs, and $\mathcal{A}^{+}_{\ell' m n}$ and $\mathcal{A}^{-}_{\ell' m n}$ are the corresponding complex QNM amplitudes. Finally, $\ell_{\text{max}}$ is the maximum harmonic degree available to us, $n_{\text{max}}$ is the maximum overtone number which we choose to be sufficiently large to capture the ringdown, and $t_0$ is the start-time of the QNM model, which must be chosen to be sufficiently late for nonlinearities to die down. The details of this construction can be found in Ref.~\cite{MaganaZertuche:2021syq}.

The remnant error is defined to measure the fidelity of a ringdown signal by characterizing how accurately such a QNM expansion can recover the properties of the remnant black hole. The QNM frequencies $\omega^{\pm}_{\ell' m n}$ and spherical-spheroidal expansion coefficients $C_{\ell\ell'm}(a\omega^{\pm}_{\ell'mn})$ are functions of $(M_f,\chi_f)$ and can be computed using the \texttt{qnm} python package~\cite{Stein:2019mop}. Given an IMR waveform mode $\h_{\ell m}(t)$, the test ringdown waveform mode $\h^{\text{ring}}_{\ell m}(t)$ is obtained by truncating until $t=t_0$. The waveform $\h^{\text{ring}}_{\ell m}(t)$ is used to construct a QNM expansion $\h^{Q}_{\ell m}(t)$ of the form of Eq.~\eqref{eq:qnm_expansion} by performing linear least-squares fitting for the QNM amplitudes $\mathcal{A}^{\pm}_{\ell' m n}$. The error between the test ringdown waveform and the QNM expansion can then be thought of as a function $\mathbb{E}_{\ell m}$\footnote{The subscript is used to indicate that the function is defined for a fixed test waveform mode $\h_{\ell m}$} of the remnant parameters $(M_f,\chi_f)$,
\begin{equation}
    \mathbb{E}_{\ell m}(M_f,\chi_f) \equiv \frac{1}{2}\frac{\|\h^{\text{ring}}_{\ell m}-\h^{Q}_{\ell m}(M_f,\chi_f)\|^2}{\|\h^{\text{ring}}_{\ell m}\|^2}.
    \label{eq:Remnant_Error_Function}
\end{equation}

\begin{figure}[bhtp]
\includegraphics[width=0.50\textwidth]{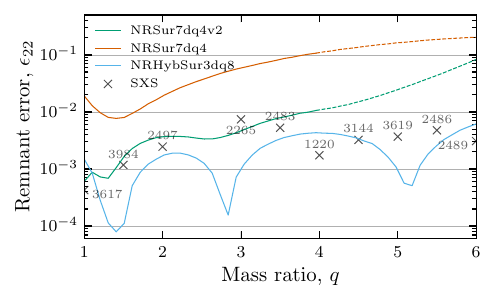}
\caption{
Remnant errors $\epsilon_{22}$ for non-spinning systems with $q\in[1,6]$ for \newsur (green), \oldsur (orange), and \texttt{NRHybSur3dq8} (blue), reproducing the analysis of Fig.~11 in Ref.~\cite{Finch:2021iip}. The remnant surrogate \texttt{NRSur3dq8Remnant} is used to compute the true remnant parameters. The dashed lines for \newsur and \oldsur over $q\in[4,6]$ indicate that these surrogate models were not trained in this region of parameter space. This analysis reproduces the previously identified limitation in surrogate ringdown accuracy, which restricts the accuracy with which remnant properties can be inferred from the ringdown signal. The multi-domain model substantially reduces these errors, achieving remnant accuracies comparable to those obtained directly from NR waveforms (cross markers; labels indicate the SXS IDs of the corresponding NR simulation) and thereby largely resolving this limitation.
}
\label{fig:Remnant_Error}
\end{figure}

\begin{figure*}[bhtp]
\includegraphics[width=\textwidth]{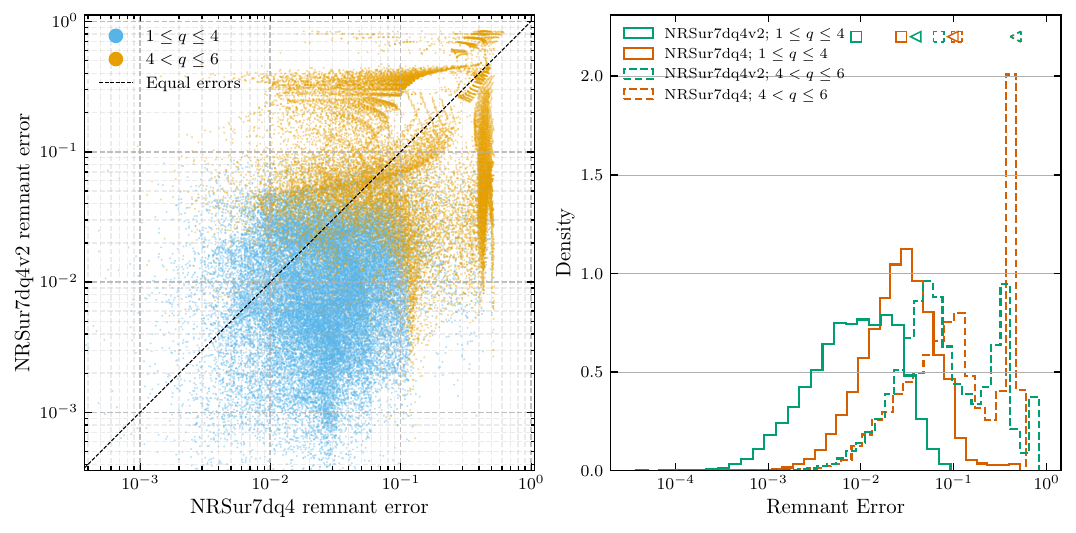}
\caption{
Distribution of remnant errors $\epsilon_{22}$ for \newsur and \oldsur over a test dataset of 55,539 points uniformly sampled over the aligned-spin subspace with $\chi_{1z},\chi_{2z}\in[-0.8,0.8]$ and mass ratios $q\in[1,6]$. The remnant surrogate \texttt{NRSur3dq8Remnant} is used to compute the true remnant parameters. \textbf{Left:} Error-error scatter plot of the remnant errors produced by \newsur and \oldsur over the test data set. The diagonal indicates equal remnant errors, with points above the diagonal being more accurately modeled by \oldsur and those below the diagonal being more accurately modeled by \newsur. The points are divided based on mass ratio, with those within the training region of $q\in[1,4]$ in blue and those outside this region in yellow. To better resolve the structure of the primary distribution, the plot is truncated to omit the sparse low-error tails of the distribution. \textbf{Right:} Remnant error distributions for \newsur (green) and \oldsur (orange) in the training region (solid) and extrapolation region (dashed). The median and 95\textsuperscript{th} percentile remnant errors of each distribution are indicated by the square and triangle markers, respectively.
}
\label{fig:Remnant_Error_Aligned_Spins}
\end{figure*}

By using an optimization algorithm, we can find the remnant parameters $(M_f^*,\chi_f^*)$ that minimize this error. Finally, given the true remnant parameters $(M_f^{\text{true}},\chi_f^{\text{true}})$, we define the remnant error as:
\begin{equation}
    \epsilon_{\ell m} \equiv \sqrt{\left(\frac{M_f^*-M_f^{\text{true}}}{M_f^{\text{true}}}\right)^2 + (\chi_f^*-\chi_f^{\text{true}})^2}.
\end{equation}
Note that for surrogate waveforms, remnant surrogate models such as \texttt{NRSur3dq8Remnant}~\cite{Varma:2018aht} or \texttt{NRSur7dq4Remnant}~\cite{Varma:2019csw} can be used to estimate the true remnant parameters, and for numerical relativity waveforms, the true values can be obtained directly from the simulations.\\

To compute $\epsilon_{22}$ using this construction, we employ a few key simplifications. We evaluate the retrograde modes only when the true remnant spin is anti-aligned with the orbital angular momentum, as their excitement is strongly correlated with the misalignment between the orbital angular momentum and the remnant spin~\cite{Li:2021wgz, Hamilton:2023znn, Zhu:2023fnf}. Furthermore, we ignore spherical-spheroidal mixing to significantly improve the conditioning of the least-squares fit, at little cost to the overall fit accuracy.\\

We first attempt to reproduce Fig.~11 from Ref.~\cite{Finch:2021iip} by evaluating $\epsilon_{22}$ for \newsur, \oldsur and \texttt{NRHybSur3dq8} over the non-spinning subspace of parameter space from $q\in[1,6]$, using \texttt{NRSur3dq8Remnant} to obtain the true remnant parameters for the surrogate waveforms. We also pick 10 non-spinning quasi-circular NR simulations\footnote{We do not pick the same NR waveforms as in~\cite{Finch:2021iip} as those simulations have been deprecated since.} from the SXS catalog~\cite{Scheel:2025jct} with mass ratios in the range $q\in[1,6]$ and compute their remnant errors against the remnant parameters obtained directly from the simulations. The QNM model is chosen to start at $t_0=0M$, with $n_{\text{max}}=7$ overtones, with the least-squares error minimization performed using basin-hopping optimization with the L-BFGS-B algorithm, with the optimizer domain constrained to $\chi_f\in[0,1]$ and $M_f\in[0,1.2M]$. The resultant remnant errors, summarized in Fig.~\ref{fig:Remnant_Error}, indicate that \newsur produces an order of magnitude improvement in remnant error over \oldsur, resulting in remnant errors that are comparable to those of the NR simulations themselves within the trained parameter space of mass ratios $q\in[1,4]$.

To further explore the accuracy of the ringdown waveforms produced by \newsur, we perform remnant error testing over a wider region of parameter space, specifically the aligned-spin subspace with $\chi_{1z},\chi_{2z}\in[-0.8,0.8]$ and mass ratios $q\in[1,6]$. We uniformly sample $51$ values of $q$ along with $33$ values each for $\chi_{1z}$ and $\chi_{2z}$, resulting in a grid of $55{,}539$ test points. We then compute the remnant errors $\epsilon_{22}$ for \newsur and \oldsur at these points, using \texttt{NRSur3dq8Remnant} to obtain the true remnant parameters. As before, the QNM model is chosen to start at $t_0=0M$, with $n_{\text{max}}=7$ overtones, with the least-squares error minimization performed using basin-hopping optimization with the L-BFGS-B algorithm, with the optimizer domain constrained to $(M_f,\chi_f)\in[0,1.2M]\times[0,1]$. See Appendix~\ref{app:unbound_mass} for a discussion on the effects of the remnant mass optimizer domain on remnant errors. 
\begin{figure*}[th]
\includegraphics[width=\textwidth]{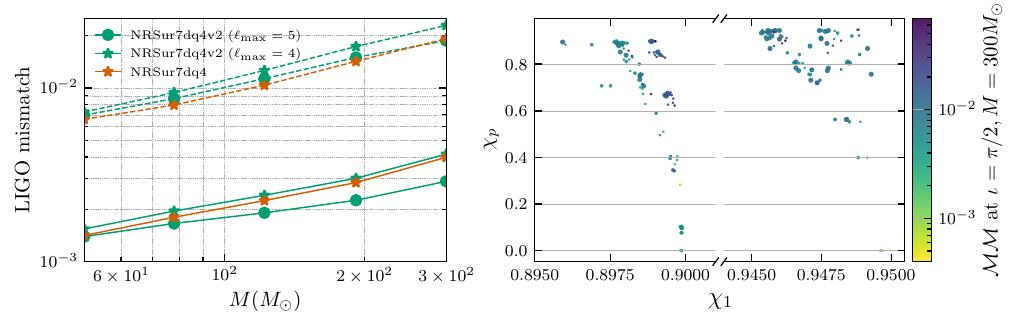}
\caption{Distribution of Advanced LIGO noise-weighted mismatches of \newsur and \oldsur against NR simulations over $202$ parameter points outside the training region. This parameter space covers mass ratio $q\in[1,4]$ and spin magnitudes $\chi_1\in[0.8,0.95]$ and $\chi_2\in[0,0.95]$. Each cross-validation waveform is used to evaluate mismatches at $27$ sky locations, and the mismatches are optimized over shifts in time, phase and polarization angle. \textbf{Left:} Mismatches for \newsur (green) and \oldsur (orange) as a function of total mass. 
The median (95\textsuperscript{th} percentile) mismatches at each total mass are indicated by the solid (dashed) lines. Mismatches calculated using all modes with $\ell\leq 4$ ($\ell\leq5$) are indicated by star (circle) markers. \textbf{Right:} Mismatches of \newsur over the parameter space covered by $202$ NR simulations outside the training region using all modes with $\ell\leq5$. The vertical axis shows the effective spin-precession parameter $\chi_p$, while the horizontal axis utilizes a broken scale to show the spin magnitude $\chi_1$ clustered about $\chi_1\approx\left\{0.9,0.95\right\}$. The marker size indicates the mass ratio $q$, and the marker color indicates the mismatch between \newsur and the NR simulation, at $\iota=\pi/2$ for a system with total mass $M=300M_{\odot}$.
}
\label{fig:MM_Extrap}
\end{figure*}

The distributions of remnant errors over these test points are summarized in
Fig.~\ref{fig:Remnant_Error_Aligned_Spins}. The left panel consists of a scatter
plot of the remnant errors of the test dataset produced by \newsur
and \oldsur, with the diagonal line indicating equal remnant errors.
We see that within the training region of mass ratios $q\in[1,4]$,
\newsur produces smaller remnant errors than \oldsur,
although the trend is difficult to infer in the extrapolation region of mass
ratios $q\in[4,6]$, where both \oldsur and \newsur
produce clusters of high remnant errors. The right panel consists of histograms
of the remnant error distributions produced by \newsur and
\oldsur. Again, we see an improvement in remnant
errors produced by \newsur over \oldsur within the
training region, with the median remnant error (indicated by square markers) of
\oldsur being $\sim\!3$ times as large as that for
\newsur. The difference in remnant error in the extrapolation
region is less pronounced in comparison, with \oldsur's median
remnant error being $\sim\!2$ times as large as that of \newsur.

\subsection{Extrapolation accuracy: Noise-weighted mismatches}
Finally, parameter estimation studies of GW events are bound to explore parameter space regions that are not covered by the training set of the surrogate model. Therefore, it is important to understand the extrapolation behavior of the surrogate model, and we focus on extrapolation in spin magnitudes. To this end, we use the Advanced LIGO design sensitivity noise curve to compute noise-weighted mismatches as described in Sec.~\ref{sec:mismatches} between our surrogate waveforms and NR waveforms. Specifically, we choose $202$ NR waveforms in the SXS public catalog with identifiers \SXSIDsExtrap. These correspond to systems with $q\in[1,4]$, spin magnitudes $\chi_1\in[0.8,0.95]$ and $\chi_2\in[0,0.95]$, and we compute the mismatches between these waveforms and the surrogate waveforms of \newsur and \oldsur. 

These mismatches are summarized in Fig.~\ref{fig:MM_Extrap}. In the left panel, we find that upon restricting to modes with $\ell\leq 4$ for both surrogate models, \newsur produces mismatches that are comparable to those of \oldsur across the entire total mass range. Including the $\ell=5$ modes of \newsur results in marginal improvements in mismatches over \oldsur over the entire total mass range. In the right panel, the mismatches of \newsur show no evident trend in effective spin-precession parameter $\chi_p$, spin magnitude $\chi_1$, or mass ratio $q$. 

\section{Evaluation Cost and Basis Truncation}
\label{sec:discussion}

\begin{figure}
    \includegraphics[width=0.495\textwidth]{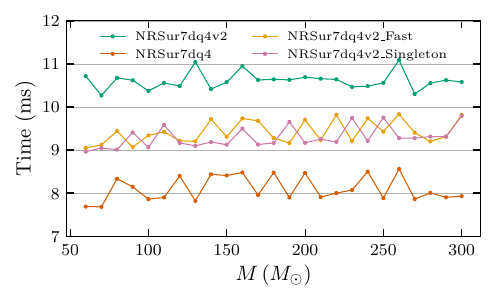}
    \caption{Evaluation times\protect\hyperlink{timing_foot}{\textsuperscript{\thefootnote}} of \oldsur (orange), and \newsur (green), as well as basis truncated versions \texttt{NRSur7dq4v2\_Fast} (yellow), and \texttt{NRSur7dq4v2\_singleton} (pink) as a function of total mass for a binary black hole system with mass ratio $q=2$ and dimensionless spins $\bm{\chi}_1=(0.2,0.3,-0.4)$ and $\bm{\chi}_2=(0.1,-0.3,0.2)$. The markers represent the minimum evaluation time over 2,000 evaluations at each total mass.
    }
    \label{fig:Timing_Comparison}
\end{figure}

The \newsur model's increase in accuracy comes at a cost of increased evaluation time due to the increased number of data pieces and EI nodes per data piece.
We quantify the increase in evaluation time of \newsur compared to \oldsur by timing the evaluation of both surrogate models for a binary black hole system with mass ratio $q=2$ and dimensionless spins $\bm{\chi}_1=(0.2,0.3,-0.4)$ and $\bm{\chi}_2=(0.1,-0.3,0.2)$, with total masses ranging from $60M_{\odot}$ to $300M_{\odot}$. For each of these systems, we make $2000$ evaluations of both \oldsur and \newsur at a starting frequency of $20$ Hz and a sampling rate of $4096$ Hz and time these evaluations. We only evaluate modes up to $\ell\leq 4$ for this test to ensure a fair comparison, as \oldsur only models modes up to $\ell=4$. The results of this test performed on a single-core of an AMD Ryzen 9 5900HX Processor @ 4.60GHz (max boost clock) are summarized in 
Fig.~\ref{fig:Timing_Comparison}, where we find that \newsur is
slower than \oldsur by a constant $\sim2.5$ ms overhead across all
tested masses. 
It is important to note that
the absolute evaluation times shown here are significantly faster than
historical benchmarks for \oldsur. Prior to this work,
\oldsur was considerably slower to evaluate. Both the legacy model
and the newer surrogates introduced here now evaluate significantly faster
thanks to recent optimizations to the \texttt{gwsurrogate}
package~\cite{Field:2025isp} detailed in
Appendix~\ref{app:gw_surrogate_optimizations}.

Although the $\sim2.5$ ms overhead is marginal, users requiring maximum speed may still wish to alleviate it. To this end, we highlight a new feature of \newsur arising directly from the nature of the surrogate modeling process. We first note that the reduced basis obtained from the SVD and the Empirical Interpolation method are both hierarchical~\cite{Field:2013cfa}, in the sense that the first $m$ basis elements and empirical interpolation nodes are a subset of the first $m+1$ basis elements and empirical interpolation nodes. This observation implies that every surrogate model contains a nested hierarchy of surrogates of lower basis sizes. 

This property is used explicitly in the implementation of \newsur, which now allows for user-level basis truncation in any of the coorbital surrogate's data pieces. The result is a degree of user-level control that can allow for faster surrogate evaluation times in cases where the user is willing to trade off a small amount of accuracy for speed. Furthermore, since inspiral and ringdown are modeled by distinct data pieces, ringdown studies such as remnant error testing can be performed by significantly truncating the basis size of inspiral data pieces.
\stepcounter{footnote}
\footnotetext[\value{footnote}]{\protect\hypertarget{timing_foot}{}The code used for these timing tests can be found in a \href{https://github.com/Abhishek-Ravishankar/gwsurrogate/tree/5864af24b496bb55cd98d5dcf4d4295a5ecf4ca4}{specific development branch} of a fork of \texttt{gwsurrogate}.}
\begin{figure}
\includegraphics[width=0.5\textwidth]{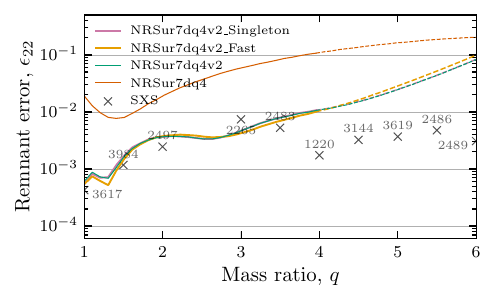}
\caption{
Remnant errors $\epsilon_{22}$ for non-spinning systems for $q\in[1,6]$ for \oldsur (orange), \newsur (green), as well as the basis truncated version \texttt{NRSur7dq4v2\_Fast} (yellow) and \texttt{NRSur7dq4v2\_Singleton} (pink). The remnant surrogate \texttt{NRSur3dq8Remnant} is used to compute the true remnant parameters. The dashed lines for $q\in[4,6]$ indicate that the models are not trained in this region of parameter space. The labels on the cross markers indicate the SXS ID of the corresponding NR simulation. Despite their reduced size and faster evaluation, \texttt{NRSur7dq4v2\_Fast} and \texttt{NRSur7dq4v2\_Singleton} achieve remnant accuracies nearly identical to that of the full model \newsur, causing the corresponding curves to largely overlap.
}
\label{fig:Remnant_Error_BasisRestrict}
\end{figure}

To showcase this feature, we consider the case of remnant error testing where the accuracy of the ringdown waveform is of primary importance. We significantly reduce the basis sizes of all the inspiral data pieces to be $1$ while keeping the basis sizes of the ringdown data pieces unchanged. The resultant remnant errors $\epsilon_{22}$ of \texttt{NRSur7dq4v2\_Singleton}, this reduced version of \newsur, for non-spinning systems for $q\in[1,6]$ are summarized in Fig.~\ref{fig:Remnant_Error_BasisRestrict}. We see that even with this drastic truncation in inspiral bases, the remnant errors produced by \texttt{NRSur7dq4v2\_Singleton} (pink) are nearly identical to those produced by the full \newsur model (green). This example highlights the ringdown waveform accuracy obtained through the multi-domain surrogate modeling process, and it underlines the utility of the user-level basis truncation feature for ringdown studies. The practical benefit of this feature is highlighted in Fig.~\ref{fig:Timing_Comparison}, which demonstrates that \texttt{NRSur7dq4v2\_Singleton} yields faster evaluation times, operating $\sim1.5\text{ms}$ faster than the full \newsur. 
Furthermore, we stress that with multi-domain surrogate models, one has the capability of speeding up the model's evaluation times, while preserving the accuracy of specific parts of the waveform. This is not possible with single-domain models, as any basis truncation leads to a loss in accuracy across the entire output waveform.

\begin{figure}[t]
  \centering%
    \includegraphics[width=0.5\textwidth]{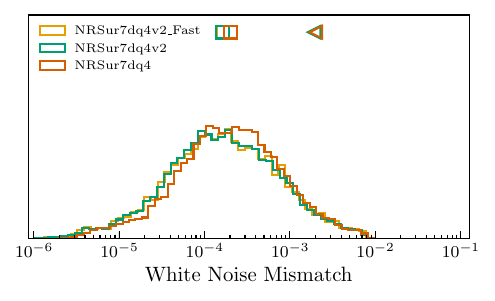}%
  \caption{Distributions of white noise-weighted mismatches of \texttt{NRSur7dq4v2\_Fast} (yellow), \newsur (green) and \oldsur (orange) against NR simulations over the cross-validation dataset. Mismatches are optimized over time, phase and polarization angle shifts, using all modes with $\ell\leq 5$ for mass ratio $q\in[1,4]$ and spin magnitudes $\chi_{1,2}\leq 0.8$. Each cross-validation waveform is used to evaluate mismatches at $5$ sky locations. The median and 95\textsuperscript{th} percentile mismatches of each distribution are indicated by the square and triangle markers, respectively.
  }
  
  \label{fig:MM_Full_Pre_Fast}
\end{figure}

A more broadly useful application of this feature is strictly truncating the data
pieces' bases based on the cross-validation testing described in
Appendix~\ref{app:cv_basis_size_testing}. To demonstrate this, we provide a
readily available model, \texttt{NRSur7dq4v2\_Fast}, with less than half the total EI nodes of the full model.
Despite its reduced size,
this version produces accuracy on par with the full \newsur model,
yielding comparable white noise mismatches (Fig.~\ref{fig:MM_Full_Pre_Fast}) and
non-spinning remnant errors (yellow curve,
Fig.~\ref{fig:Remnant_Error_BasisRestrict}). Crucially,
\texttt{NRSur7dq4v2\_Fast} achieves a $\sim1.5$ ms reduction in evaluation time
compared to the full model (Fig.~\ref{fig:Timing_Comparison}), roughly halving
the computational overhead introduced relative to \oldsur. By
exposing this truncation feature, users can construct similarly
customized, lightweight versions of \newsur suited to their
specific accuracy, speed, and memory requirements.

In practice, we leave the decision of when to employ a truncated surrogate to the user, as accuracy requirements are often task-dependent. By computing mismatches between the full \newsur model and various basis-truncated versions over a specific parameter range of interest, a user can empirically determine the minimum basis sizes required to stay within an acceptable error tolerance. While a more exhaustive study of this feature for parameter estimation is left for future work, this manual tuning offers immediate flexibility for targeted studies.

\section{Conclusion}
\label{sec:conclusion}

We present a multi-domain methodology for surrogate modeling that improves waveform accuracy in targeted time subdomains. We also introduce a diagnostic to quantify waveform smoothness, and use it to inform the surrogate construction process, ensuring that the resulting surrogate waveforms are as smooth as the fiducial waveforms over the full time domain.

We implement this methodology in the construction of \newsur, a new surrogate
model trained on $1533$ simulations with mass ratios $q\in[1,4]$ and spin
magnitudes $\chi_{1,2}\leq 0.8$. In contrast to \oldsur, which includes modes
through $\ell=4$, \newsur models the full family of spin-weighted spherical
harmonic modes with $\ell\leq5$. We perform cross-validation tests to assess
the accuracy and smoothness of \newsur relative to \oldsur and the fiducial NR
waveforms. We find that \newsur achieves accuracy and smoothness generally
comparable to \oldsur, while improving targeted regions of the waveform. In
particular, \newsur produces up to an order-of-magnitude improvement in
time-domain mismatches in the inspiral of some higher modes; see
Fig.~\ref{fig:MM_Time_Domain_Full_Pre}. We also find that the expanded mode
content improves extrapolation outside the training region relative to
\oldsur. Further, using QNM fitting to the ringdown waveform to extract
remnant parameters, we find that \newsur produces an order-of-magnitude
improvement in remnant error over \oldsur in the non-spinning subspace of
parameter space, and a $\sim\!3$ times improvement in the median remnant error
over \oldsur in the aligned-spin subspace within the training region
$q\in[1,4]$; see Figs.~\ref{fig:Remnant_Error},~\ref{fig:Remnant_Error_Aligned_Spins}.

Finally, we highlight a new feature of \newsur arising directly from the
surrogate modeling process: user-level control over model complexity. This
feature allows users to tune the tradeoff between surrogate accuracy and
evaluation time by constructing more compact versions of \newsur. The full
\newsur model is $\sim2.5$ ms slower to evaluate than \oldsur, due to both
the larger number of data pieces introduced by the multi-domain methodology
and the larger basis sizes used for each data piece, as informed by the new
cross-validation testing procedure described in
Appendix~\ref{app:cv_basis_size_testing}. However, using model-complexity
reduction, we construct \texttt{NRSur7dq4v2\_Fast}, which reduces this
additional evaluation cost to $\sim1$ ms slower than \oldsur 
while retaining accuracy comparable
to \oldsur; see Figs.~\ref{fig:Remnant_Error_BasisRestrict},~\ref{fig:MM_Full_Pre_Fast}. By comparing reduced models against the full surrogate, users can
assess accuracy loss and tailor the model to their needs while achieving
faster evaluation times. The tools to do so are made available with \newsur in
the \texttt{gwsurrogate} Python package.

In addition, we have optimized the \texttt{gwsurrogate} package to achieve a
fourfold speedup for all precessing surrogates compared to their previous
implementations. With these optimizations, \newsur can be evaluated in
$10$--$11$ ms on our testing hardware, an AMD Ryzen 9 5900HX Processor at
$4.60$ GHz; see Fig.~\ref{fig:Timing_Comparison}.

Looking forward, the multi-domain framework developed here
provides a robust and scalable path for improving surrogate accuracy in
targeted regions of the waveform, helping meet the accuracy requirements of
next-generation gravitational-wave detectors.

\begin{acknowledgments}
The authors would like to thank Ritesh Bacchar, Eliot Finch, Tousif Islam, and Peter James Nee for useful discussions. The authors also thank Lorenzo Pompili for helpful comments on this manuscript.
V.V.~acknowledges support from NSF Grant No. PHY-2309301 and UMass Dartmouth’s
Marine and Undersea Technology (MUST) Research Program funded by the Office of
Naval Research (ONR) under Grant No. N00014-23-1–2141.
K.M. is supported by NASA through the NASA Hubble Fellowship grant \#HST-HF2-51562.001-A awarded by the Space Telescope Science Institute, which is operated by the Association of Universities for Research in Astronomy, Incorporated, under NASA contract NAS5-26555.
S.E.F.~acknowledges support from NSF Grants PHY-2110496 and AST-2407454.
L.C.S. was supported in part by NSF CAREER award PHY--2047382 and a Sloan Foundation Research Fellowship.
The authors acknowledge that computations were performed on CARNiE at the Center for Scientific Computing and Visualization Research (CSCVR) of UMassD, which is supported by the ONR/DURIP Grant No. N00014181255, and on the UMass-URI UNITY HPC/AI supercomputer supported by the Massachusetts Green High Performance Computing Center (MGHPCC).
This material is based upon work supported by NSF's LIGO Laboratory which is a
major facility fully funded by the NSF.
\end{acknowledgments}

\appendix
\section{General domain decomposition methodology}
\label{app:general_domain_decomposition}
In Sec.~\ref{sec:multi_domain_method}, we described the multi-domain surrogate methodology for a two-subdomain setup. In systems where the signal exhibits highly varied morphology over its domain, the modeling process can benefit from additional subdomains to independently capture distinct signal features. Such complexity is common in zoom-whirl orbits or dynamical captures followed by eccentric inspirals and mergers. To provide a framework for these broader use cases, we now present the method in full generality. As before, we use the subscript $_S$ to denote a surrogate model, and we write the gravitational waveform produced by a fiducial model as $h(t;\bm{\lambda})$, where $t$ is defined on the domain $\TD=[t_I, t_F]$. Throughout the rest of this appendix, $t$ represents time, although more generally it may be any strictly monotonic function of time.\footnote{For example, when modeling eccentric systems it can be helpful to re-parameterize the system as a function the relativistic anomaly, which is a strictly monotonic (and hence, invertible) function of time~\cite{Nee:2025nmh}.}

Consider a partition $\left\{\tdb{k}\in[t_I, t_F]\mid k\in\left\{0,1,\ldots,\NDomains\right\}\right\}$ of the domain $\TD$, where
\begin{equation}
    t_I=\tdb{0} < \tdb{1} < \ldots < \tdb{\NDomains} = t_F.
\end{equation}
We will refer to the sub-interval
\begin{equation}
    \TD_i=[\tdb{i-1}, \tdb{i}],\qquad i\in\left\{1,\ldots,\NDomains\right\}
\end{equation}
as the $i$\textsuperscript{th} subdomain. Note that this set of subdomains $\{\TD_i\}_{i=1}^{\NDomains}$ forms a closed cover of the domain $\TD=[t_I, t_F]$. 

In order to construct a surrogate model of this fiducial waveform family, we first define a new cover $\{\tilde{\TD}_i\}_{i=1}^{\NDomains}$ from the original cover $\{\TD_i\}_{i=1}^{\NDomains}$ by extending the elements of the cover to form a overlap regions of widths $\Delta t_i>0$ 
between adjacent subdomains:
\begin{align}
    &\begin{aligned}
        \tdb{i}^+ &= \tdb{i} + \frac{\Delta t_i}{2}\\
        \tdb{i}^- &= \tdb{i} - \frac{\Delta t_i}{2}
    \end{aligned}\ ,
    \qquad i\in\{1,\ldots,\NDomains-1\}\\
    &\tilde{\TD}_i = \begin{cases}
        [\tdb{0}, \tdb{1}^+], & i=1\\
        [\tdb{i-1}^-, \tdb{i}^+], & i\in\{2,\ldots,\NDomains-1\}\\
        [\tdb{\NDomains-1}^-, \tdb{\NDomains}], & i=\NDomains
    \end{cases}.
\end{align}
We note here that $\Delta t_i=0$ is also a possible choice, resulting in non-overlapping subdomains as used in Ref.~\cite{Rink:2024swg}. This choice, however, is not recommended as it can lead to non-smooth surrogate waveforms at the subdomain boundaries. This in turn pollutes the frequency-domain waveform with high frequency artifacts. 

Having defined this new cover of the time domain, we can now use it to construct a multi-domain surrogate model. As mentioned in Ref.~\cite{Field:2013cfa}, the fiducial waveform model is evaluated at a sufficiently minimal number of parameter values $\bm{\lambda}_i$ to produce a dataset $\TDataset=\left\{h(t;\bm{\lambda}_i)\right\}_{i=1}^{\NDataset}$ over a densely sampled time grid, where $\NDataset$ is the number of parameter values. To independently model the fiducial waveform over each subdomain $\tilde{\TD}_i$, we define a set of \emph{decomposition mask} functions
 $\{\!\DMask{~^{i}}(t)\}_{i=1}^{\NDomains}$, where the function $\DMask{~^{i}}(t)$ is supported on the subdomain $\tilde{\TD}_i$:
\begin{equation}
 \begin{gathered}
    \DMask{~^{i}} :\TD\rightarrow [0,1]\text{, s.t. }\\
    \text{supp}\, \DMask{~^{i}} \equiv \left\{t\in \TD : \DMask{~^{i}} (t)\neq 0\right\}\subseteq \tilde{\TD}_i
\end{gathered}
\end{equation}
These functions allow for the construction of $\NDomains$ datasets $\TDataset_i=\{\!\DMask{~^{i}}(t)h(t;\bm{\lambda}_j)\}_{j=1}^{\NDataset}$, which is each supported entirely on the subdomain $\tilde{\TD}_i$. For each of these datasets $\TDataset_i$, we can construct an independent surrogate model $(\!\DMask{~^{i}} h)_S(t;\bm{\lambda})$. The process of surrogate construction ensures that the surrogate waveform is also supported on the subdomain $\tilde{\TD}_i$. Finally, we combine these $\NDomains$ surrogate waveforms $\{(\!\DMask{~^{i}} h)_S(t;\bm{\lambda})\}_{i=1}^{\NDomains}$ to form the multi-domain surrogate waveform as:
\begin{equation}
    h_S(t;\bm{\lambda}) \equiv \sum_{i=1}^{\NDomains} \!\BMask{~^{i}}(t)(\!\DMask{~^{i}} h)_S(t;\bm{\lambda}) \,.
\end{equation}
Here, $\{\!\BMask{~^{i}}(t)\}_{i=1}^{\NDomains}$ is a set of functions that we refer to as \emph{blending mask} functions,
which are defined to have the same properties as the masking functions:
\begin{equation}
 \begin{gathered}
    \BMask{~^{i}} :\TD\rightarrow [0,1]\text{, s.t. }\\
    \text{supp}\ \BMask{~^{i}} \equiv \left\{t\in \TD : \BMask{~^{i}} (t)\neq 0\right\}\subseteq \tilde{\TD}_i
 \end{gathered}
\end{equation}
The functions $\DMask{~^{i}}(t)$ and $\BMask{~^{i}}(t)$ are so far left unspecified, but their choices can not be completely independent. To see this, consider the case of evaluating the multi-domain surrogate waveform at a parameter value $\bm{\lambda}$. 
Assuming that the surrogate models over the individual subdomains are accurate, we expect the following approximate equalities to hold:
\begin{align}
    (\!\DMask{~^{i}} h)_S(t;\bm{\lambda}) &\approx \DMask{~^{i}}(t)h(t;\bm{\lambda}),\quad \forall i\in\{1,\ldots,\NDomains\}\\
    \implies h_S(t;\bm{\lambda})&\approx \sum_{i=1}^{\NDomains} \BMask{~^{i}}(t)\DMask{~^{i}}(t)h(t;\bm{\lambda})
\end{align}
Since we want the multi-domain surrogate waveforms to be an accurate representation of the fiducial waveforms, this results in the following constraint on the decomposition and blending masks:
\begin{align}
    h_S(t;\bm{\lambda}) \approx h(t;\bm{\lambda})\implies \sum_{i=1}^{\NDomains} \BMask{~^{i}}(t)\DMask{~^{i}}(t) \equiv 1 \,.
\end{align}
Finally, as we would like for the final surrogate waveform to be smooth, the element-wise products $\{\!\BMask{~^{i}} \DMask{~^{i}}\}_{i=1}^{\NDomains}$ of the decomposition and blending masks form a smooth partition of unity subordinate to the cover $\{\tilde{\TD}_i\}_{i=1}^{\NDomains}$.

\section{Cross validation basis size testing}
\label{app:cv_basis_size_testing}

In Sec.~\ref{sec:surrogate_modeling_process}, we describe the process of surrogate modeling, beginning with the construction of a reduced basis from the training dataset, which serves as a low-dimensional representation of the waveform data. To ensure the accuracy of this representation, the reduced basis size $m$ is chosen such that Eq.~\eqref{eq:SVD_err_term}, the error produced by the approximation, is sufficiently small. The choice of $m$ is crucial for the effectiveness of the downstream surrogate modeling process, as the size and shape of the reduced basis determine the number and placement of EI nodes, which in turn affects the number and locations of parametric fits. A basis size that is too small can lead to an inaccurate surrogate model, as the errors associated with projection onto the reduced basis will be large. On the other hand, a basis size that is too large can also result in inaccuracies due to overfitting, as the surrogate model can start fitting to noise in the training data rather than the underlying signal. Overfitting can also occur as a consequence of decaying singular values at larger basis sizes, which can lead to a poorly conditioned interpolation matrix and hence, inaccurate empirical interpolation. Finally, a larger basis size also leads to increased surrogate evaluation times, which is undesirable for most applications.

Historically, this choice has been made by examining the decaying singular value spectrum of the data piece obtained through reduced basis construction with SVD and selecting a cutoff at a basis size with sufficiently small singular values. While this method does allow for bounding the projection error it cannot account for fitting to noise. Furthermore, optimizing for singular value decay focuses exclusively on basis truncation error, which leaves out sources of error arising from interpolation and parametric fits, thereby failing to provide a complete measure of the surrogate's predictive accuracy.

To address this issue, we implement a more systematic method of choosing the basis size for each data piece based on cross-validation. For every data piece, $5\%$ of the training dataset is randomly selected and set aside as a validation dataset, while the rest of the training dataset is used to construct a surrogate model with a very large basis size of, say, $75$. The surrogate model is then evaluated at the parameter values of the validation dataset, and the RMS errors between the surrogate and fiducial waveforms are computed. Since the basis size can be restricted without retraining as discussed in Sec.~\ref{sec:discussion}, the basis element with the smallest singular value is then dropped along with the corresponding empirical interpolation node to obtain a new lower-dimensional surrogate model. This surrogate is again evaluated at the validation parameter values to compute errors, and this process of basis truncation and error computation is repeated until the basis size is reduced to $1$. This constitutes a single trial, and this is repeated to obtain $20$ trials with different random selections of the validation dataset, and the error distributions are collated across trials to obtain a distribution of errors at each basis size. 
\begin{figure}
    \includegraphics[width=0.5\textwidth]{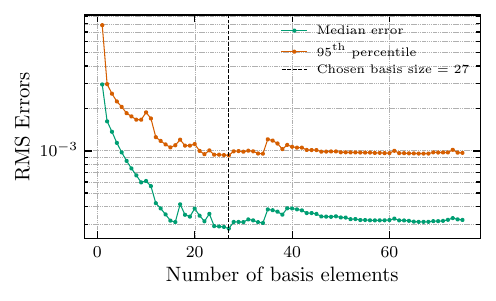}
    \caption{Cross validation basis size testing for $\mathrm{Im}\!\left(\!\I{\mathpzc{h}}^{-}_{22}\right)$. Median (orange) and 95\textsuperscript{th} percentile (green) cross validation root mean square errors as a function of basis size. Based on the observed error distributions, a basis size of $27$ is chosen for this data piece, which is indicated by the vertical dashed black line.}
    \label{fig:CV_Basis_Testing}
\end{figure}

An example of the collated errors produced by this method for $\IM\!\left(\!\I{\h}^{-}_{22}\right)$ are shown in Fig.~\ref{fig:CV_Basis_Testing}. Based on the minima of the median and 95\textsuperscript{th} percentile error curves, a basis size of $27$ is chosen for this data piece. Although cross-validation still involves an element of ad hoc selection, it naturally corrects for overfitting and improves stability against noise. Most importantly, by centering the selection criterion on the surrogate's output error instead of the basis representation error, the resulting basis size accounts for all stages of the modeling workflow. This ensures that the final model is better optimized against all potential sources of error. We also note that this method, while discussed in the context of a reduced basis obtained through SVD, can be applied to any hierarchical basis selection method, such as the greedy reduced basis algorithm~\cite{Field:2011mf}.

\section{Remnant error calculation with optimizer domain unbound in remnant mass}
\label{app:unbound_mass}

Section~\ref{subsec:rem_err} discusses the process by which the ringdown waveform of the surrogate model is tested for accuracy by recovering the remnant parameters from the surrogate ringdown waveform using a QNM expansion and comparing the recovered remnant parameters to the true remnant parameters to obtain a remnant error. We now go on to discuss choices made in the calculation of this quantity, focusing on the optimizer's remnant mass domain bound. 

The QNM expansion is constructed by performing a least-squares fit for the QNM amplitudes, where the remnant parameters $(M_f,\chi_f)$ are used to compute the QNM frequencies and spherical-spheroidal expansion coefficients. The least-squares error is minimized using an optimization algorithm over a chosen domain of remnant parameters to obtain the remnant parameters $(M_f^*,\chi_f^*)$ that best fit the ringdown waveform. A natural choice for the remnant spin domain is $\chi_f\in[0,1]$, as this is the range of physically allowed black hole spins. The choice of the remnant mass optimization domain can have a significant impact on the remnant error obtained, as it can affect the convergence of the optimizer to a minimum of the least-squares error function.

\begin{figure}[bt]
\includegraphics[width=0.5\textwidth]{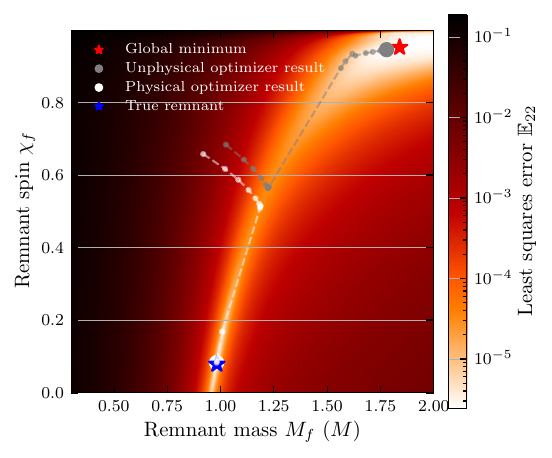}
\caption{
Error landscape $\mathbb{E}_{22}$ of the QNM amplitude linear least-squares fit (as defined in Eq.~\ref{eq:Remnant_Error_Function}), applied to \oldsur for an aligned-spin system with $q=3.6$, $\chi_{1z}=-0.8$, and $\chi_{2z}=-0.45$. The error landscape is plotted as a function of remnant mass $M_f$ and remnant spin $\chi_f$, with the true remnant parameters indicated by the blue star. Within a domain of $M_f\in[0,2M]$ and $\chi_f\in[0,1]$, a highly unphysical global minimum is observed at $(M_f,\chi_f)=(1.8402M,0.9520)$, indicated by the red star. An L-BFGS optimizer initialized as done in the main text, at $(M_f,\chi_f)=(0.99M,0.7)$, converges to this unphysical minimum through the gray curve and results in a remnant error of $\epsilon_{22}=1.186$. Modifying the initialization of the optimizer to be at $(M_f,\chi_f)=(0.8M,0.7)$ results in convergence via the white curve to the true minimum at $(M_f,\chi_f)=(0.7989M,0.7001)$ and a remnant error of $\epsilon_{22}=0.0046$.
}
\label{fig:Remnant_Error_Landscape}
\end{figure}

One might presume that the remnant mass domain could be left unbounded above, under the assumption that a high-fidelity surrogate would automatically recover values within the total mass of the binary. However, at some parameter points closer to the boundary of the training region, the least squares error landscapes for \oldsur and \newsur with an unbounded remnant mass domain of $M_f\in[0,\infty)$ can result in highly unphysical minima. An example of this is shown in Fig.~\ref{fig:Remnant_Error_Landscape}, where a spurious minimum at $(M^*_f,\chi^*_f)=(1.8402M,0.9520)$ is observed for \oldsur, occurring alongside the physically relevant minimum at $(M^*_f,\chi^*_f)=(0.7989M,0.7001)$. In this case, the unphysical minimum is also global, and the L-BFGS-B optimizer converges to it instead of the physical minimum when initialized as done in the main text at $(M^*_f,\chi^*_f)=(0.99M,0.7)$. This results in a large remnant error of $\epsilon_{22}=1.186$, which is not representative of the true accuracy of the surrogate ringdown waveform, since this spurious minimum need not solely be a consequence of waveform systematics, but can also arise from choices made in the QNM fitting procedure.

\begin{figure}[bt]
\includegraphics[width=0.5\textwidth]{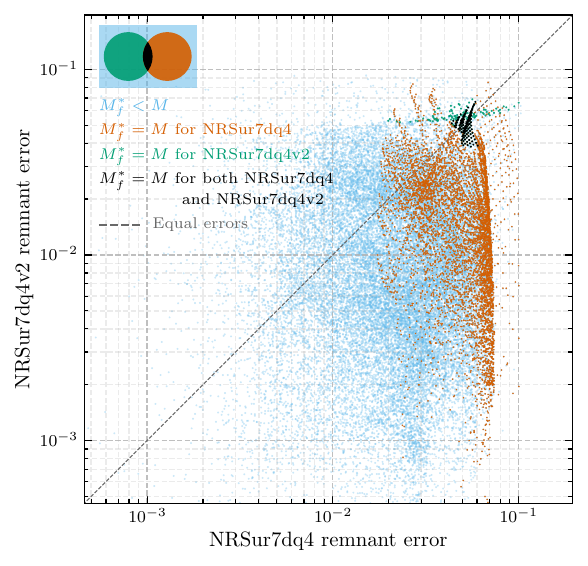}
\caption{
Error-error scatter plot of remnant errors $\epsilon_{22}$ for \newsur and \oldsur over the same aligned-spin samples as Fig.~\ref{fig:Remnant_Error_Aligned_Spins}. The QNM model is also identical to that of Sec.~\ref{subsec:rem_err}, except for an optimizer remnant mass domain of $M_f\in[0,M]$ instead of $M_f\in[0,1.2M]$. The points shown in blue represent parameter values where the optimizer converges to a remnant mass within the interior of the optimizer domain, i.e., $0 < M^*_f < M$, for both \newsur and \oldsur. The points where the optimizer result in a boundary remnant mass of $M^*_f=M$ for \newsur (\oldsur) are shaded in green (orange). Finally, the points where the optimizer result in a boundary remnant mass of $M^*_f=M$ for both \newsur and \oldsur are shaded in black. The relation between these colors is pictorially represented by the Venn diagram and its associated text in the legend. To better resolve the structure of the primary distribution, the plot is truncated to omit the sparse low-error tails of the distribution. It is evident from the plot that \oldsur is significantly more prone to convergence to the remnant mass boundary than \newsur. This railing against the boundary artificially suppresses the remnant errors for \oldsur and results in a dense front of orange points at $\epsilon_{22}\approx0.07$.
}
\label{fig:Remnant_Error_UnitMassBound}
\end{figure}

While exploring the source of such unphysical minima could motivate leaving the remnant mass domain unbounded, such an investigation is outside the scope of this work and we instead impose an upper bound on the remnant mass. While a limit of $M_f=M$ appears natural, minor waveform systematics in quantities such as the $(2,2)$ mode frequency can shift the best-fit remnant mass slightly beyond this theoretical limit. Although such a minimum is unphysical, it remains the closest to the true remnant parameters and accurately reflects the surrogate's fidelity. Restricting the optimizer to $M_f\in[0,M]$ can artificially constrain the search to a boundary value rather than a true interior minimum of the least-squares error function. This effect is visible in the remnant errors of \oldsur and \newsur shown in Fig.~\ref{fig:Remnant_Error_UnitMassBound}. The orange and green points in the left panel indicate where the optimizer terminates at the $M_f=M$ boundary for \oldsur and \newsur respectively. Notably, \oldsur is significantly more prone to this convergence to the boundary than \newsur. Imposing this bound forces the optimizer to choose a physical but non-optimal remnant mass, which artificially suppresses the reported errors for \oldsur. In contrast, \newsur remains largely unaffected. The dense front of orange points in Fig.~\ref{fig:Remnant_Error_UnitMassBound} is a direct artifact of the domain boundary, which diffuses in Fig.~\ref{fig:Remnant_Error_Aligned_Spins} when the upper bound of $M_f$ is increased beyond $M$.

The need to avoid highly unphysical minima while still allowing for the recovery of slightly unphysical results motivates the use of a relaxed but still bounded optimizer domain of $M_f\in[0,1.2M]$ for remnant error testing in the main text. We find that it drastically reduces the number of points where the optimizer converges to its remnant mass boundary for \oldsur and allows for the calculated remnant errors to be more representative of the true surrogate accuracy. We also note that this choice of upper bound is somewhat ad hoc, and further investigation into the source of unphysical minima in the least-squares error landscape could motivate a more informed choice of optimizer domain.

\section{Optimizations in \texttt{gwsurrogate}}
\label{app:gw_surrogate_optimizations}

\begin{figure*}[th]
\includegraphics[width=\textwidth]{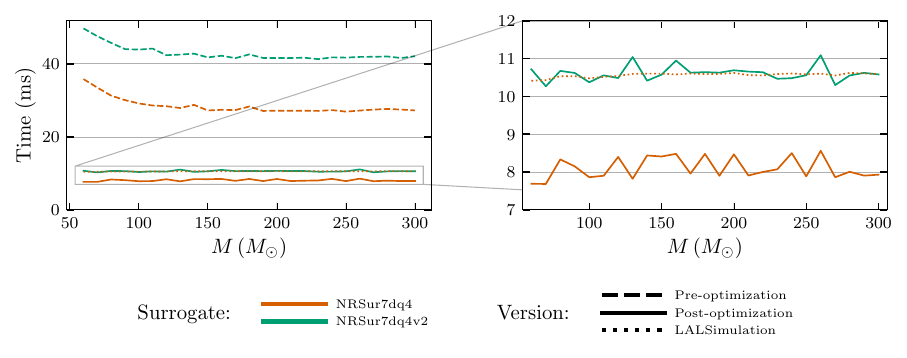}
\caption{Evaluation times as a function of total mass before and after \texttt{gwsurrogate} optimizations for a binary black hole system with mass ratio $q=2$ and dimensionless spins $\bm{\chi}_1=(0.2,0.3,-0.4)$ and $\bm{\chi}_2=(0.1,-0.3,0.2)$. \oldsur is shown in orange and \newsur in green, with dashed (solid) lines representing the pre-optimization (post-optimization) implementation in \texttt{gwsurrogate}, and dotted lines representing the implementation in \texttt{LALSimulation}. The markers represent the minimum evaluation time over 2,000 evaluations at each total mass. \textbf{Left:} Evaluation times for all implementations of both models. \textbf{Right:} A zoomed-in view of the \texttt{LALSimulation} and optimized \texttt{gwsurrogate} implementations.
}
\label{fig:timing_optimizations}
\end{figure*}

With \texttt{gwsurrogate} being a public interface to the surrogate models, we
recently performed a series of performance improvements to the code. None of
these change the output of \texttt{gwsurrogate}.
Profiling the code was crucial, but so too is understanding the computational
complexity of each stage of surrogate evaluation, to understand if a speed
bottleneck is fundamental or not.

Some of the optimizations are less trivial than others, and so we will provide a
varying degree of detail. The most significant optimizations are:
\begin{itemize}
\item moving \texttt{min/max} calls that checked bounds on the time array during
  time interpolation into the \texttt{C} code for the cubic spline. This provided the
  biggest improvement;
\item using an optimized cubic spline implementation that is able to natively
  handle both \texttt{double} and \texttt{complex<double>} types in a single
  pass over the data, as well as handling multiple data arrays in a single pass
  (i.e., different $\ell,m$ modes at once). This reduced memory allocation
  costs, various other overhead related to spline construction and evaluation,
  and allows for SIMD vectorization of the evaluation;
\item using an optimized \texttt{C} implementation to compute the Wigner D
  matrices, exploiting symmetry and hard-coding values for $\ell\le4$;
\item using tools like \texttt{np.asarray}, \texttt{np.empty}, avoiding
  \texttt{tuple([...])}, and performing in-place modifications to reduce memory
  allocations and default-initialization overhead;
\item moving the ODE integration and fit evaluation into \texttt{C}, in turn
  eliminating hundreds of calls from \texttt{C} back into python code;
\item fusing the waveform rotation and Wigner D matrix computation in a manner
  that allows for vectorization and looping over data in a single pass.
\end{itemize}
While in general these changes are not insightful or revolutionary,
together they amount to a respectable $\approx4\times$ speedup in evaluation of
the presented surrogate.

We note that these improvements were performed entirely using Claude Code. The
authors ran \texttt{gwsurrogate} with profiling enabled and then instructed
Claude Code to generate the appropriate changes. Unit tests for the components
being changed were crucial to get reliable results from Claude Code. Before this
work \texttt{gwsurrogate} primarily had regression tests where surrogate
evaluations are compared to known values. The functional workflow was as follows:
\begin{enumerate}
\item Run profiling and assess what part of the surrogate generation is
  dominating the runtime cost.
\item Instruct Claude Code to generate unit tests for that portion of the code,
  with guidance on edge cases that must be tested.
\item Instruct Claude Code to generate the code changes that we expect to
  reduce runtime, verifying them against the unit tests.
\item Verify runtime changes and possibly have Claude Code make further
  refinements on the current part of the code being optimized.
\item Verify the changes against the regression tests. If any fail, have Claude Code update the unit tests to catch the failure, and then revise the code optimizations to pass both test suites.
\end{enumerate}
This iterative work flow allowed the optimizations to be performed in under 12
hours of human time with minimal knowledge of the \texttt{gwsurrogate} code
base. 

The results of these optimizations are shown in Fig.~\ref{fig:timing_optimizations}, which plots the evaluation times (calculated as specified in Sec.~\ref{sec:discussion}) as a function of total mass for both \newsur and \oldsur, before and after optimization, alongside the \oldsur implementation in \texttt{LALSimulation}. The optimizations yield speedups of ${\approx}\,3$--$4.5\times$ for \oldsur and ${\approx}\,3.5$--$5\times$ for \newsur. Prior to optimization, the \texttt{gwsurrogate} evaluation times were heavily dominated by the cost of spline interpolation onto the target time grid, leading to slower performance at lower total masses. Post-optimization, the computational cost stems primarily from Wigner D matrix computation for waveform rotation, spline interpolation, and dynamics surrogate evaluation. This distributed workload results in a mostly constant evaluation time across all tested masses, explaining why the relative speedups are larger at lower total masses.

Finally, the optimized \texttt{gwsurrogate} implementations of \newsur and \oldsur require ${\approx}\,10.5$ ms and ${\approx}\,8$ ms to evaluate, respectively. By comparison, the \texttt{LALSimulation} implementation of \oldsur takes ${\approx}\,10.5$ ms. As a result, the \texttt{gwsurrogate} implementation now performs slightly faster than the \texttt{LALSimulation} implementation of \oldsur. Furthermore, \newsur is on par with the \texttt{LALSimulation} implementation of \oldsur, all while providing improved accuracy.

\bibliography{References}

\end{document}

%% file: macros/sxsid_macro.tex
\newcommand{\SXSIDs}{SXS:BBH:\{325-329, 331-609, 634-1100, 1346-1350, 1446-1447, 1452-1454, 1456-1459, 1461-1462, 1465-1474, 1478-1509, 1514-2082\}}

%% file: macros/sxsid_extrap_macro.tex
\newcommand{\SXSIDsExtrap}{SXS:BBH:\{4009-4020, 4025-4065, 4067-4166, 4285-4286, 4435-4435\}}

%% file: figs/Domain_Decomp_Diag.pdf_tex
\begingroup%
  \makeatletter%
  \providecommand\color[2][]{%
    \errmessage{(Inkscape) Color is used for the text in Inkscape, but the package 'color.sty' is not loaded}%
    \renewcommand\color[2][]{}%
  }%
  \providecommand\transparent[1]{%
    \errmessage{(Inkscape) Transparency is used (non-zero) for the text in Inkscape, but the package 'transparent.sty' is not loaded}%
    \renewcommand\transparent[1]{}%
  }%
  \providecommand\rotatebox[2]{#2}%
  \newcommand*\fsize{\dimexpr\f@size pt\relax}%
  \newcommand*\lineheight[1]{\fontsize{\fsize}{#1\fsize}\selectfont}%
  \ifx\svgwidth\undefined%
    \setlength{\unitlength}{502.04284187bp}%
    \ifx\svgscale\undefined%
      \relax%
    \else%
      \setlength{\unitlength}{\unitlength * \real{\svgscale}}%
    \fi%
  \else%
    \setlength{\unitlength}{\svgwidth}%
  \fi%
  \global\let\svgwidth\undefined%
  \global\let\svgscale\undefined%
  \makeatother%
  \begin{picture}(1,0.96369466)%
    \lineheight{1}%
    \setlength\tabcolsep{0pt}%
    \put(0,0){\includegraphics[width=\unitlength,page=1]{Domain_Decomp_Diag.pdf}}%
    \put(0.02418482,0.2535237){\color[rgb]{0,0,0}\makebox(0,0)[lt]{\lineheight{1.25}\smash{\begin{tabular}[t]{l}$\Scale[1.2]{h_S(t,\bm{\lambda})}$\end{tabular}}}}%
    \put(0,0){\includegraphics[width=\unitlength,page=2]{Domain_Decomp_Diag.pdf}}%
    \put(0.46074927,0.53498547){\color[rgb]{0.39215686,0.56078431,1}\makebox(0,0)[lt]{\lineheight{1.25}\smash{\begin{tabular}[t]{l}$\Scale[1.2]{\BMask{\I}(t)}$\end{tabular}}}}%
    \put(0.46074927,0.0722224){\color[rgb]{0.8627451,0.14901961,0.49803922}\makebox(0,0)[lt]{\lineheight{1.25}\smash{\begin{tabular}[t]{l}$\Scale[1.2]{\BMask{\R}(t)}$\end{tabular}}}}%
    \put(0.45516924,0.34793812){\color[rgb]{0.39215686,0.56078431,1}\makebox(0,0)[lt]{\lineheight{1.25}\smash{\begin{tabular}[t]{l}$\Scale[1.2]{\I{h_S}(t,\bm{\lambda})}$\end{tabular}}}}%
    \put(0.45583747,0.17426319){\color[rgb]{0.8627451,0.14901961,0.49803922}\makebox(0,0)[lt]{\lineheight{1.25}\smash{\begin{tabular}[t]{l}$\Scale[1.2]{\R{h_S}(t,\bm{\lambda})}$\end{tabular}}}}%
    \put(0,0){\includegraphics[width=\unitlength,page=3]{Domain_Decomp_Diag.pdf}}%
    \put(0.45789923,0.78371578){\color[rgb]{0.39215686,0.56078431,1}\makebox(0,0)[lt]{\lineheight{1.25}\smash{\begin{tabular}[t]{l}$\Scale[1.2]{\I{h}(t,\bm{\lambda})}$\end{tabular}}}}%
    \put(0.45734452,0.64207495){\color[rgb]{0.8627451,0.14901961,0.49803922}\makebox(0,0)[lt]{\lineheight{1.25}\smash{\begin{tabular}[t]{l}$\Scale[1.2]{\R{h}(t,\bm{\lambda})}$\end{tabular}}}}%
    \put(0,0){\includegraphics[width=\unitlength,page=4]{Domain_Decomp_Diag.pdf}}%
    \put(0.02466634,0.70248175){\color[rgb]{0,0,0}\makebox(0,0)[lt]{\lineheight{1.25}\smash{\begin{tabular}[t]{l}$\Scale[1.2]{h(t,\bm{\lambda})}$\end{tabular}}}}%
    \put(0.02119668,0.88543548){\color[rgb]{0.39215686,0.56078431,1}\makebox(0,0)[lt]{\lineheight{1.25}\smash{\begin{tabular}[t]{l}$\Scale[1.2]{\DMask{\I}(t)}$\end{tabular}}}}%
    \put(0.0212148,0.59190732){\color[rgb]{0.8627451,0.14901961,0.49803922}\makebox(0,0)[lt]{\lineheight{1.25}\smash{\begin{tabular}[t]{l}$\Scale[1.2]{\DMask{\R}(t)}$\end{tabular}}}}%
    \put(0,0){\includegraphics[width=\unitlength,page=5]{Domain_Decomp_Diag.pdf}}%
  \end{picture}%
\endgroup%

%% file: paper.bbl
\begin{thebibliography}{97}%
\makeatletter
\providecommand \@ifxundefined [1]{%
 \@ifx{#1\undefined}
}%
\providecommand \@ifnum [1]{%
 \ifnum #1\expandafter \@firstoftwo
 \else \expandafter \@secondoftwo
 \fi
}%
\providecommand \@ifx [1]{%
 \ifx #1\expandafter \@firstoftwo
 \else \expandafter \@secondoftwo
 \fi
}%
\providecommand \natexlab [1]{#1}%
\providecommand \enquote  [1]{``#1''}%
\providecommand \bibnamefont  [1]{#1}%
\providecommand \bibfnamefont [1]{#1}%
\providecommand \citenamefont [1]{#1}%
\providecommand \href@noop [0]{\@secondoftwo}%
\providecommand \href [0]{\begingroup \@sanitize@url \@href}%
\providecommand \@href[1]{\@@startlink{#1}\@@href}%
\providecommand \@@href[1]{\endgroup#1\@@endlink}%
\providecommand \@sanitize@url [0]{\catcode `\\12\catcode `\$12\catcode
  `\&12\catcode `\#12\catcode `\^12\catcode `\_12\catcode `\%12\relax}%
\providecommand \@@startlink[1]{}%
\providecommand \@@endlink[0]{}%
\providecommand \url  [0]{\begingroup\@sanitize@url \@url }%
\providecommand \@url [1]{\endgroup\@href {#1}{\urlprefix }}%
\providecommand \urlprefix  [0]{URL }%
\providecommand \Eprint [0]{\href }%
\providecommand \doibase [0]{http://dx.doi.org/}%
\providecommand \selectlanguage [0]{\@gobble}%
\providecommand \bibinfo  [0]{\@secondoftwo}%
\providecommand \bibfield  [0]{\@secondoftwo}%
\providecommand \translation [1]{[#1]}%
\providecommand \BibitemOpen [0]{}%
\providecommand \bibitemStop [0]{}%
\providecommand \bibitemNoStop [0]{.\EOS\space}%
\providecommand \EOS [0]{\spacefactor3000\relax}%
\providecommand \BibitemShut  [1]{\csname bibitem#1\endcsname}%
\let\auto@bib@innerbib\@empty
\bibitem [{\citenamefont {Aasi}\ \emph {et~al.}(2015)\citenamefont {Aasi} \emph
  {et~al.}}]{TheLIGOScientific:2014jea}%
  \BibitemOpen
  \bibfield  {author} {\bibinfo {author} {\bibfnamefont {J.}~\bibnamefont
  {Aasi}} \emph {et~al.} (\bibinfo {collaboration} {LIGO Scientific}),\
  }\bibfield  {title} {\enquote {\bibinfo {title} {{Advanced LIGO}},}\ }\href
  {\doibase 10.1088/0264-9381/32/7/074001} {\bibfield  {journal} {\bibinfo
  {journal} {Class. Quant. Grav.}\ }\textbf {\bibinfo {volume} {32}},\ \bibinfo
  {pages} {074001} (\bibinfo {year} {2015})},\ \Eprint
  {http://arxiv.org/abs/1411.4547} {arXiv:1411.4547 [gr-qc]} \BibitemShut
  {NoStop}%
\bibitem [{\citenamefont {Acernese}\ \emph {et~al.}(2015)\citenamefont
  {Acernese} \emph {et~al.}}]{TheVirgo:2014hva}%
  \BibitemOpen
  \bibfield  {author} {\bibinfo {author} {\bibfnamefont {F.}~\bibnamefont
  {Acernese}} \emph {et~al.} (\bibinfo {collaboration} {Virgo}),\ }\bibfield
  {title} {\enquote {\bibinfo {title} {{Advanced Virgo: a second-generation
  interferometric gravitational wave detector}},}\ }\href {\doibase
  10.1088/0264-9381/32/2/024001} {\bibfield  {journal} {\bibinfo  {journal}
  {Class. Quant. Grav.}\ }\textbf {\bibinfo {volume} {32}},\ \bibinfo {pages}
  {024001} (\bibinfo {year} {2015})},\ \Eprint {http://arxiv.org/abs/1408.3978}
  {arXiv:1408.3978 [gr-qc]} \BibitemShut {NoStop}%
\bibitem [{\citenamefont {Akutsu}\ \emph {et~al.}(2021)\citenamefont {Akutsu}
  \emph {et~al.}}]{KAGRA:2020tym}%
  \BibitemOpen
  \bibfield  {author} {\bibinfo {author} {\bibfnamefont {T.}~\bibnamefont
  {Akutsu}} \emph {et~al.} (\bibinfo {collaboration} {KAGRA}),\ }\bibfield
  {title} {\enquote {\bibinfo {title} {{Overview of KAGRA: Detector design and
  construction history}},}\ }\href {\doibase 10.1093/ptep/ptaa125} {\bibfield
  {journal} {\bibinfo  {journal} {PTEP}\ }\textbf {\bibinfo {volume} {2021}},\
  \bibinfo {pages} {05A101} (\bibinfo {year} {2021})},\ \Eprint
  {http://arxiv.org/abs/2005.05574} {arXiv:2005.05574 [physics.ins-det]}
  \BibitemShut {NoStop}%
\bibitem [{\citenamefont {Abbott}\ \emph
  {et~al.}(2016{\natexlab{a}})\citenamefont {Abbott} \emph
  {et~al.}}]{Abbott:2016blz}%
  \BibitemOpen
  \bibfield  {author} {\bibinfo {author} {\bibfnamefont {B.~P.}\ \bibnamefont
  {Abbott}} \emph {et~al.} (\bibinfo {collaboration} {LIGO Scientific,
  Virgo}),\ }\bibfield  {title} {\enquote {\bibinfo {title} {{Observation of
  Gravitational Waves from a Binary Black Hole Merger}},}\ }\href {\doibase
  10.1103/PhysRevLett.116.061102} {\bibfield  {journal} {\bibinfo  {journal}
  {Phys. Rev. Lett.}\ }\textbf {\bibinfo {volume} {116}},\ \bibinfo {pages}
  {061102} (\bibinfo {year} {2016}{\natexlab{a}})},\ \Eprint
  {http://arxiv.org/abs/1602.03837} {arXiv:1602.03837 [gr-qc]} \BibitemShut
  {NoStop}%
\bibitem [{\citenamefont {Abbott}\ \emph
  {et~al.}(2016{\natexlab{b}})\citenamefont {Abbott} \emph
  {et~al.}}]{Abbott:2016nmj}%
  \BibitemOpen
  \bibfield  {author} {\bibinfo {author} {\bibfnamefont {B.~P.}\ \bibnamefont
  {Abbott}} \emph {et~al.} (\bibinfo {collaboration} {LIGO Scientific,
  Virgo}),\ }\bibfield  {title} {\enquote {\bibinfo {title} {{GW151226:
  Observation of Gravitational Waves from a 22-Solar-Mass Binary Black Hole
  Coalescence}},}\ }\href {\doibase 10.1103/PhysRevLett.116.241103} {\bibfield
  {journal} {\bibinfo  {journal} {Phys. Rev. Lett.}\ }\textbf {\bibinfo
  {volume} {116}},\ \bibinfo {pages} {241103} (\bibinfo {year}
  {2016}{\natexlab{b}})},\ \Eprint {http://arxiv.org/abs/1606.04855}
  {arXiv:1606.04855 [gr-qc]} \BibitemShut {NoStop}%
\bibitem [{\citenamefont {Abbott}\ \emph {et~al.}(2019)\citenamefont {Abbott}
  \emph {et~al.}}]{LIGOScientific:2018mvr}%
  \BibitemOpen
  \bibfield  {author} {\bibinfo {author} {\bibfnamefont {B.~P.}\ \bibnamefont
  {Abbott}} \emph {et~al.} (\bibinfo {collaboration} {LIGO Scientific,
  Virgo}),\ }\bibfield  {title} {\enquote {\bibinfo {title} {{GWTC-1: A
  Gravitational-Wave Transient Catalog of Compact Binary Mergers Observed by
  LIGO and Virgo during the First and Second Observing Runs}},}\ }\href
  {\doibase 10.1103/PhysRevX.9.031040} {\bibfield  {journal} {\bibinfo
  {journal} {Phys. Rev.}\ }\textbf {\bibinfo {volume} {X9}},\ \bibinfo {pages}
  {031040} (\bibinfo {year} {2019})},\ \Eprint
  {http://arxiv.org/abs/1811.12907} {arXiv:1811.12907 [astro-ph.HE]}
  \BibitemShut {NoStop}%
\bibitem [{\citenamefont {Abbott}\ \emph {et~al.}(2021)\citenamefont {Abbott}
  \emph {et~al.}}]{LIGOScientific:2020ibl}%
  \BibitemOpen
  \bibfield  {author} {\bibinfo {author} {\bibfnamefont {R.}~\bibnamefont
  {Abbott}} \emph {et~al.} (\bibinfo {collaboration} {LIGO Scientific,
  Virgo}),\ }\bibfield  {title} {\enquote {\bibinfo {title} {{GWTC-2: Compact
  Binary Coalescences Observed by LIGO and Virgo During the First Half of the
  Third Observing Run}},}\ }\href {\doibase 10.1103/PhysRevX.11.021053}
  {\bibfield  {journal} {\bibinfo  {journal} {Phys. Rev. X}\ }\textbf {\bibinfo
  {volume} {11}},\ \bibinfo {pages} {021053} (\bibinfo {year} {2021})},\
  \Eprint {http://arxiv.org/abs/2010.14527} {arXiv:2010.14527 [gr-qc]}
  \BibitemShut {NoStop}%
\bibitem [{\citenamefont {Abbott}\ \emph {et~al.}(2024)\citenamefont {Abbott}
  \emph {et~al.}}]{LIGOScientific:2021usb}%
  \BibitemOpen
  \bibfield  {author} {\bibinfo {author} {\bibfnamefont {R.}~\bibnamefont
  {Abbott}} \emph {et~al.} (\bibinfo {collaboration} {LIGO Scientific,
  VIRGO}),\ }\bibfield  {title} {\enquote {\bibinfo {title} {{GWTC-2.1: Deep
  extended catalog of compact binary coalescences observed by LIGO and Virgo
  during the first half of the third observing run}},}\ }\href {\doibase
  10.1103/PhysRevD.109.022001} {\bibfield  {journal} {\bibinfo  {journal}
  {Phys. Rev. D}\ }\textbf {\bibinfo {volume} {109}},\ \bibinfo {pages}
  {022001} (\bibinfo {year} {2024})},\ \Eprint
  {http://arxiv.org/abs/2108.01045} {arXiv:2108.01045 [gr-qc]} \BibitemShut
  {NoStop}%
\bibitem [{\citenamefont {Abbott}\ \emph {et~al.}(2023)\citenamefont {Abbott}
  \emph {et~al.}}]{KAGRA:2021vkt}%
  \BibitemOpen
  \bibfield  {author} {\bibinfo {author} {\bibfnamefont {R.}~\bibnamefont
  {Abbott}} \emph {et~al.} (\bibinfo {collaboration} {KAGRA, VIRGO, LIGO
  Scientific}),\ }\bibfield  {title} {\enquote {\bibinfo {title} {{GWTC-3:
  Compact Binary Coalescences Observed by LIGO and Virgo during the Second Part
  of the Third Observing Run}},}\ }\href {\doibase 10.1103/PhysRevX.13.041039}
  {\bibfield  {journal} {\bibinfo  {journal} {Phys. Rev. X}\ }\textbf {\bibinfo
  {volume} {13}},\ \bibinfo {pages} {041039} (\bibinfo {year} {2023})},\
  \Eprint {http://arxiv.org/abs/2111.03606} {arXiv:2111.03606 [gr-qc]}
  \BibitemShut {NoStop}%
\bibitem [{\citenamefont {Abac}\ \emph
  {et~al.}(2025{\natexlab{a}})\citenamefont {Abac} \emph
  {et~al.}}]{LIGOScientific:2025slb}%
  \BibitemOpen
  \bibfield  {author} {\bibinfo {author} {\bibfnamefont {A.~G.}\ \bibnamefont
  {Abac}} \emph {et~al.} (\bibinfo {collaboration} {LIGO Scientific, VIRGO,
  KAGRA}),\ }\bibfield  {title} {\enquote {\bibinfo {title} {{GWTC-4.0:
  Updating the Gravitational-Wave Transient Catalog with Observations from the
  First Part of the Fourth LIGO-Virgo-KAGRA Observing Run}},}\ }\href@noop {}
  {\  (\bibinfo {year} {2025}{\natexlab{a}})},\ \Eprint
  {http://arxiv.org/abs/2508.18082} {arXiv:2508.18082 [gr-qc]} \BibitemShut
  {NoStop}%
\bibitem [{\citenamefont {Abac}\ \emph
  {et~al.}(2025{\natexlab{b}})\citenamefont {Abac} \emph
  {et~al.}}]{LIGOScientific:2025snk}%
  \BibitemOpen
  \bibfield  {author} {\bibinfo {author} {\bibfnamefont {A.~G.}\ \bibnamefont
  {Abac}} \emph {et~al.} (\bibinfo {collaboration} {LIGO Scientific, VIRGO,
  KAGRA}),\ }\bibfield  {title} {\enquote {\bibinfo {title} {{Open Data from
  LIGO, Virgo, and KAGRA through the First Part of the Fourth Observing
  Run}},}\ }\href@noop {} {\  (\bibinfo {year} {2025}{\natexlab{b}})},\ \Eprint
  {http://arxiv.org/abs/2508.18079} {arXiv:2508.18079 [gr-qc]} \BibitemShut
  {NoStop}%
\bibitem [{\citenamefont {Abbott}\ \emph {et~al.}(2017)\citenamefont {Abbott}
  \emph {et~al.}}]{LIGOScientific:2016wof}%
  \BibitemOpen
  \bibfield  {author} {\bibinfo {author} {\bibfnamefont {Benjamin~P}\
  \bibnamefont {Abbott}} \emph {et~al.} (\bibinfo {collaboration} {LIGO
  Scientific}),\ }\bibfield  {title} {\enquote {\bibinfo {title} {{Exploring
  the Sensitivity of Next Generation Gravitational Wave Detectors}},}\ }\href
  {\doibase 10.1088/1361-6382/aa51f4} {\bibfield  {journal} {\bibinfo
  {journal} {Class. Quant. Grav.}\ }\textbf {\bibinfo {volume} {34}},\ \bibinfo
  {pages} {044001} (\bibinfo {year} {2017})},\ \Eprint
  {http://arxiv.org/abs/1607.08697} {arXiv:1607.08697 [astro-ph.IM]}
  \BibitemShut {NoStop}%
\bibitem [{\citenamefont {Punturo}\ \emph {et~al.}(2010)\citenamefont {Punturo}
  \emph {et~al.}}]{Punturo:2010zz}%
  \BibitemOpen
  \bibfield  {author} {\bibinfo {author} {\bibfnamefont {M.}~\bibnamefont
  {Punturo}} \emph {et~al.},\ }\bibfield  {title} {\enquote {\bibinfo {title}
  {{The Einstein Telescope: A third-generation gravitational wave
  observatory}},}\ }\bibfield  {booktitle} {\emph {\bibinfo {booktitle}
  {{Proceedings, 14th Workshop on Gravitational wave data analysis (GWDAW-14):
  Rome, Italy, January 26-29, 2010}}},\ }\href {\doibase
  10.1088/0264-9381/27/19/194002} {\bibfield  {journal} {\bibinfo  {journal}
  {Class. Quant. Grav.}\ }\textbf {\bibinfo {volume} {27}},\ \bibinfo {pages}
  {194002} (\bibinfo {year} {2010})}\BibitemShut {NoStop}%
\bibitem [{\citenamefont {Maggiore}\ \emph {et~al.}(2020)\citenamefont
  {Maggiore} \emph {et~al.}}]{Maggiore:2019uih}%
  \BibitemOpen
  \bibfield  {author} {\bibinfo {author} {\bibfnamefont {Michele}\ \bibnamefont
  {Maggiore}} \emph {et~al.},\ }\bibfield  {title} {\enquote {\bibinfo {title}
  {{Science Case for the Einstein Telescope}},}\ }\href {\doibase
  10.1088/1475-7516/2020/03/050} {\bibfield  {journal} {\bibinfo  {journal}
  {JCAP}\ }\textbf {\bibinfo {volume} {03}},\ \bibinfo {pages} {050} (\bibinfo
  {year} {2020})},\ \Eprint {http://arxiv.org/abs/1912.02622} {arXiv:1912.02622
  [astro-ph.CO]} \BibitemShut {NoStop}%
\bibitem [{\citenamefont {Reitze}\ \emph {et~al.}(2019)\citenamefont {Reitze}
  \emph {et~al.}}]{Reitze:2019iox}%
  \BibitemOpen
  \bibfield  {author} {\bibinfo {author} {\bibfnamefont {David}\ \bibnamefont
  {Reitze}} \emph {et~al.},\ }\bibfield  {title} {\enquote {\bibinfo {title}
  {{Cosmic Explorer: The U.S. Contribution to Gravitational-Wave Astronomy
  beyond LIGO}},}\ }\href@noop {} {\bibfield  {journal} {\bibinfo  {journal}
  {Bull. Am. Astron. Soc.}\ }\textbf {\bibinfo {volume} {51}},\ \bibinfo
  {pages} {035} (\bibinfo {year} {2019})},\ \Eprint
  {http://arxiv.org/abs/1907.04833} {arXiv:1907.04833 [astro-ph.IM]}
  \BibitemShut {NoStop}%
\bibitem [{\citenamefont {Amaro-Seoane}\ \emph {et~al.}(2017)\citenamefont
  {Amaro-Seoane} \emph {et~al.}}]{LISA:2017pwj}%
  \BibitemOpen
  \bibfield  {author} {\bibinfo {author} {\bibfnamefont {Pau}\ \bibnamefont
  {Amaro-Seoane}} \emph {et~al.} (\bibinfo {collaboration} {LISA}),\ }\bibfield
   {title} {\enquote {\bibinfo {title} {{Laser Interferometer Space
  Antenna}},}\ }\href@noop {} {\  (\bibinfo {year} {2017})},\ \Eprint
  {http://arxiv.org/abs/1702.00786} {arXiv:1702.00786 [astro-ph.IM]}
  \BibitemShut {NoStop}%
\bibitem [{\citenamefont {Flaminio}(2020)}]{Flaminio:2020lqk}%
  \BibitemOpen
  \bibfield  {author} {\bibinfo {author} {\bibfnamefont {Raffaele}\
  \bibnamefont {Flaminio}},\ }\bibfield  {title} {\enquote {\bibinfo {title}
  {{Status and plans of the Virgo gravitational wave detector}},}\ }\href
  {\doibase 10.1117/12.2565418} {\bibfield  {journal} {\bibinfo  {journal}
  {Proc. SPIE Int. Soc. Opt. Eng.}\ }\textbf {\bibinfo {volume} {11445}},\
  \bibinfo {pages} {1144511} (\bibinfo {year} {2020})}\BibitemShut {NoStop}%
\bibitem [{\citenamefont {Fritschel}\ \emph {et~al.}(2023)\citenamefont
  {Fritschel}, \citenamefont {Kuns}, \citenamefont {Driggers}, \citenamefont
  {Effler}, \citenamefont {Lantz}, \citenamefont {Ottaway}, \citenamefont
  {Ballmer}, \citenamefont {Dooley}, \citenamefont {Adhikari}, \citenamefont
  {Evans}, \citenamefont {Farr}, \citenamefont {Gonzalez}, \citenamefont
  {Schmidt},\ and\ \citenamefont {Raja}}]{Fritschel:2023}%
  \BibitemOpen
  \bibfield  {author} {\bibinfo {author} {\bibfnamefont {Peter}\ \bibnamefont
  {Fritschel}}, \bibinfo {author} {\bibfnamefont {Kevin}\ \bibnamefont {Kuns}},
  \bibinfo {author} {\bibfnamefont {Jenne}\ \bibnamefont {Driggers}}, \bibinfo
  {author} {\bibfnamefont {Anamaria}\ \bibnamefont {Effler}}, \bibinfo {author}
  {\bibfnamefont {Brian}\ \bibnamefont {Lantz}}, \bibinfo {author}
  {\bibfnamefont {David}\ \bibnamefont {Ottaway}}, \bibinfo {author}
  {\bibfnamefont {Stefan}\ \bibnamefont {Ballmer}}, \bibinfo {author}
  {\bibfnamefont {Katherine}\ \bibnamefont {Dooley}}, \bibinfo {author}
  {\bibfnamefont {Rana}\ \bibnamefont {Adhikari}}, \bibinfo {author}
  {\bibfnamefont {Matthew}\ \bibnamefont {Evans}}, \bibinfo {author}
  {\bibfnamefont {Benjamin}\ \bibnamefont {Farr}}, \bibinfo {author}
  {\bibfnamefont {Gabriela}\ \bibnamefont {Gonzalez}}, \bibinfo {author}
  {\bibfnamefont {Patricia}\ \bibnamefont {Schmidt}}, \ and\ \bibinfo {author}
  {\bibfnamefont {Sendhill}\ \bibnamefont {Raja}} (\bibinfo {collaboration}
  {{LIGO} Collaboration}),\ }\href {https://dcc.ligo.org/LIGO-T2200287/public}
  {\enquote {\bibinfo {title} {Report of the lsc post-o5 study group},}\ }
  (\bibinfo {year} {2023}),\ \bibinfo {note} {{LIGO} Document
  T2200287-v3}\BibitemShut {NoStop}%
\bibitem [{\citenamefont {Luo}\ \emph {et~al.}(2016)\citenamefont {Luo} \emph
  {et~al.}}]{TianQin:2015yph}%
  \BibitemOpen
  \bibfield  {author} {\bibinfo {author} {\bibfnamefont {Jun}\ \bibnamefont
  {Luo}} \emph {et~al.} (\bibinfo {collaboration} {TianQin}),\ }\bibfield
  {title} {\enquote {\bibinfo {title} {{TianQin: a space-borne gravitational
  wave detector}},}\ }\href {\doibase 10.1088/0264-9381/33/3/035010} {\bibfield
   {journal} {\bibinfo  {journal} {Class. Quant. Grav.}\ }\textbf {\bibinfo
  {volume} {33}},\ \bibinfo {pages} {035010} (\bibinfo {year} {2016})},\
  \Eprint {http://arxiv.org/abs/1512.02076} {arXiv:1512.02076 [astro-ph.IM]}
  \BibitemShut {NoStop}%
\bibitem [{\citenamefont {Wu}\ \emph {et~al.}(2021)\citenamefont {Wu} \emph
  {et~al.}}]{TaijiScientific:2021qgx}%
  \BibitemOpen
  \bibfield  {author} {\bibinfo {author} {\bibfnamefont {Yue-Liang}\
  \bibnamefont {Wu}} \emph {et~al.} (\bibinfo {collaboration} {Taiji
  Scientific}),\ }\bibfield  {title} {\enquote {\bibinfo {title}
  {{China{\textquoteright}s first step towards probing the expanding universe
  and the nature of gravity using a space borne gravitational wave antenna}},}\
  }\href {\doibase 10.1038/s42005-021-00529-z} {\bibfield  {journal} {\bibinfo
  {journal} {Commun. Phys.}\ }\textbf {\bibinfo {volume} {4}},\ \bibinfo
  {pages} {34} (\bibinfo {year} {2021})}\BibitemShut {NoStop}%
\bibitem [{\citenamefont {Kawamura}\ \emph {et~al.}(2006)\citenamefont
  {Kawamura} \emph {et~al.}}]{Kawamura:2006up}%
  \BibitemOpen
  \bibfield  {author} {\bibinfo {author} {\bibfnamefont {S.}~\bibnamefont
  {Kawamura}} \emph {et~al.},\ }\bibfield  {title} {\enquote {\bibinfo {title}
  {{The Japanese space gravitational wave antenna DECIGO}},}\ }\href {\doibase
  10.1088/0264-9381/23/8/S17} {\bibfield  {journal} {\bibinfo  {journal}
  {Class. Quant. Grav.}\ }\textbf {\bibinfo {volume} {23}},\ \bibinfo {pages}
  {S125--S132} (\bibinfo {year} {2006})}\BibitemShut {NoStop}%
\bibitem [{\citenamefont {Ajith}\ \emph {et~al.}(2025)\citenamefont {Ajith}
  \emph {et~al.}}]{Ajith:2024mie}%
  \BibitemOpen
  \bibfield  {author} {\bibinfo {author} {\bibfnamefont {Parameswaran}\
  \bibnamefont {Ajith}} \emph {et~al.},\ }\bibfield  {title} {\enquote
  {\bibinfo {title} {{The Lunar Gravitational-wave Antenna: mission studies and
  science case}},}\ }\href {\doibase 10.1088/1475-7516/2025/01/108} {\bibfield
  {journal} {\bibinfo  {journal} {JCAP}\ }\textbf {\bibinfo {volume} {01}},\
  \bibinfo {pages} {108} (\bibinfo {year} {2025})},\ \Eprint
  {http://arxiv.org/abs/2404.09181} {arXiv:2404.09181 [gr-qc]} \BibitemShut
  {NoStop}%
\bibitem [{\citenamefont {Buonanno}\ and\ \citenamefont
  {Damour}(1999)}]{Buonanno:1998gg}%
  \BibitemOpen
  \bibfield  {author} {\bibinfo {author} {\bibfnamefont {A.}~\bibnamefont
  {Buonanno}}\ and\ \bibinfo {author} {\bibfnamefont {T.}~\bibnamefont
  {Damour}},\ }\bibfield  {title} {\enquote {\bibinfo {title} {{Effective
  one-body approach to general relativistic two-body dynamics}},}\ }\href
  {\doibase 10.1103/PhysRevD.59.084006} {\bibfield  {journal} {\bibinfo
  {journal} {Phys. Rev.}\ }\textbf {\bibinfo {volume} {D59}},\ \bibinfo {pages}
  {084006} (\bibinfo {year} {1999})},\ \Eprint
  {http://arxiv.org/abs/gr-qc/9811091} {arXiv:gr-qc/9811091 [gr-qc]}
  \BibitemShut {NoStop}%
\bibitem [{\citenamefont {Buonanno}\ and\ \citenamefont
  {Damour}(2000)}]{Buonanno:2000ef}%
  \BibitemOpen
  \bibfield  {author} {\bibinfo {author} {\bibfnamefont {Alessandra}\
  \bibnamefont {Buonanno}}\ and\ \bibinfo {author} {\bibfnamefont {Thibault}\
  \bibnamefont {Damour}},\ }\bibfield  {title} {\enquote {\bibinfo {title}
  {{Transition from inspiral to plunge in binary black hole coalescences}},}\
  }\href {\doibase 10.1103/PhysRevD.62.064015} {\bibfield  {journal} {\bibinfo
  {journal} {Phys. Rev.}\ }\textbf {\bibinfo {volume} {D62}},\ \bibinfo {pages}
  {064015} (\bibinfo {year} {2000})},\ \Eprint
  {http://arxiv.org/abs/gr-qc/0001013} {arXiv:gr-qc/0001013 [gr-qc]}
  \BibitemShut {NoStop}%
\bibitem [{\citenamefont {Buonanno}\ \emph {et~al.}(2006)\citenamefont
  {Buonanno}, \citenamefont {Chen},\ and\ \citenamefont
  {Damour}}]{Buonanno:2005xu}%
  \BibitemOpen
  \bibfield  {author} {\bibinfo {author} {\bibfnamefont {Alessandra}\
  \bibnamefont {Buonanno}}, \bibinfo {author} {\bibfnamefont {Yanbei}\
  \bibnamefont {Chen}}, \ and\ \bibinfo {author} {\bibfnamefont {Thibault}\
  \bibnamefont {Damour}},\ }\bibfield  {title} {\enquote {\bibinfo {title}
  {{Transition from inspiral to plunge in precessing binaries of spinning black
  holes}},}\ }\href {\doibase 10.1103/PhysRevD.74.104005} {\bibfield  {journal}
  {\bibinfo  {journal} {Phys. Rev. D}\ }\textbf {\bibinfo {volume} {74}},\
  \bibinfo {pages} {104005} (\bibinfo {year} {2006})},\ \Eprint
  {http://arxiv.org/abs/gr-qc/0508067} {arXiv:gr-qc/0508067} \BibitemShut
  {NoStop}%
\bibitem [{\citenamefont {Damour}\ \emph {et~al.}(2000)\citenamefont {Damour},
  \citenamefont {Jaranowski},\ and\ \citenamefont {Schaefer}}]{Damour:2000we}%
  \BibitemOpen
  \bibfield  {author} {\bibinfo {author} {\bibfnamefont {Thibault}\
  \bibnamefont {Damour}}, \bibinfo {author} {\bibfnamefont {Piotr}\
  \bibnamefont {Jaranowski}}, \ and\ \bibinfo {author} {\bibfnamefont
  {Gerhard}\ \bibnamefont {Schaefer}},\ }\bibfield  {title} {\enquote {\bibinfo
  {title} {{On the determination of the last stable orbit for circular general
  relativistic binaries at the third postNewtonian approximation}},}\ }\href
  {\doibase 10.1103/PhysRevD.62.084011} {\bibfield  {journal} {\bibinfo
  {journal} {Phys. Rev. D}\ }\textbf {\bibinfo {volume} {62}},\ \bibinfo
  {pages} {084011} (\bibinfo {year} {2000})},\ \Eprint
  {http://arxiv.org/abs/gr-qc/0005034} {arXiv:gr-qc/0005034} \BibitemShut
  {NoStop}%
\bibitem [{\citenamefont {Damour}(2001)}]{Damour:2001tu}%
  \BibitemOpen
  \bibfield  {author} {\bibinfo {author} {\bibfnamefont {Thibault}\
  \bibnamefont {Damour}},\ }\bibfield  {title} {\enquote {\bibinfo {title}
  {{Coalescence of two spinning black holes: an effective one-body
  approach}},}\ }\href {\doibase 10.1103/PhysRevD.64.124013} {\bibfield
  {journal} {\bibinfo  {journal} {Phys. Rev. D}\ }\textbf {\bibinfo {volume}
  {64}},\ \bibinfo {pages} {124013} (\bibinfo {year} {2001})},\ \Eprint
  {http://arxiv.org/abs/gr-qc/0103018} {arXiv:gr-qc/0103018} \BibitemShut
  {NoStop}%
\bibitem [{\citenamefont {Buonanno}\ and\ \citenamefont
  {Sathyaprakash}(2014)}]{Buonanno:2014aza}%
  \BibitemOpen
  \bibfield  {author} {\bibinfo {author} {\bibfnamefont {Alessandra}\
  \bibnamefont {Buonanno}}\ and\ \bibinfo {author} {\bibfnamefont {B.~S.}\
  \bibnamefont {Sathyaprakash}},\ }\enquote {\bibinfo {title} {{Sources of
  Gravitational Waves: Theory and Observations}},}\ \ (\bibinfo {year} {2014})\
  \Eprint {http://arxiv.org/abs/1410.7832} {arXiv:1410.7832 [gr-qc]}
  \BibitemShut {NoStop}%
\bibitem [{\citenamefont {Damour}(2008)}]{Damour:2008yg}%
  \BibitemOpen
  \bibfield  {author} {\bibinfo {author} {\bibfnamefont {Thibault}\
  \bibnamefont {Damour}},\ }\bibfield  {title} {\enquote {\bibinfo {title}
  {{Introductory lectures on the Effective One Body formalism}},}\ }\href
  {\doibase 10.1142/S0217751X08039992} {\bibfield  {journal} {\bibinfo
  {journal} {Int. J. Mod. Phys. A}\ }\textbf {\bibinfo {volume} {23}},\
  \bibinfo {pages} {1130--1148} (\bibinfo {year} {2008})},\ \Eprint
  {http://arxiv.org/abs/0802.4047} {arXiv:0802.4047 [gr-qc]} \BibitemShut
  {NoStop}%
\bibitem [{\citenamefont {Taracchini}\ \emph {et~al.}(2014)\citenamefont
  {Taracchini} \emph {et~al.}}]{Taracchini:2013rva}%
  \BibitemOpen
  \bibfield  {author} {\bibinfo {author} {\bibfnamefont {Andrea}\ \bibnamefont
  {Taracchini}} \emph {et~al.},\ }\bibfield  {title} {\enquote {\bibinfo
  {title} {{Effective-one-body model for black-hole binaries with generic mass
  ratios and spins}},}\ }\href {\doibase 10.1103/PhysRevD.89.061502} {\bibfield
   {journal} {\bibinfo  {journal} {Phys. Rev.}\ }\textbf {\bibinfo {volume}
  {D89}},\ \bibinfo {pages} {061502} (\bibinfo {year} {2014})},\ \Eprint
  {http://arxiv.org/abs/1311.2544} {arXiv:1311.2544 [gr-qc]} \BibitemShut
  {NoStop}%
\bibitem [{\citenamefont {Pan}\ \emph {et~al.}(2014)\citenamefont {Pan},
  \citenamefont {Buonanno}, \citenamefont {Taracchini}, \citenamefont {Kidder},
  \citenamefont {Mroué}, \citenamefont {Pfeiffer}, \citenamefont {Scheel},\
  and\ \citenamefont {Szilágyi}}]{Pan:2013rra}%
  \BibitemOpen
  \bibfield  {author} {\bibinfo {author} {\bibfnamefont {Yi}~\bibnamefont
  {Pan}}, \bibinfo {author} {\bibfnamefont {Alessandra}\ \bibnamefont
  {Buonanno}}, \bibinfo {author} {\bibfnamefont {Andrea}\ \bibnamefont
  {Taracchini}}, \bibinfo {author} {\bibfnamefont {Lawrence~E.}\ \bibnamefont
  {Kidder}}, \bibinfo {author} {\bibfnamefont {Abdul~H.}\ \bibnamefont
  {Mroué}}, \bibinfo {author} {\bibfnamefont {Harald~P.}\ \bibnamefont
  {Pfeiffer}}, \bibinfo {author} {\bibfnamefont {Mark~A.}\ \bibnamefont
  {Scheel}}, \ and\ \bibinfo {author} {\bibfnamefont {Béla}\ \bibnamefont
  {Szilágyi}},\ }\bibfield  {title} {\enquote {\bibinfo {title}
  {{Inspiral-merger-ringdown waveforms of spinning, precessing black-hole
  binaries in the effective-one-body formalism}},}\ }\href {\doibase
  10.1103/PhysRevD.89.084006} {\bibfield  {journal} {\bibinfo  {journal} {Phys.
  Rev.}\ }\textbf {\bibinfo {volume} {D89}},\ \bibinfo {pages} {084006}
  (\bibinfo {year} {2014})},\ \Eprint {http://arxiv.org/abs/1307.6232}
  {arXiv:1307.6232 [gr-qc]} \BibitemShut {NoStop}%
\bibitem [{\citenamefont {Ossokine}\ \emph {et~al.}(2020)\citenamefont
  {Ossokine} \emph {et~al.}}]{Ossokine:2020kjp}%
  \BibitemOpen
  \bibfield  {author} {\bibinfo {author} {\bibfnamefont {Serguei}\ \bibnamefont
  {Ossokine}} \emph {et~al.},\ }\bibfield  {title} {\enquote {\bibinfo {title}
  {{Multipolar Effective-One-Body Waveforms for Precessing Binary Black Holes:
  Construction and Validation}},}\ }\href {\doibase
  10.1103/PhysRevD.102.044055} {\bibfield  {journal} {\bibinfo  {journal}
  {Phys. Rev. D}\ }\textbf {\bibinfo {volume} {102}},\ \bibinfo {pages}
  {044055} (\bibinfo {year} {2020})},\ \Eprint
  {http://arxiv.org/abs/2004.09442} {arXiv:2004.09442 [gr-qc]} \BibitemShut
  {NoStop}%
\bibitem [{\citenamefont {Ramos-Buades}\ \emph {et~al.}(2023)\citenamefont
  {Ramos-Buades}, \citenamefont {Buonanno}, \citenamefont {Estell{\'e}s},
  \citenamefont {Khalil}, \citenamefont {Mihaylov}, \citenamefont {Ossokine},
  \citenamefont {Pompili},\ and\ \citenamefont
  {Shiferaw}}]{Ramos-Buades:2023ehm}%
  \BibitemOpen
  \bibfield  {author} {\bibinfo {author} {\bibfnamefont {Antoni}\ \bibnamefont
  {Ramos-Buades}}, \bibinfo {author} {\bibfnamefont {Alessandra}\ \bibnamefont
  {Buonanno}}, \bibinfo {author} {\bibfnamefont {H{\'e}ctor}\ \bibnamefont
  {Estell{\'e}s}}, \bibinfo {author} {\bibfnamefont {Mohammed}\ \bibnamefont
  {Khalil}}, \bibinfo {author} {\bibfnamefont {Deyan~P.}\ \bibnamefont
  {Mihaylov}}, \bibinfo {author} {\bibfnamefont {Serguei}\ \bibnamefont
  {Ossokine}}, \bibinfo {author} {\bibfnamefont {Lorenzo}\ \bibnamefont
  {Pompili}}, \ and\ \bibinfo {author} {\bibfnamefont {Mahlet}\ \bibnamefont
  {Shiferaw}},\ }\bibfield  {title} {\enquote {\bibinfo {title} {{Next
  generation of accurate and efficient multipolar precessing-spin
  effective-one-body waveforms for binary black holes}},}\ }\href {\doibase
  10.1103/PhysRevD.108.124037} {\bibfield  {journal} {\bibinfo  {journal}
  {Phys. Rev. D}\ }\textbf {\bibinfo {volume} {108}},\ \bibinfo {pages}
  {124037} (\bibinfo {year} {2023})},\ \Eprint
  {http://arxiv.org/abs/2303.18046} {arXiv:2303.18046 [gr-qc]} \BibitemShut
  {NoStop}%
\bibitem [{\citenamefont {Nagar}\ \emph {et~al.}(2018)\citenamefont {Nagar}
  \emph {et~al.}}]{Nagar:2018zoe}%
  \BibitemOpen
  \bibfield  {author} {\bibinfo {author} {\bibfnamefont {Alessandro}\
  \bibnamefont {Nagar}} \emph {et~al.},\ }\bibfield  {title} {\enquote
  {\bibinfo {title} {{Time-domain effective-one-body gravitational waveforms
  for coalescing compact binaries with nonprecessing spins, tides and self-spin
  effects}},}\ }\href {\doibase 10.1103/PhysRevD.98.104052} {\bibfield
  {journal} {\bibinfo  {journal} {Phys. Rev. D}\ }\textbf {\bibinfo {volume}
  {98}},\ \bibinfo {pages} {104052} (\bibinfo {year} {2018})},\ \Eprint
  {http://arxiv.org/abs/1806.01772} {arXiv:1806.01772 [gr-qc]} \BibitemShut
  {NoStop}%
\bibitem [{\citenamefont {Albanesi}\ \emph {et~al.}(2025)\citenamefont
  {Albanesi}, \citenamefont {Gamba}, \citenamefont {Bernuzzi}, \citenamefont
  {Fontbut{\'e}}, \citenamefont {Gonzalez},\ and\ \citenamefont
  {Nagar}}]{Albanesi:2025txj}%
  \BibitemOpen
  \bibfield  {author} {\bibinfo {author} {\bibfnamefont {Simone}\ \bibnamefont
  {Albanesi}}, \bibinfo {author} {\bibfnamefont {Rossella}\ \bibnamefont
  {Gamba}}, \bibinfo {author} {\bibfnamefont {Sebastiano}\ \bibnamefont
  {Bernuzzi}}, \bibinfo {author} {\bibfnamefont {Joan}\ \bibnamefont
  {Fontbut{\'e}}}, \bibinfo {author} {\bibfnamefont {Alejandra}\ \bibnamefont
  {Gonzalez}}, \ and\ \bibinfo {author} {\bibfnamefont {Alessandro}\
  \bibnamefont {Nagar}},\ }\bibfield  {title} {\enquote {\bibinfo {title}
  {{Effective-one-body modeling for generic compact binaries with arbitrary
  orbits}},}\ }\href {\doibase 10.1103/3snf-w1x7} {\bibfield  {journal}
  {\bibinfo  {journal} {Phys. Rev. D}\ }\textbf {\bibinfo {volume} {112}},\
  \bibinfo {pages} {L121503} (\bibinfo {year} {2025})},\ \Eprint
  {http://arxiv.org/abs/2503.14580} {arXiv:2503.14580 [gr-qc]} \BibitemShut
  {NoStop}%
\bibitem [{\citenamefont {Khan}\ \emph {et~al.}(2016)\citenamefont {Khan},
  \citenamefont {Husa}, \citenamefont {Hannam}, \citenamefont {Ohme},
  \citenamefont {Pürrer}, \citenamefont {Jiménez~Forteza},\ and\
  \citenamefont {Bohé}}]{Khan:2015jqa}%
  \BibitemOpen
  \bibfield  {author} {\bibinfo {author} {\bibfnamefont {Sebastian}\
  \bibnamefont {Khan}}, \bibinfo {author} {\bibfnamefont {Sascha}\ \bibnamefont
  {Husa}}, \bibinfo {author} {\bibfnamefont {Mark}\ \bibnamefont {Hannam}},
  \bibinfo {author} {\bibfnamefont {Frank}\ \bibnamefont {Ohme}}, \bibinfo
  {author} {\bibfnamefont {Michael}\ \bibnamefont {Pürrer}}, \bibinfo {author}
  {\bibfnamefont {Xisco}\ \bibnamefont {Jiménez~Forteza}}, \ and\ \bibinfo
  {author} {\bibfnamefont {Alejandro}\ \bibnamefont {Bohé}},\ }\bibfield
  {title} {\enquote {\bibinfo {title} {{Frequency-domain gravitational waves
  from nonprecessing black-hole binaries. II. A phenomenological model for the
  advanced detector era}},}\ }\href {\doibase 10.1103/PhysRevD.93.044007}
  {\bibfield  {journal} {\bibinfo  {journal} {Phys. Rev.}\ }\textbf {\bibinfo
  {volume} {D93}},\ \bibinfo {pages} {044007} (\bibinfo {year} {2016})},\
  \Eprint {http://arxiv.org/abs/1508.07253} {arXiv:1508.07253 [gr-qc]}
  \BibitemShut {NoStop}%
\bibitem [{\citenamefont {Husa}\ \emph {et~al.}(2016)\citenamefont {Husa},
  \citenamefont {Khan}, \citenamefont {Hannam}, \citenamefont {P{\"u}rrer},
  \citenamefont {Ohme}, \citenamefont {Jim{\'e}nez~Forteza},\ and\
  \citenamefont {Boh{\'e}}}]{Husa:2015iqa}%
  \BibitemOpen
  \bibfield  {author} {\bibinfo {author} {\bibfnamefont {Sascha}\ \bibnamefont
  {Husa}}, \bibinfo {author} {\bibfnamefont {Sebastian}\ \bibnamefont {Khan}},
  \bibinfo {author} {\bibfnamefont {Mark}\ \bibnamefont {Hannam}}, \bibinfo
  {author} {\bibfnamefont {Michael}\ \bibnamefont {P{\"u}rrer}}, \bibinfo
  {author} {\bibfnamefont {Frank}\ \bibnamefont {Ohme}}, \bibinfo {author}
  {\bibfnamefont {Xisco}\ \bibnamefont {Jim{\'e}nez~Forteza}}, \ and\ \bibinfo
  {author} {\bibfnamefont {Alejandro}\ \bibnamefont {Boh{\'e}}},\ }\bibfield
  {title} {\enquote {\bibinfo {title} {{Frequency-domain gravitational waves
  from nonprecessing black-hole binaries. I. New numerical waveforms and
  anatomy of the signal}},}\ }\href {\doibase 10.1103/PhysRevD.93.044006}
  {\bibfield  {journal} {\bibinfo  {journal} {Phys. Rev.}\ }\textbf {\bibinfo
  {volume} {D93}},\ \bibinfo {pages} {044006} (\bibinfo {year} {2016})},\
  \Eprint {http://arxiv.org/abs/1508.07250} {arXiv:1508.07250 [gr-qc]}
  \BibitemShut {NoStop}%
\bibitem [{\citenamefont {Hannam}\ \emph {et~al.}(2014)\citenamefont {Hannam},
  \citenamefont {Schmidt}, \citenamefont {Bohé}, \citenamefont {Haegel},
  \citenamefont {Husa}, \citenamefont {Ohme}, \citenamefont {Pratten},\ and\
  \citenamefont {Pürrer}}]{Hannam:2013oca}%
  \BibitemOpen
  \bibfield  {author} {\bibinfo {author} {\bibfnamefont {Mark}\ \bibnamefont
  {Hannam}}, \bibinfo {author} {\bibfnamefont {Patricia}\ \bibnamefont
  {Schmidt}}, \bibinfo {author} {\bibfnamefont {Alejandro}\ \bibnamefont
  {Bohé}}, \bibinfo {author} {\bibfnamefont {Leïla}\ \bibnamefont {Haegel}},
  \bibinfo {author} {\bibfnamefont {Sascha}\ \bibnamefont {Husa}}, \bibinfo
  {author} {\bibfnamefont {Frank}\ \bibnamefont {Ohme}}, \bibinfo {author}
  {\bibfnamefont {Geraint}\ \bibnamefont {Pratten}}, \ and\ \bibinfo {author}
  {\bibfnamefont {Michael}\ \bibnamefont {Pürrer}},\ }\bibfield  {title}
  {\enquote {\bibinfo {title} {{Simple Model of Complete Precessing
  Black-Hole-Binary Gravitational Waveforms}},}\ }\href {\doibase
  10.1103/PhysRevLett.113.151101} {\bibfield  {journal} {\bibinfo  {journal}
  {Phys. Rev. Lett.}\ }\textbf {\bibinfo {volume} {113}},\ \bibinfo {pages}
  {151101} (\bibinfo {year} {2014})},\ \Eprint {http://arxiv.org/abs/1308.3271}
  {arXiv:1308.3271 [gr-qc]} \BibitemShut {NoStop}%
\bibitem [{\citenamefont {Pratten}\ \emph {et~al.}(2021)\citenamefont {Pratten}
  \emph {et~al.}}]{Pratten:2020ceb}%
  \BibitemOpen
  \bibfield  {author} {\bibinfo {author} {\bibfnamefont {Geraint}\ \bibnamefont
  {Pratten}} \emph {et~al.},\ }\bibfield  {title} {\enquote {\bibinfo {title}
  {{Computationally efficient models for the dominant and subdominant harmonic
  modes of precessing binary black holes}},}\ }\href {\doibase
  10.1103/PhysRevD.103.104056} {\bibfield  {journal} {\bibinfo  {journal}
  {Phys. Rev. D}\ }\textbf {\bibinfo {volume} {103}},\ \bibinfo {pages}
  {104056} (\bibinfo {year} {2021})},\ \Eprint
  {http://arxiv.org/abs/2004.06503} {arXiv:2004.06503 [gr-qc]} \BibitemShut
  {NoStop}%
\bibitem [{\citenamefont {Pratten}\ \emph {et~al.}(2020)\citenamefont
  {Pratten}, \citenamefont {Husa}, \citenamefont {Garcia-Quiros}, \citenamefont
  {Colleoni}, \citenamefont {Ramos-Buades}, \citenamefont {Estelles},\ and\
  \citenamefont {Jaume}}]{Pratten:2020fqn}%
  \BibitemOpen
  \bibfield  {author} {\bibinfo {author} {\bibfnamefont {Geraint}\ \bibnamefont
  {Pratten}}, \bibinfo {author} {\bibfnamefont {Sascha}\ \bibnamefont {Husa}},
  \bibinfo {author} {\bibfnamefont {Cecilio}\ \bibnamefont {Garcia-Quiros}},
  \bibinfo {author} {\bibfnamefont {Marta}\ \bibnamefont {Colleoni}}, \bibinfo
  {author} {\bibfnamefont {Antoni}\ \bibnamefont {Ramos-Buades}}, \bibinfo
  {author} {\bibfnamefont {Hector}\ \bibnamefont {Estelles}}, \ and\ \bibinfo
  {author} {\bibfnamefont {Rafel}\ \bibnamefont {Jaume}},\ }\bibfield  {title}
  {\enquote {\bibinfo {title} {{Setting the cornerstone for a family of models
  for gravitational waves from compact binaries: The dominant harmonic for
  nonprecessing quasicircular black holes}},}\ }\href {\doibase
  10.1103/PhysRevD.102.064001} {\bibfield  {journal} {\bibinfo  {journal}
  {Phys. Rev. D}\ }\textbf {\bibinfo {volume} {102}},\ \bibinfo {pages}
  {064001} (\bibinfo {year} {2020})},\ \Eprint
  {http://arxiv.org/abs/2001.11412} {arXiv:2001.11412 [gr-qc]} \BibitemShut
  {NoStop}%
\bibitem [{\citenamefont {Ajith}\ \emph {et~al.}(2007)\citenamefont {Ajith}
  \emph {et~al.}}]{Ajith:2007qp}%
  \BibitemOpen
  \bibfield  {author} {\bibinfo {author} {\bibfnamefont {Parameswaran}\
  \bibnamefont {Ajith}} \emph {et~al.},\ }\bibfield  {title} {\enquote
  {\bibinfo {title} {{Phenomenological template family for black-hole
  coalescence waveforms}},}\ }\bibfield  {booktitle} {\emph {\bibinfo
  {booktitle} {{Gravitational wave data analysis. Proceedings: 11th Workshop,
  GWDAW-11, Potsdam, Germany, Dec 18-21, 2006}}},\ }\href {\doibase
  10.1088/0264-9381/24/19/S31} {\bibfield  {journal} {\bibinfo  {journal}
  {Class. Quant. Grav.}\ }\textbf {\bibinfo {volume} {24}},\ \bibinfo {pages}
  {S689--S700} (\bibinfo {year} {2007})},\ \Eprint
  {http://arxiv.org/abs/0704.3764} {arXiv:0704.3764 [gr-qc]} \BibitemShut
  {NoStop}%
\bibitem [{\citenamefont {Ajith}\ \emph {et~al.}(2008)\citenamefont {Ajith}
  \emph {et~al.}}]{Ajith:2007kx}%
  \BibitemOpen
  \bibfield  {author} {\bibinfo {author} {\bibfnamefont {P.}~\bibnamefont
  {Ajith}} \emph {et~al.},\ }\bibfield  {title} {\enquote {\bibinfo {title} {{A
  Template bank for gravitational waveforms from coalescing binary black holes.
  I. Non-spinning binaries}},}\ }\href {\doibase 10.1103/PhysRevD.77.104017}
  {\bibfield  {journal} {\bibinfo  {journal} {Phys. Rev. D}\ }\textbf {\bibinfo
  {volume} {77}},\ \bibinfo {pages} {104017} (\bibinfo {year} {2008})},\
  \bibinfo {note} {[Erratum: Phys.Rev.D 79, 129901 (2009)]},\ \Eprint
  {http://arxiv.org/abs/0710.2335} {arXiv:0710.2335 [gr-qc]} \BibitemShut
  {NoStop}%
\bibitem [{\citenamefont {Ajith}\ \emph {et~al.}(2011)\citenamefont {Ajith}
  \emph {et~al.}}]{Ajith:2009bn}%
  \BibitemOpen
  \bibfield  {author} {\bibinfo {author} {\bibfnamefont {P.}~\bibnamefont
  {Ajith}} \emph {et~al.},\ }\bibfield  {title} {\enquote {\bibinfo {title}
  {{Inspiral-merger-ringdown waveforms for black-hole binaries with
  non-precessing spins}},}\ }\href {\doibase 10.1103/PhysRevLett.106.241101}
  {\bibfield  {journal} {\bibinfo  {journal} {Phys. Rev. Lett.}\ }\textbf
  {\bibinfo {volume} {106}},\ \bibinfo {pages} {241101} (\bibinfo {year}
  {2011})},\ \Eprint {http://arxiv.org/abs/0909.2867} {arXiv:0909.2867 [gr-qc]}
  \BibitemShut {NoStop}%
\bibitem [{\citenamefont {Santamaria}\ \emph {et~al.}(2010)\citenamefont
  {Santamaria} \emph {et~al.}}]{Santamaria:2010yb}%
  \BibitemOpen
  \bibfield  {author} {\bibinfo {author} {\bibfnamefont {L.}~\bibnamefont
  {Santamaria}} \emph {et~al.},\ }\bibfield  {title} {\enquote {\bibinfo
  {title} {{Matching post-Newtonian and numerical relativity waveforms:
  systematic errors and a new phenomenological model for non-precessing black
  hole binaries}},}\ }\href {\doibase 10.1103/PhysRevD.82.064016} {\bibfield
  {journal} {\bibinfo  {journal} {Phys. Rev. D}\ }\textbf {\bibinfo {volume}
  {82}},\ \bibinfo {pages} {064016} (\bibinfo {year} {2010})},\ \Eprint
  {http://arxiv.org/abs/1005.3306} {arXiv:1005.3306 [gr-qc]} \BibitemShut
  {NoStop}%
\bibitem [{\citenamefont {London}\ \emph {et~al.}(2018)\citenamefont {London},
  \citenamefont {Khan}, \citenamefont {Fauchon-Jones}, \citenamefont {García},
  \citenamefont {Hannam}, \citenamefont {Husa}, \citenamefont
  {Jiménez-Forteza}, \citenamefont {Kalaghatgi}, \citenamefont {Ohme},\ and\
  \citenamefont {Pannarale}}]{London:2017bcn}%
  \BibitemOpen
  \bibfield  {author} {\bibinfo {author} {\bibfnamefont {Lionel}\ \bibnamefont
  {London}}, \bibinfo {author} {\bibfnamefont {Sebastian}\ \bibnamefont
  {Khan}}, \bibinfo {author} {\bibfnamefont {Edward}\ \bibnamefont
  {Fauchon-Jones}}, \bibinfo {author} {\bibfnamefont {Cecilio}\ \bibnamefont
  {García}}, \bibinfo {author} {\bibfnamefont {Mark}\ \bibnamefont {Hannam}},
  \bibinfo {author} {\bibfnamefont {Sascha}\ \bibnamefont {Husa}}, \bibinfo
  {author} {\bibfnamefont {Xisco}\ \bibnamefont {Jiménez-Forteza}}, \bibinfo
  {author} {\bibfnamefont {Chinmay}\ \bibnamefont {Kalaghatgi}}, \bibinfo
  {author} {\bibfnamefont {Frank}\ \bibnamefont {Ohme}}, \ and\ \bibinfo
  {author} {\bibfnamefont {Francesco}\ \bibnamefont {Pannarale}},\ }\bibfield
  {title} {\enquote {\bibinfo {title} {{First higher-multipole model of
  gravitational waves from spinning and coalescing black-hole binaries}},}\
  }\href {\doibase 10.1103/PhysRevLett.120.161102} {\bibfield  {journal}
  {\bibinfo  {journal} {Phys. Rev. Lett.}\ }\textbf {\bibinfo {volume} {120}},\
  \bibinfo {pages} {161102} (\bibinfo {year} {2018})},\ \Eprint
  {http://arxiv.org/abs/1708.00404} {arXiv:1708.00404 [gr-qc]} \BibitemShut
  {NoStop}%
\bibitem [{\citenamefont {Khan}\ \emph {et~al.}(2019)\citenamefont {Khan},
  \citenamefont {Chatziioannou}, \citenamefont {Hannam},\ and\ \citenamefont
  {Ohme}}]{Khan:2018fmp}%
  \BibitemOpen
  \bibfield  {author} {\bibinfo {author} {\bibfnamefont {Sebastian}\
  \bibnamefont {Khan}}, \bibinfo {author} {\bibfnamefont {Katerina}\
  \bibnamefont {Chatziioannou}}, \bibinfo {author} {\bibfnamefont {Mark}\
  \bibnamefont {Hannam}}, \ and\ \bibinfo {author} {\bibfnamefont {Frank}\
  \bibnamefont {Ohme}},\ }\bibfield  {title} {\enquote {\bibinfo {title}
  {{Phenomenological model for the gravitational-wave signal from precessing
  binary black holes with two-spin effects}},}\ }\href {\doibase
  10.1103/PhysRevD.100.024059} {\bibfield  {journal} {\bibinfo  {journal}
  {Phys. Rev. D}\ }\textbf {\bibinfo {volume} {100}},\ \bibinfo {pages}
  {024059} (\bibinfo {year} {2019})},\ \Eprint
  {http://arxiv.org/abs/1809.10113} {arXiv:1809.10113 [gr-qc]} \BibitemShut
  {NoStop}%
\bibitem [{\citenamefont {Khan}\ \emph {et~al.}(2020)\citenamefont {Khan},
  \citenamefont {Ohme}, \citenamefont {Chatziioannou},\ and\ \citenamefont
  {Hannam}}]{Khan:2019kot}%
  \BibitemOpen
  \bibfield  {author} {\bibinfo {author} {\bibfnamefont {Sebastian}\
  \bibnamefont {Khan}}, \bibinfo {author} {\bibfnamefont {Frank}\ \bibnamefont
  {Ohme}}, \bibinfo {author} {\bibfnamefont {Katerina}\ \bibnamefont
  {Chatziioannou}}, \ and\ \bibinfo {author} {\bibfnamefont {Mark}\
  \bibnamefont {Hannam}},\ }\bibfield  {title} {\enquote {\bibinfo {title}
  {{Including higher order multipoles in gravitational-wave models for
  precessing binary black holes}},}\ }\href {\doibase
  10.1103/PhysRevD.101.024056} {\bibfield  {journal} {\bibinfo  {journal}
  {Phys. Rev.}\ }\textbf {\bibinfo {volume} {D101}},\ \bibinfo {pages} {024056}
  (\bibinfo {year} {2020})},\ \Eprint {http://arxiv.org/abs/1911.06050}
  {arXiv:1911.06050 [gr-qc]} \BibitemShut {NoStop}%
\bibitem [{\citenamefont {Dietrich}\ \emph
  {et~al.}(2019{\natexlab{a}})\citenamefont {Dietrich}, \citenamefont {Khan},
  \citenamefont {Dudi}, \citenamefont {Kapadia}, \citenamefont {Kumar},
  \citenamefont {Nagar}, \citenamefont {Ohme}, \citenamefont {Pannarale},
  \citenamefont {Samajdar}, \citenamefont {Bernuzzi}, \citenamefont {Carullo},
  \citenamefont {Del~Pozzo}, \citenamefont {Haney}, \citenamefont {Markakis},
  \citenamefont {P\"urrer}, \citenamefont {Riemenschneider}, \citenamefont
  {Setyawati}, \citenamefont {Tsang},\ and\ \citenamefont {Van
  Den~Broeck}}]{Dietrich:2018nrt}%
  \BibitemOpen
  \bibfield  {author} {\bibinfo {author} {\bibfnamefont {Tim}\ \bibnamefont
  {Dietrich}}, \bibinfo {author} {\bibfnamefont {Sebastian}\ \bibnamefont
  {Khan}}, \bibinfo {author} {\bibfnamefont {Reetika}\ \bibnamefont {Dudi}},
  \bibinfo {author} {\bibfnamefont {Shasvath~J.}\ \bibnamefont {Kapadia}},
  \bibinfo {author} {\bibfnamefont {Prayush}\ \bibnamefont {Kumar}}, \bibinfo
  {author} {\bibfnamefont {Alessandro}\ \bibnamefont {Nagar}}, \bibinfo
  {author} {\bibfnamefont {Frank}\ \bibnamefont {Ohme}}, \bibinfo {author}
  {\bibfnamefont {Francesco}\ \bibnamefont {Pannarale}}, \bibinfo {author}
  {\bibfnamefont {Anuradha}\ \bibnamefont {Samajdar}}, \bibinfo {author}
  {\bibfnamefont {Sebastiano}\ \bibnamefont {Bernuzzi}}, \bibinfo {author}
  {\bibfnamefont {Gregorio}\ \bibnamefont {Carullo}}, \bibinfo {author}
  {\bibfnamefont {Walter}\ \bibnamefont {Del~Pozzo}}, \bibinfo {author}
  {\bibfnamefont {Maria}\ \bibnamefont {Haney}}, \bibinfo {author}
  {\bibfnamefont {Charalampos}\ \bibnamefont {Markakis}}, \bibinfo {author}
  {\bibfnamefont {Michael}\ \bibnamefont {P\"urrer}}, \bibinfo {author}
  {\bibfnamefont {Gunnar}\ \bibnamefont {Riemenschneider}}, \bibinfo {author}
  {\bibfnamefont {Yoshinta~Eka}\ \bibnamefont {Setyawati}}, \bibinfo {author}
  {\bibfnamefont {Ka~Wa}\ \bibnamefont {Tsang}}, \ and\ \bibinfo {author}
  {\bibfnamefont {Chris}\ \bibnamefont {Van Den~Broeck}},\ }\bibfield  {title}
  {\enquote {\bibinfo {title} {Matter imprints in waveform models for neutron
  star binaries: Tidal and self-spin effects},}\ }\href {\doibase
  10.1103/PhysRevD.99.024029} {\bibfield  {journal} {\bibinfo  {journal} {Phys.
  Rev. D}\ }\textbf {\bibinfo {volume} {99}},\ \bibinfo {pages} {024029}
  (\bibinfo {year} {2019}{\natexlab{a}})},\ \Eprint
  {http://arxiv.org/abs/1804.02235} {arXiv:1804.02235 [gr-qc]} \BibitemShut
  {NoStop}%
\bibitem [{\citenamefont {Dietrich}\ \emph
  {et~al.}(2019{\natexlab{b}})\citenamefont {Dietrich}, \citenamefont
  {Samajdar}, \citenamefont {Khan}, \citenamefont {Johnson-McDaniel},
  \citenamefont {Dudi},\ and\ \citenamefont {Tichy}}]{Dietrich:2019nrt}%
  \BibitemOpen
  \bibfield  {author} {\bibinfo {author} {\bibfnamefont {Tim}\ \bibnamefont
  {Dietrich}}, \bibinfo {author} {\bibfnamefont {Anuradha}\ \bibnamefont
  {Samajdar}}, \bibinfo {author} {\bibfnamefont {Sebastian}\ \bibnamefont
  {Khan}}, \bibinfo {author} {\bibfnamefont {Nathan~K.}\ \bibnamefont
  {Johnson-McDaniel}}, \bibinfo {author} {\bibfnamefont {Reetika}\ \bibnamefont
  {Dudi}}, \ and\ \bibinfo {author} {\bibfnamefont {Wolfgang}\ \bibnamefont
  {Tichy}},\ }\bibfield  {title} {\enquote {\bibinfo {title} {Improving the
  nrtidal model for binary neutron star systems},}\ }\href {\doibase
  10.1103/PhysRevD.100.044003} {\bibfield  {journal} {\bibinfo  {journal}
  {Phys. Rev. D}\ }\textbf {\bibinfo {volume} {100}},\ \bibinfo {pages}
  {044003} (\bibinfo {year} {2019}{\natexlab{b}})},\ \Eprint
  {http://arxiv.org/abs/1905.06011} {arXiv:1905.06011 [gr-qc]} \BibitemShut
  {NoStop}%
\bibitem [{\citenamefont {Thompson}\ \emph {et~al.}(2020)\citenamefont
  {Thompson}, \citenamefont {Fauchon-Jones}, \citenamefont {Khan},
  \citenamefont {Nitoglia}, \citenamefont {Pannarale}, \citenamefont
  {Dietrich},\ and\ \citenamefont {Hannam}}]{Thompson:2020nei}%
  \BibitemOpen
  \bibfield  {author} {\bibinfo {author} {\bibfnamefont {Jonathan~E.}\
  \bibnamefont {Thompson}}, \bibinfo {author} {\bibfnamefont {Edward}\
  \bibnamefont {Fauchon-Jones}}, \bibinfo {author} {\bibfnamefont {Sebastian}\
  \bibnamefont {Khan}}, \bibinfo {author} {\bibfnamefont {Elisa}\ \bibnamefont
  {Nitoglia}}, \bibinfo {author} {\bibfnamefont {Francesco}\ \bibnamefont
  {Pannarale}}, \bibinfo {author} {\bibfnamefont {Tim}\ \bibnamefont
  {Dietrich}}, \ and\ \bibinfo {author} {\bibfnamefont {Mark}\ \bibnamefont
  {Hannam}},\ }\bibfield  {title} {\enquote {\bibinfo {title} {{Modeling the
  gravitational wave signature of neutron star black hole coalescences:
  PhenomNSBH}},}\ }\href {\doibase 10.1103/PhysRevD.101.124059} {\bibfield
  {journal} {\bibinfo  {journal} {Phys. Rev. D}\ }\textbf {\bibinfo {volume}
  {101}},\ \bibinfo {pages} {124059} (\bibinfo {year} {2020})},\ \Eprint
  {http://arxiv.org/abs/2002.08383} {arXiv:2002.08383 [gr-qc]} \BibitemShut
  {NoStop}%
\bibitem [{\citenamefont {Garc\'\i{}a-Quir\'os}\ \emph
  {et~al.}(2020)\citenamefont {Garc\'\i{}a-Quir\'os}, \citenamefont {Colleoni},
  \citenamefont {Husa}, \citenamefont {Estell\'es}, \citenamefont {Pratten},
  \citenamefont {Ramos-Buades}, \citenamefont {Mateu-Lucena},\ and\
  \citenamefont {Jaume}}]{Garcia-Quiros:2020qpx}%
  \BibitemOpen
  \bibfield  {author} {\bibinfo {author} {\bibfnamefont {Cecilio}\ \bibnamefont
  {Garc\'\i{}a-Quir\'os}}, \bibinfo {author} {\bibfnamefont {Marta}\
  \bibnamefont {Colleoni}}, \bibinfo {author} {\bibfnamefont {Sascha}\
  \bibnamefont {Husa}}, \bibinfo {author} {\bibfnamefont {H\'ector}\
  \bibnamefont {Estell\'es}}, \bibinfo {author} {\bibfnamefont {Geraint}\
  \bibnamefont {Pratten}}, \bibinfo {author} {\bibfnamefont {Antoni}\
  \bibnamefont {Ramos-Buades}}, \bibinfo {author} {\bibfnamefont {Maite}\
  \bibnamefont {Mateu-Lucena}}, \ and\ \bibinfo {author} {\bibfnamefont
  {Rafel}\ \bibnamefont {Jaume}},\ }\bibfield  {title} {\enquote {\bibinfo
  {title} {{Multimode frequency-domain model for the gravitational wave signal
  from nonprecessing black-hole binaries}},}\ }\href {\doibase
  10.1103/PhysRevD.102.064002} {\bibfield  {journal} {\bibinfo  {journal}
  {Phys. Rev. D}\ }\textbf {\bibinfo {volume} {102}},\ \bibinfo {pages}
  {064002} (\bibinfo {year} {2020})},\ \Eprint
  {http://arxiv.org/abs/2001.10914} {arXiv:2001.10914 [gr-qc]} \BibitemShut
  {NoStop}%
\bibitem [{\citenamefont {Garc\'\i{}a-Quir\'os}\ \emph
  {et~al.}(2021)\citenamefont {Garc\'\i{}a-Quir\'os}, \citenamefont {Husa},
  \citenamefont {Mateu-Lucena},\ and\ \citenamefont
  {Borchers}}]{Garcia-Quiros:2020qlt}%
  \BibitemOpen
  \bibfield  {author} {\bibinfo {author} {\bibfnamefont {Cecilio}\ \bibnamefont
  {Garc\'\i{}a-Quir\'os}}, \bibinfo {author} {\bibfnamefont {Sascha}\
  \bibnamefont {Husa}}, \bibinfo {author} {\bibfnamefont {Maite}\ \bibnamefont
  {Mateu-Lucena}}, \ and\ \bibinfo {author} {\bibfnamefont {Angela}\
  \bibnamefont {Borchers}},\ }\bibfield  {title} {\enquote {\bibinfo {title}
  {{Accelerating the evaluation of
  inspiral\textendash{}merger\textendash{}ringdown waveforms with adapted
  grids}},}\ }\href {\doibase 10.1088/1361-6382/abc36e} {\bibfield  {journal}
  {\bibinfo  {journal} {Class. Quant. Grav.}\ }\textbf {\bibinfo {volume}
  {38}},\ \bibinfo {pages} {015006} (\bibinfo {year} {2021})},\ \Eprint
  {http://arxiv.org/abs/2001.10897} {arXiv:2001.10897 [gr-qc]} \BibitemShut
  {NoStop}%
\bibitem [{\citenamefont {Blackman}\ \emph {et~al.}(2015)\citenamefont
  {Blackman}, \citenamefont {Field}, \citenamefont {Galley}, \citenamefont
  {Szilágyi}, \citenamefont {Scheel}, \citenamefont {Tiglio},\ and\
  \citenamefont {Hemberger}}]{Blackman:2015pia}%
  \BibitemOpen
  \bibfield  {author} {\bibinfo {author} {\bibfnamefont {Jonathan}\
  \bibnamefont {Blackman}}, \bibinfo {author} {\bibfnamefont {Scott~E.}\
  \bibnamefont {Field}}, \bibinfo {author} {\bibfnamefont {Chad~R.}\
  \bibnamefont {Galley}}, \bibinfo {author} {\bibfnamefont {Béla}\
  \bibnamefont {Szilágyi}}, \bibinfo {author} {\bibfnamefont {Mark~A.}\
  \bibnamefont {Scheel}}, \bibinfo {author} {\bibfnamefont {Manuel}\
  \bibnamefont {Tiglio}}, \ and\ \bibinfo {author} {\bibfnamefont {Daniel~A.}\
  \bibnamefont {Hemberger}},\ }\bibfield  {title} {\enquote {\bibinfo {title}
  {{Fast and Accurate Prediction of Numerical Relativity Waveforms from Binary
  Black Hole Coalescences Using Surrogate Models}},}\ }\href {\doibase
  10.1103/PhysRevLett.115.121102} {\bibfield  {journal} {\bibinfo  {journal}
  {Phys. Rev. Lett.}\ }\textbf {\bibinfo {volume} {115}},\ \bibinfo {pages}
  {121102} (\bibinfo {year} {2015})},\ \Eprint
  {http://arxiv.org/abs/1502.07758} {arXiv:1502.07758 [gr-qc]} \BibitemShut
  {NoStop}%
\bibitem [{\citenamefont {Blackman}\ \emph
  {et~al.}(2017{\natexlab{a}})\citenamefont {Blackman}, \citenamefont {Field},
  \citenamefont {Scheel}, \citenamefont {Galley}, \citenamefont {Hemberger},
  \citenamefont {Schmidt},\ and\ \citenamefont {Smith}}]{Blackman:2017dfb}%
  \BibitemOpen
  \bibfield  {author} {\bibinfo {author} {\bibfnamefont {Jonathan}\
  \bibnamefont {Blackman}}, \bibinfo {author} {\bibfnamefont {Scott~E.}\
  \bibnamefont {Field}}, \bibinfo {author} {\bibfnamefont {Mark~A.}\
  \bibnamefont {Scheel}}, \bibinfo {author} {\bibfnamefont {Chad~R.}\
  \bibnamefont {Galley}}, \bibinfo {author} {\bibfnamefont {Daniel~A.}\
  \bibnamefont {Hemberger}}, \bibinfo {author} {\bibfnamefont {Patricia}\
  \bibnamefont {Schmidt}}, \ and\ \bibinfo {author} {\bibfnamefont {Rory}\
  \bibnamefont {Smith}},\ }\bibfield  {title} {\enquote {\bibinfo {title} {{A
  Surrogate Model of Gravitational Waveforms from Numerical Relativity
  Simulations of Precessing Binary Black Hole Mergers}},}\ }\href {\doibase
  10.1103/PhysRevD.95.104023} {\bibfield  {journal} {\bibinfo  {journal} {Phys.
  Rev.}\ }\textbf {\bibinfo {volume} {D95}},\ \bibinfo {pages} {104023}
  (\bibinfo {year} {2017}{\natexlab{a}})},\ \Eprint
  {http://arxiv.org/abs/1701.00550} {arXiv:1701.00550 [gr-qc]} \BibitemShut
  {NoStop}%
\bibitem [{\citenamefont {Blackman}\ \emph
  {et~al.}(2017{\natexlab{b}})\citenamefont {Blackman}, \citenamefont {Field},
  \citenamefont {Scheel}, \citenamefont {Galley}, \citenamefont {Ott},
  \citenamefont {Boyle}, \citenamefont {Kidder}, \citenamefont {Pfeiffer},\
  and\ \citenamefont {Szilágyi}}]{Blackman:2017pcm}%
  \BibitemOpen
  \bibfield  {author} {\bibinfo {author} {\bibfnamefont {Jonathan}\
  \bibnamefont {Blackman}}, \bibinfo {author} {\bibfnamefont {Scott~E.}\
  \bibnamefont {Field}}, \bibinfo {author} {\bibfnamefont {Mark~A.}\
  \bibnamefont {Scheel}}, \bibinfo {author} {\bibfnamefont {Chad~R.}\
  \bibnamefont {Galley}}, \bibinfo {author} {\bibfnamefont {Christian~D.}\
  \bibnamefont {Ott}}, \bibinfo {author} {\bibfnamefont {Michael}\ \bibnamefont
  {Boyle}}, \bibinfo {author} {\bibfnamefont {Lawrence~E.}\ \bibnamefont
  {Kidder}}, \bibinfo {author} {\bibfnamefont {Harald~P.}\ \bibnamefont
  {Pfeiffer}}, \ and\ \bibinfo {author} {\bibfnamefont {Béla}\ \bibnamefont
  {Szilágyi}},\ }\bibfield  {title} {\enquote {\bibinfo {title} {{Numerical
  relativity waveform surrogate model for generically precessing binary black
  hole mergers}},}\ }\href {\doibase 10.1103/PhysRevD.96.024058} {\bibfield
  {journal} {\bibinfo  {journal} {Phys. Rev.}\ }\textbf {\bibinfo {volume}
  {D96}},\ \bibinfo {pages} {024058} (\bibinfo {year} {2017}{\natexlab{b}})},\
  \Eprint {http://arxiv.org/abs/1705.07089} {arXiv:1705.07089 [gr-qc]}
  \BibitemShut {NoStop}%
\bibitem [{\citenamefont {Varma}\ \emph
  {et~al.}(2019{\natexlab{a}})\citenamefont {Varma}, \citenamefont {Field},
  \citenamefont {Scheel}, \citenamefont {Blackman}, \citenamefont {Kidder},\
  and\ \citenamefont {Pfeiffer}}]{Varma:2018mmi}%
  \BibitemOpen
  \bibfield  {author} {\bibinfo {author} {\bibfnamefont {Vijay}\ \bibnamefont
  {Varma}}, \bibinfo {author} {\bibfnamefont {Scott~E.}\ \bibnamefont {Field}},
  \bibinfo {author} {\bibfnamefont {Mark~A.}\ \bibnamefont {Scheel}}, \bibinfo
  {author} {\bibfnamefont {Jonathan}\ \bibnamefont {Blackman}}, \bibinfo
  {author} {\bibfnamefont {Lawrence~E.}\ \bibnamefont {Kidder}}, \ and\
  \bibinfo {author} {\bibfnamefont {Harald~P.}\ \bibnamefont {Pfeiffer}},\
  }\bibfield  {title} {\enquote {\bibinfo {title} {{Surrogate model of
  hybridized numerical relativity binary black hole waveforms}},}\ }\href
  {\doibase 10.1103/PhysRevD.99.064045} {\bibfield  {journal} {\bibinfo
  {journal} {Phys. Rev.}\ }\textbf {\bibinfo {volume} {D99}},\ \bibinfo {pages}
  {064045} (\bibinfo {year} {2019}{\natexlab{a}})},\ \Eprint
  {http://arxiv.org/abs/1812.07865} {arXiv:1812.07865 [gr-qc]} \BibitemShut
  {NoStop}%
\bibitem [{\citenamefont {Varma}\ \emph
  {et~al.}(2019{\natexlab{b}})\citenamefont {Varma}, \citenamefont {Field},
  \citenamefont {Scheel}, \citenamefont {Blackman}, \citenamefont {Gerosa},
  \citenamefont {Stein}, \citenamefont {Kidder},\ and\ \citenamefont
  {Pfeiffer}}]{Varma:2019csw}%
  \BibitemOpen
  \bibfield  {author} {\bibinfo {author} {\bibfnamefont {Vijay}\ \bibnamefont
  {Varma}}, \bibinfo {author} {\bibfnamefont {Scott~E.}\ \bibnamefont {Field}},
  \bibinfo {author} {\bibfnamefont {Mark~A.}\ \bibnamefont {Scheel}}, \bibinfo
  {author} {\bibfnamefont {Jonathan}\ \bibnamefont {Blackman}}, \bibinfo
  {author} {\bibfnamefont {Davide}\ \bibnamefont {Gerosa}}, \bibinfo {author}
  {\bibfnamefont {Leo~C.}\ \bibnamefont {Stein}}, \bibinfo {author}
  {\bibfnamefont {Lawrence~E.}\ \bibnamefont {Kidder}}, \ and\ \bibinfo
  {author} {\bibfnamefont {Harald~P.}\ \bibnamefont {Pfeiffer}},\ }\bibfield
  {title} {\enquote {\bibinfo {title} {{Surrogate models for precessing binary
  black hole simulations with unequal masses}},}\ }\href {\doibase
  10.1103/PhysRevResearch.1.033015} {\bibfield  {journal} {\bibinfo  {journal}
  {Phys. Rev. Research.}\ }\textbf {\bibinfo {volume} {1}},\ \bibinfo {pages}
  {033015} (\bibinfo {year} {2019}{\natexlab{b}})},\ \Eprint
  {http://arxiv.org/abs/1905.09300} {arXiv:1905.09300 [gr-qc]} \BibitemShut
  {NoStop}%
\bibitem [{\citenamefont {Islam}\ \emph
  {et~al.}(2021{\natexlab{a}})\citenamefont {Islam}, \citenamefont {Varma},
  \citenamefont {Lodman}, \citenamefont {Field}, \citenamefont {Khanna},
  \citenamefont {Scheel}, \citenamefont {Pfeiffer}, \citenamefont {Gerosa},\
  and\ \citenamefont {Kidder}}]{Islam:2021mha}%
  \BibitemOpen
  \bibfield  {author} {\bibinfo {author} {\bibfnamefont {Tousif}\ \bibnamefont
  {Islam}}, \bibinfo {author} {\bibfnamefont {Vijay}\ \bibnamefont {Varma}},
  \bibinfo {author} {\bibfnamefont {Jackie}\ \bibnamefont {Lodman}}, \bibinfo
  {author} {\bibfnamefont {Scott~E.}\ \bibnamefont {Field}}, \bibinfo {author}
  {\bibfnamefont {Gaurav}\ \bibnamefont {Khanna}}, \bibinfo {author}
  {\bibfnamefont {Mark~A.}\ \bibnamefont {Scheel}}, \bibinfo {author}
  {\bibfnamefont {Harald~P.}\ \bibnamefont {Pfeiffer}}, \bibinfo {author}
  {\bibfnamefont {Davide}\ \bibnamefont {Gerosa}}, \ and\ \bibinfo {author}
  {\bibfnamefont {Lawrence~E.}\ \bibnamefont {Kidder}},\ }\bibfield  {title}
  {\enquote {\bibinfo {title} {{Eccentric binary black hole surrogate models
  for the gravitational waveform and remnant properties: comparable mass,
  nonspinning case}},}\ }\href {\doibase 10.1103/PhysRevD.103.064022}
  {\bibfield  {journal} {\bibinfo  {journal} {Phys. Rev. D}\ }\textbf {\bibinfo
  {volume} {103}},\ \bibinfo {pages} {064022} (\bibinfo {year}
  {2021}{\natexlab{a}})},\ \Eprint {http://arxiv.org/abs/2101.11798}
  {arXiv:2101.11798 [gr-qc]} \BibitemShut {NoStop}%
\bibitem [{\citenamefont {Gramaxo~Freitas}\ \emph {et~al.}(2025)\citenamefont
  {Gramaxo~Freitas}, \citenamefont {Theodoropoulos}, \citenamefont
  {Villanueva}, \citenamefont {Fernandes}, \citenamefont {Nunes}, \citenamefont
  {Font}, \citenamefont {Onofre}, \citenamefont {Torres-Forn{\'e}},\ and\
  \citenamefont {Martin-Guerrero}}]{GramaxoFreitas:2024bpk}%
  \BibitemOpen
  \bibfield  {author} {\bibinfo {author} {\bibfnamefont {Osvaldo}\ \bibnamefont
  {Gramaxo~Freitas}}, \bibinfo {author} {\bibfnamefont {Anastasios}\
  \bibnamefont {Theodoropoulos}}, \bibinfo {author} {\bibfnamefont {Nino}\
  \bibnamefont {Villanueva}}, \bibinfo {author} {\bibfnamefont {Tiago}\
  \bibnamefont {Fernandes}}, \bibinfo {author} {\bibfnamefont {Solange}\
  \bibnamefont {Nunes}}, \bibinfo {author} {\bibfnamefont {Jos{\'e}~A.}\
  \bibnamefont {Font}}, \bibinfo {author} {\bibfnamefont {Antonio}\
  \bibnamefont {Onofre}}, \bibinfo {author} {\bibfnamefont {Alejandro}\
  \bibnamefont {Torres-Forn{\'e}}}, \ and\ \bibinfo {author} {\bibfnamefont
  {Jos{\'e}~D.}\ \bibnamefont {Martin-Guerrero}},\ }\bibfield  {title}
  {\enquote {\bibinfo {title} {{Deep learning powered numerical relativity
  surrogate for binary black hole waveforms}},}\ }\href {\doibase
  10.1103/7bkx-hs53} {\bibfield  {journal} {\bibinfo  {journal} {Phys. Rev. D}\
  }\textbf {\bibinfo {volume} {112}},\ \bibinfo {pages} {043026} (\bibinfo
  {year} {2025})},\ \Eprint {http://arxiv.org/abs/2412.06946} {arXiv:2412.06946
  [gr-qc]} \BibitemShut {NoStop}%
\bibitem [{\citenamefont {Nee}\ \emph {et~al.}(2025)\citenamefont {Nee} \emph
  {et~al.}}]{Nee:2025nmh}%
  \BibitemOpen
  \bibfield  {author} {\bibinfo {author} {\bibfnamefont {Peter~James}\
  \bibnamefont {Nee}} \emph {et~al.},\ }\bibfield  {title} {\enquote {\bibinfo
  {title} {{Eccentric binary black holes: A new framework for numerical
  relativity waveform surrogates}},}\ }\href@noop {} {\  (\bibinfo {year}
  {2025})},\ \Eprint {http://arxiv.org/abs/2510.00106} {arXiv:2510.00106
  [gr-qc]} \BibitemShut {NoStop}%
\bibitem [{\citenamefont {Islam}\ \emph {et~al.}(2025)\citenamefont {Islam},
  \citenamefont {Vajpeyi}, \citenamefont {Shaik}, \citenamefont {Haster},
  \citenamefont {Varma}, \citenamefont {Field}, \citenamefont {Lange},
  \citenamefont {O'Shaughnessy},\ and\ \citenamefont {Smith}}]{Islam:2023zzj}%
  \BibitemOpen
  \bibfield  {author} {\bibinfo {author} {\bibfnamefont {Tousif}\ \bibnamefont
  {Islam}}, \bibinfo {author} {\bibfnamefont {Avi}\ \bibnamefont {Vajpeyi}},
  \bibinfo {author} {\bibfnamefont {Feroz~H.}\ \bibnamefont {Shaik}}, \bibinfo
  {author} {\bibfnamefont {Carl-Johan}\ \bibnamefont {Haster}}, \bibinfo
  {author} {\bibfnamefont {Vijay}\ \bibnamefont {Varma}}, \bibinfo {author}
  {\bibfnamefont {Scott~E.}\ \bibnamefont {Field}}, \bibinfo {author}
  {\bibfnamefont {Jacob}\ \bibnamefont {Lange}}, \bibinfo {author}
  {\bibfnamefont {Richard}\ \bibnamefont {O'Shaughnessy}}, \ and\ \bibinfo
  {author} {\bibfnamefont {Rory}\ \bibnamefont {Smith}},\ }\bibfield  {title}
  {\enquote {\bibinfo {title} {{Analysis of GWTC-3 with fully precessing
  numerical relativity surrogate models}},}\ }\href {\doibase
  10.1103/48ck-2fff} {\bibfield  {journal} {\bibinfo  {journal} {Phys. Rev. D}\
  }\textbf {\bibinfo {volume} {112}},\ \bibinfo {pages} {044001} (\bibinfo
  {year} {2025})},\ \Eprint {http://arxiv.org/abs/2309.14473} {arXiv:2309.14473
  [gr-qc]} \BibitemShut {NoStop}%
\bibitem [{\citenamefont {Abbott}\ \emph
  {et~al.}(2020{\natexlab{a}})\citenamefont {Abbott} \emph
  {et~al.}}]{LIGOScientific:2020ufj}%
  \BibitemOpen
  \bibfield  {author} {\bibinfo {author} {\bibfnamefont {R.}~\bibnamefont
  {Abbott}} \emph {et~al.} (\bibinfo {collaboration} {LIGO Scientific,
  Virgo}),\ }\bibfield  {title} {\enquote {\bibinfo {title} {{Properties and
  Astrophysical Implications of the 150 M$_\odot$ Binary Black Hole Merger
  GW190521}},}\ }\href {\doibase 10.3847/2041-8213/aba493} {\bibfield
  {journal} {\bibinfo  {journal} {Astrophys. J. Lett.}\ }\textbf {\bibinfo
  {volume} {900}},\ \bibinfo {pages} {L13} (\bibinfo {year}
  {2020}{\natexlab{a}})},\ \Eprint {http://arxiv.org/abs/2009.01190}
  {arXiv:2009.01190 [astro-ph.HE]} \BibitemShut {NoStop}%
\bibitem [{\citenamefont {Abbott}\ \emph
  {et~al.}(2020{\natexlab{b}})\citenamefont {Abbott} \emph
  {et~al.}}]{LIGOScientific:2020iuh}%
  \BibitemOpen
  \bibfield  {author} {\bibinfo {author} {\bibfnamefont {R.}~\bibnamefont
  {Abbott}} \emph {et~al.} (\bibinfo {collaboration} {LIGO Scientific,
  Virgo}),\ }\bibfield  {title} {\enquote {\bibinfo {title} {{GW190521: A
  Binary Black Hole Merger with a Total Mass of $150 M_{\odot}$}},}\ }\href
  {\doibase 10.1103/PhysRevLett.125.101102} {\bibfield  {journal} {\bibinfo
  {journal} {Phys. Rev. Lett.}\ }\textbf {\bibinfo {volume} {125}},\ \bibinfo
  {pages} {101102} (\bibinfo {year} {2020}{\natexlab{b}})},\ \Eprint
  {http://arxiv.org/abs/2009.01075} {arXiv:2009.01075 [gr-qc]} \BibitemShut
  {NoStop}%
\bibitem [{\citenamefont {Islam}\ \emph
  {et~al.}(2021{\natexlab{b}})\citenamefont {Islam}, \citenamefont {Field},
  \citenamefont {Haster},\ and\ \citenamefont {Smith}}]{Islam:2020reh}%
  \BibitemOpen
  \bibfield  {author} {\bibinfo {author} {\bibfnamefont {Tousif}\ \bibnamefont
  {Islam}}, \bibinfo {author} {\bibfnamefont {Scott~E.}\ \bibnamefont {Field}},
  \bibinfo {author} {\bibfnamefont {Carl-Johan}\ \bibnamefont {Haster}}, \ and\
  \bibinfo {author} {\bibfnamefont {Rory}\ \bibnamefont {Smith}},\ }\bibfield
  {title} {\enquote {\bibinfo {title} {{Improved analysis of GW190412 with a
  precessing numerical relativity surrogate waveform model}},}\ }\href
  {\doibase 10.1103/PhysRevD.103.104027} {\bibfield  {journal} {\bibinfo
  {journal} {Phys. Rev. D}\ }\textbf {\bibinfo {volume} {103}},\ \bibinfo
  {pages} {104027} (\bibinfo {year} {2021}{\natexlab{b}})},\ \Eprint
  {http://arxiv.org/abs/2010.04848} {arXiv:2010.04848 [gr-qc]} \BibitemShut
  {NoStop}%
\bibitem [{\citenamefont {Hannam}\ \emph {et~al.}(2022)\citenamefont {Hannam}
  \emph {et~al.}}]{Hannam:2021pit}%
  \BibitemOpen
  \bibfield  {author} {\bibinfo {author} {\bibfnamefont {Mark}\ \bibnamefont
  {Hannam}} \emph {et~al.},\ }\bibfield  {title} {\enquote {\bibinfo {title}
  {{General-relativistic precession in a black-hole binary}},}\ }\href
  {\doibase 10.1038/s41586-022-05212-z} {\bibfield  {journal} {\bibinfo
  {journal} {Nature}\ }\textbf {\bibinfo {volume} {610}},\ \bibinfo {pages}
  {652--655} (\bibinfo {year} {2022})},\ \Eprint
  {http://arxiv.org/abs/2112.11300} {arXiv:2112.11300 [gr-qc]} \BibitemShut
  {NoStop}%
\bibitem [{\citenamefont {Abac}\ \emph
  {et~al.}(2025{\natexlab{c}})\citenamefont {Abac} \emph
  {et~al.}}]{LIGOScientific:2025rid}%
  \BibitemOpen
  \bibfield  {author} {\bibinfo {author} {\bibfnamefont {A.~G.}\ \bibnamefont
  {Abac}} \emph {et~al.} (\bibinfo {collaboration} {LIGO Scientific, Virgo,
  KAGRA}),\ }\bibfield  {title} {\enquote {\bibinfo {title} {{GW250114: Testing
  Hawking{\textquoteright}s Area Law and the Kerr Nature of Black Holes}},}\
  }\href {\doibase 10.1103/kw5g-d732} {\bibfield  {journal} {\bibinfo
  {journal} {Phys. Rev. Lett.}\ }\textbf {\bibinfo {volume} {135}},\ \bibinfo
  {pages} {111403} (\bibinfo {year} {2025}{\natexlab{c}})},\ \Eprint
  {http://arxiv.org/abs/2509.08054} {arXiv:2509.08054 [gr-qc]} \BibitemShut
  {NoStop}%
\bibitem [{\citenamefont {Abac}\ \emph {et~al.}(2026)\citenamefont {Abac} \emph
  {et~al.}}]{LIGOScientific:2025wao}%
  \BibitemOpen
  \bibfield  {author} {\bibinfo {author} {\bibfnamefont {A.~G.}\ \bibnamefont
  {Abac}} \emph {et~al.} (\bibinfo {collaboration} {LIGO Scientific, Virgo,
  KAGRA}),\ }\bibfield  {title} {\enquote {\bibinfo {title} {{Black Hole
  Spectroscopy and Tests of General Relativity with GW250114}},}\ }\href
  {\doibase 10.1103/6c61-fm1n} {\bibfield  {journal} {\bibinfo  {journal}
  {Phys. Rev. Lett.}\ }\textbf {\bibinfo {volume} {136}},\ \bibinfo {pages}
  {041403} (\bibinfo {year} {2026})},\ \Eprint
  {http://arxiv.org/abs/2509.08099} {arXiv:2509.08099 [gr-qc]} \BibitemShut
  {NoStop}%
\bibitem [{\citenamefont {Finch}\ and\ \citenamefont
  {Moore}(2021)}]{Finch:2021iip}%
  \BibitemOpen
  \bibfield  {author} {\bibinfo {author} {\bibfnamefont {Eliot}\ \bibnamefont
  {Finch}}\ and\ \bibinfo {author} {\bibfnamefont {Christopher~J.}\
  \bibnamefont {Moore}},\ }\bibfield  {title} {\enquote {\bibinfo {title}
  {{Modeling the ringdown from precessing black hole binaries}},}\ }\href
  {\doibase 10.1103/PhysRevD.103.084048} {\bibfield  {journal} {\bibinfo
  {journal} {Phys. Rev. D}\ }\textbf {\bibinfo {volume} {103}},\ \bibinfo
  {pages} {084048} (\bibinfo {year} {2021})},\ \Eprint
  {http://arxiv.org/abs/2102.07794} {arXiv:2102.07794 [gr-qc]} \BibitemShut
  {NoStop}%
\bibitem [{\citenamefont {Siegel}\ \emph {et~al.}(2025)\citenamefont {Siegel},
  \citenamefont {Khusid}, \citenamefont {Isi},\ and\ \citenamefont
  {Farr}}]{Siegel:2025xgb}%
  \BibitemOpen
  \bibfield  {author} {\bibinfo {author} {\bibfnamefont {Harrison}\
  \bibnamefont {Siegel}}, \bibinfo {author} {\bibfnamefont {Nicole~M.}\
  \bibnamefont {Khusid}}, \bibinfo {author} {\bibfnamefont {Maximiliano}\
  \bibnamefont {Isi}}, \ and\ \bibinfo {author} {\bibfnamefont {Will~M.}\
  \bibnamefont {Farr}},\ }\bibfield  {title} {\enquote {\bibinfo {title}
  {{GW231123 ringdown: interpretation as multimodal Kerr signal}},}\
  }\href@noop {} {\  (\bibinfo {year} {2025})},\ \Eprint
  {http://arxiv.org/abs/2511.02691} {arXiv:2511.02691 [gr-qc]} \BibitemShut
  {NoStop}%
\bibitem [{\citenamefont {Abac}\ \emph
  {et~al.}(2025{\natexlab{d}})\citenamefont {Abac} \emph
  {et~al.}}]{LIGOScientific:2025rsn}%
  \BibitemOpen
  \bibfield  {author} {\bibinfo {author} {\bibfnamefont {A.~G.}\ \bibnamefont
  {Abac}} \emph {et~al.} (\bibinfo {collaboration} {LIGO Scientific, VIRGO,
  KAGRA}),\ }\bibfield  {title} {\enquote {\bibinfo {title} {{GW231123: A
  Binary Black Hole Merger with Total Mass 190{\textendash}265 $M_{\odot}$}},}\
  }\href {\doibase 10.3847/2041-8213/ae0c9c} {\bibfield  {journal} {\bibinfo
  {journal} {Astrophys. J. Lett.}\ }\textbf {\bibinfo {volume} {993}},\
  \bibinfo {pages} {L25} (\bibinfo {year} {2025}{\natexlab{d}})},\ \Eprint
  {http://arxiv.org/abs/2507.08219} {arXiv:2507.08219 [astro-ph.HE]}
  \BibitemShut {NoStop}%
\bibitem [{\citenamefont {Rink}\ \emph {et~al.}(2024)\citenamefont {Rink},
  \citenamefont {Bachhar}, \citenamefont {Islam}, \citenamefont {Rifat},
  \citenamefont {Gonzalez-Quesada}, \citenamefont {Field}, \citenamefont
  {Khanna}, \citenamefont {Hughes},\ and\ \citenamefont
  {Varma}}]{Rink:2024swg}%
  \BibitemOpen
  \bibfield  {author} {\bibinfo {author} {\bibfnamefont {Katie}\ \bibnamefont
  {Rink}}, \bibinfo {author} {\bibfnamefont {Ritesh}\ \bibnamefont {Bachhar}},
  \bibinfo {author} {\bibfnamefont {Tousif}\ \bibnamefont {Islam}}, \bibinfo
  {author} {\bibfnamefont {Nur E.~M.}\ \bibnamefont {Rifat}}, \bibinfo {author}
  {\bibfnamefont {Kevin}\ \bibnamefont {Gonzalez-Quesada}}, \bibinfo {author}
  {\bibfnamefont {Scott~E.}\ \bibnamefont {Field}}, \bibinfo {author}
  {\bibfnamefont {Gaurav}\ \bibnamefont {Khanna}}, \bibinfo {author}
  {\bibfnamefont {Scott~A.}\ \bibnamefont {Hughes}}, \ and\ \bibinfo {author}
  {\bibfnamefont {Vijay}\ \bibnamefont {Varma}},\ }\bibfield  {title} {\enquote
  {\bibinfo {title} {{Gravitational wave surrogate model for spinning,
  intermediate mass ratio binaries based on perturbation theory and numerical
  relativity}},}\ }\href {\doibase 10.1103/PhysRevD.110.124069} {\bibfield
  {journal} {\bibinfo  {journal} {Phys. Rev. D}\ }\textbf {\bibinfo {volume}
  {110}},\ \bibinfo {pages} {124069} (\bibinfo {year} {2024})},\ \Eprint
  {http://arxiv.org/abs/2407.18319} {arXiv:2407.18319 [gr-qc]} \BibitemShut
  {NoStop}%
\bibitem [{\citenamefont {P{\"u}rrer}(2016)}]{Purrer:2015tud}%
  \BibitemOpen
  \bibfield  {author} {\bibinfo {author} {\bibfnamefont {Michael}\ \bibnamefont
  {P{\"u}rrer}},\ }\bibfield  {title} {\enquote {\bibinfo {title} {{Frequency
  domain reduced order model of aligned-spin effective-one-body waveforms with
  generic mass-ratios and spins}},}\ }\href {\doibase
  10.1103/PhysRevD.93.064041} {\bibfield  {journal} {\bibinfo  {journal} {Phys.
  Rev. D}\ }\textbf {\bibinfo {volume} {93}},\ \bibinfo {pages} {064041}
  (\bibinfo {year} {2016})},\ \Eprint {http://arxiv.org/abs/1512.02248}
  {arXiv:1512.02248 [gr-qc]} \BibitemShut {NoStop}%
\bibitem [{\citenamefont {Cotesta}\ \emph {et~al.}(2020)\citenamefont
  {Cotesta}, \citenamefont {Marsat},\ and\ \citenamefont
  {P{\"u}rrer}}]{Cotesta:2020qhw}%
  \BibitemOpen
  \bibfield  {author} {\bibinfo {author} {\bibfnamefont {Roberto}\ \bibnamefont
  {Cotesta}}, \bibinfo {author} {\bibfnamefont {Sylvain}\ \bibnamefont
  {Marsat}}, \ and\ \bibinfo {author} {\bibfnamefont {Michael}\ \bibnamefont
  {P{\"u}rrer}},\ }\bibfield  {title} {\enquote {\bibinfo {title} {{Frequency
  domain reduced order model of aligned-spin effective-one-body waveforms with
  higher-order modes}},}\ }\href {\doibase 10.1103/PhysRevD.101.124040}
  {\bibfield  {journal} {\bibinfo  {journal} {Phys. Rev. D}\ }\textbf {\bibinfo
  {volume} {101}},\ \bibinfo {pages} {124040} (\bibinfo {year} {2020})},\
  \Eprint {http://arxiv.org/abs/2003.12079} {arXiv:2003.12079 [gr-qc]}
  \BibitemShut {NoStop}%
\bibitem [{\citenamefont {Pompili}\ \emph {et~al.}(2023)\citenamefont {Pompili}
  \emph {et~al.}}]{Pompili:2023tna}%
  \BibitemOpen
  \bibfield  {author} {\bibinfo {author} {\bibfnamefont {Lorenzo}\ \bibnamefont
  {Pompili}} \emph {et~al.},\ }\bibfield  {title} {\enquote {\bibinfo {title}
  {{Laying the foundation of the effective-one-body waveform models SEOBNRv5:
  Improved accuracy and efficiency for spinning nonprecessing binary black
  holes}},}\ }\href {\doibase 10.1103/PhysRevD.108.124035} {\bibfield
  {journal} {\bibinfo  {journal} {Phys. Rev. D}\ }\textbf {\bibinfo {volume}
  {108}},\ \bibinfo {pages} {124035} (\bibinfo {year} {2023})},\ \Eprint
  {http://arxiv.org/abs/2303.18039} {arXiv:2303.18039 [gr-qc]} \BibitemShut
  {NoStop}%
\bibitem [{\citenamefont {Maday}\ \emph {et~al.}(2009)\citenamefont {Maday},
  \citenamefont {Nguyen}, \citenamefont {Patera},\ and\ \citenamefont
  {Pau}}]{Maday:2009}%
  \BibitemOpen
  \bibfield  {author} {\bibinfo {author} {\bibfnamefont {Y.}~\bibnamefont
  {Maday}}, \bibinfo {author} {\bibfnamefont {N.~.C}\ \bibnamefont {Nguyen}},
  \bibinfo {author} {\bibfnamefont {A.~T.}\ \bibnamefont {Patera}}, \ and\
  \bibinfo {author} {\bibfnamefont {S.~H.}\ \bibnamefont {Pau}},\ }\bibfield
  {title} {\enquote {\bibinfo {title} {A general multipurpose interpolation
  procedure: the magic points},}\ }\href {\doibase 10.3934/cpaa.2009.8.383}
  {\bibfield  {journal} {\bibinfo  {journal} {Communications on Pure and
  Applied Analysis}\ }\textbf {\bibinfo {volume} {8}},\ \bibinfo {pages}
  {383--404} (\bibinfo {year} {2009})}\BibitemShut {NoStop}%
\bibitem [{\citenamefont {Chaturantabut}\ and\ \citenamefont
  {Sorensen}(2010)}]{chaturantabut2010nonlinear}%
  \BibitemOpen
  \bibfield  {author} {\bibinfo {author} {\bibfnamefont {Saifon}\ \bibnamefont
  {Chaturantabut}}\ and\ \bibinfo {author} {\bibfnamefont {Danny~C}\
  \bibnamefont {Sorensen}},\ }\bibfield  {title} {\enquote {\bibinfo {title}
  {Nonlinear model reduction via discrete empirical interpolation},}\
  }\href@noop {} {\bibfield  {journal} {\bibinfo  {journal} {SIAM Journal on
  Scientific Computing}\ }\textbf {\bibinfo {volume} {32}},\ \bibinfo {pages}
  {2737--2764} (\bibinfo {year} {2010})}\BibitemShut {NoStop}%
\bibitem [{\citenamefont {{Field}}\ \emph {et~al.}(2014)\citenamefont
  {{Field}}, \citenamefont {{Galley}}, \citenamefont {{Hesthaven}},
  \citenamefont {{Kaye}},\ and\ \citenamefont {{Tiglio}}}]{Field:2013cfa}%
  \BibitemOpen
  \bibfield  {author} {\bibinfo {author} {\bibfnamefont {S.~E.}\ \bibnamefont
  {{Field}}}, \bibinfo {author} {\bibfnamefont {C.~R.}\ \bibnamefont
  {{Galley}}}, \bibinfo {author} {\bibfnamefont {J.~S.}\ \bibnamefont
  {{Hesthaven}}}, \bibinfo {author} {\bibfnamefont {J.}~\bibnamefont {{Kaye}}},
  \ and\ \bibinfo {author} {\bibfnamefont {M.}~\bibnamefont {{Tiglio}}},\
  }\bibfield  {title} {\enquote {\bibinfo {title} {{Fast Prediction and
  Evaluation of Gravitational Waveforms Using Surrogate Models}},}\ }\href
  {\doibase 10.1103/PhysRevX.4.031006} {\bibfield  {journal} {\bibinfo
  {journal} {Phys. Rev. X}\ }\textbf {\bibinfo {volume} {4}},\ \bibinfo {eid}
  {031006} (\bibinfo {year} {2014})},\ \Eprint {http://arxiv.org/abs/1308.3565}
  {arXiv:1308.3565 [gr-qc]} \BibitemShut {NoStop}%
\bibitem [{\citenamefont {Maga{\~n}a~Zertuche}\ \emph
  {et~al.}(2025)\citenamefont {Maga{\~n}a~Zertuche} \emph
  {et~al.}}]{MaganaZertuche:2024ajz}%
  \BibitemOpen
  \bibfield  {author} {\bibinfo {author} {\bibfnamefont {Lorena}\ \bibnamefont
  {Maga{\~n}a~Zertuche}} \emph {et~al.},\ }\bibfield  {title} {\enquote
  {\bibinfo {title} {{High-precision ringdown surrogate model for nonprecessing
  binary black holes}},}\ }\href {\doibase 10.1103/q7sy-g3kl} {\bibfield
  {journal} {\bibinfo  {journal} {Phys. Rev. D}\ }\textbf {\bibinfo {volume}
  {112}},\ \bibinfo {pages} {024077} (\bibinfo {year} {2025})},\ \Eprint
  {http://arxiv.org/abs/2408.05300} {arXiv:2408.05300 [gr-qc]} \BibitemShut
  {NoStop}%
\bibitem [{\citenamefont {Chua}\ \emph {et~al.}(2021)\citenamefont {Chua},
  \citenamefont {Katz}, \citenamefont {Warburton},\ and\ \citenamefont
  {Hughes}}]{Chua:2020stf}%
  \BibitemOpen
  \bibfield  {author} {\bibinfo {author} {\bibfnamefont {Alvin J.~K.}\
  \bibnamefont {Chua}}, \bibinfo {author} {\bibfnamefont {Michael~L.}\
  \bibnamefont {Katz}}, \bibinfo {author} {\bibfnamefont {Niels}\ \bibnamefont
  {Warburton}}, \ and\ \bibinfo {author} {\bibfnamefont {Scott~A.}\
  \bibnamefont {Hughes}},\ }\bibfield  {title} {\enquote {\bibinfo {title}
  {{Rapid generation of fully relativistic extreme-mass-ratio-inspiral waveform
  templates for LISA data analysis}},}\ }\href {\doibase
  10.1103/PhysRevLett.126.051102} {\bibfield  {journal} {\bibinfo  {journal}
  {Phys. Rev. Lett.}\ }\textbf {\bibinfo {volume} {126}},\ \bibinfo {pages}
  {051102} (\bibinfo {year} {2021})},\ \Eprint
  {http://arxiv.org/abs/2008.06071} {arXiv:2008.06071 [gr-qc]} \BibitemShut
  {NoStop}%
\bibitem [{\citenamefont {Thomas}\ \emph {et~al.}(2022)\citenamefont {Thomas},
  \citenamefont {Pratten},\ and\ \citenamefont {Schmidt}}]{Thomas:2022rmc}%
  \BibitemOpen
  \bibfield  {author} {\bibinfo {author} {\bibfnamefont {Lucy~M.}\ \bibnamefont
  {Thomas}}, \bibinfo {author} {\bibfnamefont {Geraint}\ \bibnamefont
  {Pratten}}, \ and\ \bibinfo {author} {\bibfnamefont {Patricia}\ \bibnamefont
  {Schmidt}},\ }\bibfield  {title} {\enquote {\bibinfo {title} {{Accelerating
  multimodal gravitational waveforms from precessing compact binaries with
  artificial neural networks}},}\ }\href {\doibase 10.1103/PhysRevD.106.104029}
  {\bibfield  {journal} {\bibinfo  {journal} {Phys. Rev. D}\ }\textbf {\bibinfo
  {volume} {106}},\ \bibinfo {pages} {104029} (\bibinfo {year} {2022})},\
  \Eprint {http://arxiv.org/abs/2205.14066} {arXiv:2205.14066 [gr-qc]}
  \BibitemShut {NoStop}%
\bibitem [{\citenamefont {Thomas}\ \emph {et~al.}(2025)\citenamefont {Thomas},
  \citenamefont {Chatziioannou}, \citenamefont {Varma},\ and\ \citenamefont
  {Field}}]{Thomas:2025rje}%
  \BibitemOpen
  \bibfield  {author} {\bibinfo {author} {\bibfnamefont {Lucy~M.}\ \bibnamefont
  {Thomas}}, \bibinfo {author} {\bibfnamefont {Katerina}\ \bibnamefont
  {Chatziioannou}}, \bibinfo {author} {\bibfnamefont {Vijay}\ \bibnamefont
  {Varma}}, \ and\ \bibinfo {author} {\bibfnamefont {Scott~E.}\ \bibnamefont
  {Field}},\ }\bibfield  {title} {\enquote {\bibinfo {title} {{Optimizing
  neural network surrogate models: Application to black hole merger
  remnants}},}\ }\href {\doibase 10.1103/PhysRevD.111.104029} {\bibfield
  {journal} {\bibinfo  {journal} {Phys. Rev. D}\ }\textbf {\bibinfo {volume}
  {111}},\ \bibinfo {pages} {104029} (\bibinfo {year} {2025})},\ \Eprint
  {http://arxiv.org/abs/2501.16462} {arXiv:2501.16462 [gr-qc]} \BibitemShut
  {NoStop}%
\bibitem [{\citenamefont {McKechan}\ \emph {et~al.}(2010)\citenamefont
  {McKechan}, \citenamefont {Robinson},\ and\ \citenamefont
  {Sathyaprakash}}]{McKechan:2010kp}%
  \BibitemOpen
  \bibfield  {author} {\bibinfo {author} {\bibfnamefont {D.~J.~A.}\
  \bibnamefont {McKechan}}, \bibinfo {author} {\bibfnamefont {C.}~\bibnamefont
  {Robinson}}, \ and\ \bibinfo {author} {\bibfnamefont {B.~S.}\ \bibnamefont
  {Sathyaprakash}},\ }\bibfield  {title} {\enquote {\bibinfo {title} {{A
  tapering window for time-domain templates and simulated signals in the
  detection of gravitational waves from coalescing compact binaries}},}\
  }\bibfield  {booktitle} {\emph {\bibinfo {booktitle} {{Gravitational waves.
  Proceedings, 8th Edoardo Amaldi Conference, Amaldi 8, New York, USA, June
  22-26, 2009}}},\ }\href {\doibase 10.1088/0264-9381/27/8/084020} {\bibfield
  {journal} {\bibinfo  {journal} {Class. Quant. Grav.}\ }\textbf {\bibinfo
  {volume} {27}},\ \bibinfo {pages} {084020} (\bibinfo {year} {2010})},\
  \Eprint {http://arxiv.org/abs/1003.2939} {arXiv:1003.2939 [gr-qc]}
  \BibitemShut {NoStop}%
\bibitem [{\citenamefont {Jiang}\ and\ \citenamefont
  {Shu}(1996)}]{JIANG1996202}%
  \BibitemOpen
  \bibfield  {author} {\bibinfo {author} {\bibfnamefont {Guang-Shan}\
  \bibnamefont {Jiang}}\ and\ \bibinfo {author} {\bibfnamefont {Chi-Wang}\
  \bibnamefont {Shu}},\ }\bibfield  {title} {\enquote {\bibinfo {title}
  {Efficient implementation of weighted eno schemes},}\ }\href {\doibase
  https://doi.org/10.1006/jcph.1996.0130} {\bibfield  {journal} {\bibinfo
  {journal} {Journal of Computational Physics}\ }\textbf {\bibinfo {volume}
  {126}},\ \bibinfo {pages} {202--228} (\bibinfo {year} {1996})}\BibitemShut
  {NoStop}%
\bibitem [{\citenamefont {Scheel}\ \emph {et~al.}(2025)\citenamefont {Scheel}
  \emph {et~al.}}]{Scheel:2025jct}%
  \BibitemOpen
  \bibfield  {author} {\bibinfo {author} {\bibfnamefont {Mark~A.}\ \bibnamefont
  {Scheel}} \emph {et~al.},\ }\bibfield  {title} {\enquote {\bibinfo {title}
  {{The SXS collaboration{\textquoteright}s third catalog of binary black hole
  simulations}},}\ }\href {\doibase 10.1088/1361-6382/adfd34} {\bibfield
  {journal} {\bibinfo  {journal} {Class. Quant. Grav.}\ }\textbf {\bibinfo
  {volume} {42}},\ \bibinfo {pages} {195017} (\bibinfo {year} {2025})},\
  \Eprint {http://arxiv.org/abs/2505.13378} {arXiv:2505.13378 [gr-qc]}
  \BibitemShut {NoStop}%
\bibitem [{\citenamefont {Mitman}\ \emph {et~al.}(2021)\citenamefont {Mitman}
  \emph {et~al.}}]{Mitman:2020bjf}%
  \BibitemOpen
  \bibfield  {author} {\bibinfo {author} {\bibfnamefont {Keefe}\ \bibnamefont
  {Mitman}} \emph {et~al.},\ }\bibfield  {title} {\enquote {\bibinfo {title}
  {{Adding gravitational memory to waveform catalogs using BMS balance
  laws}},}\ }\href {\doibase 10.1103/PhysRevD.103.024031} {\bibfield  {journal}
  {\bibinfo  {journal} {Phys. Rev. D}\ }\textbf {\bibinfo {volume} {103}},\
  \bibinfo {pages} {024031} (\bibinfo {year} {2021})},\ \Eprint
  {http://arxiv.org/abs/2011.01309} {arXiv:2011.01309 [gr-qc]} \BibitemShut
  {NoStop}%
\bibitem [{\citenamefont {Boyle}(2013)}]{Boyle:2013nka}%
  \BibitemOpen
  \bibfield  {author} {\bibinfo {author} {\bibfnamefont {Michael}\ \bibnamefont
  {Boyle}},\ }\bibfield  {title} {\enquote {\bibinfo {title} {{Angular velocity
  of gravitational radiation from precessing binaries and the corotating
  frame}},}\ }\href {\doibase 10.1103/PhysRevD.87.104006} {\bibfield  {journal}
  {\bibinfo  {journal} {Phys. Rev.}\ }\textbf {\bibinfo {volume} {D87}},\
  \bibinfo {pages} {104006} (\bibinfo {year} {2013})},\ \Eprint
  {http://arxiv.org/abs/1302.2919} {arXiv:1302.2919 [gr-qc]} \BibitemShut
  {NoStop}%
\bibitem [{\citenamefont {{LIGO Scientific
  Collaboration}}(2018)}]{aLIGODesignNoiseCurve}%
  \BibitemOpen
  \bibfield  {author} {\bibinfo {author} {\bibnamefont {{LIGO Scientific
  Collaboration}}},\ }\href@noop {} {\emph {\bibinfo {title} {Updated Advanced
  LIGO sensitivity design curve}}},\ \bibinfo {type} {Tech. Rep.}\ (\bibinfo
  {year} {2018})\ \bibinfo {note}
  {\url{https://dcc.ligo.org/LIGO-T1800044/public}}\BibitemShut {NoStop}%
\bibitem [{\citenamefont {Giesler}\ \emph {et~al.}(2019)\citenamefont
  {Giesler}, \citenamefont {Isi}, \citenamefont {Scheel},\ and\ \citenamefont
  {Teukolsky}}]{Giesler:2019uxc}%
  \BibitemOpen
  \bibfield  {author} {\bibinfo {author} {\bibfnamefont {Matthew}\ \bibnamefont
  {Giesler}}, \bibinfo {author} {\bibfnamefont {Maximiliano}\ \bibnamefont
  {Isi}}, \bibinfo {author} {\bibfnamefont {Mark}\ \bibnamefont {Scheel}}, \
  and\ \bibinfo {author} {\bibfnamefont {Saul}\ \bibnamefont {Teukolsky}},\
  }\bibfield  {title} {\enquote {\bibinfo {title} {{Black hole ringdown: the
  importance of overtones}},}\ }\href {\doibase 10.1103/PhysRevX.9.041060}
  {\bibfield  {journal} {\bibinfo  {journal} {Phys. Rev.}\ }\textbf {\bibinfo
  {volume} {X9}},\ \bibinfo {pages} {041060} (\bibinfo {year} {2019})},\
  \Eprint {http://arxiv.org/abs/1903.08284} {arXiv:1903.08284 [gr-qc]}
  \BibitemShut {NoStop}%
\bibitem [{\citenamefont {Maga{\~n}a~Zertuche}\ \emph
  {et~al.}(2022)\citenamefont {Maga{\~n}a~Zertuche} \emph
  {et~al.}}]{MaganaZertuche:2021syq}%
  \BibitemOpen
  \bibfield  {author} {\bibinfo {author} {\bibfnamefont {Lorena}\ \bibnamefont
  {Maga{\~n}a~Zertuche}} \emph {et~al.},\ }\bibfield  {title} {\enquote
  {\bibinfo {title} {{High precision ringdown modeling: Multimode fits and BMS
  frames}},}\ }\href {\doibase 10.1103/PhysRevD.105.104015} {\bibfield
  {journal} {\bibinfo  {journal} {Phys. Rev. D}\ }\textbf {\bibinfo {volume}
  {105}},\ \bibinfo {pages} {104015} (\bibinfo {year} {2022})},\ \Eprint
  {http://arxiv.org/abs/2110.15922} {arXiv:2110.15922 [gr-qc]} \BibitemShut
  {NoStop}%
\bibitem [{\citenamefont {Stein}(2019)}]{Stein:2019mop}%
  \BibitemOpen
  \bibfield  {author} {\bibinfo {author} {\bibfnamefont {Leo~C.}\ \bibnamefont
  {Stein}},\ }\bibfield  {title} {\enquote {\bibinfo {title} {{qnm: A Python
  package for calculating Kerr quasinormal modes, separation constants, and
  spherical-spheroidal mixing coefficients}},}\ }\href {\doibase
  10.21105/joss.01683} {\bibfield  {journal} {\bibinfo  {journal} {J. Open
  Source Softw.}\ }\textbf {\bibinfo {volume} {4}},\ \bibinfo {pages} {1683}
  (\bibinfo {year} {2019})},\ \Eprint {http://arxiv.org/abs/1908.10377}
  {arXiv:1908.10377 [gr-qc]} \BibitemShut {NoStop}%
\bibitem [{\citenamefont {Varma}\ \emph
  {et~al.}(2019{\natexlab{c}})\citenamefont {Varma}, \citenamefont {Gerosa},
  \citenamefont {Stein}, \citenamefont {Hébert},\ and\ \citenamefont
  {Zhang}}]{Varma:2018aht}%
  \BibitemOpen
  \bibfield  {author} {\bibinfo {author} {\bibfnamefont {Vijay}\ \bibnamefont
  {Varma}}, \bibinfo {author} {\bibfnamefont {Davide}\ \bibnamefont {Gerosa}},
  \bibinfo {author} {\bibfnamefont {Leo~C.}\ \bibnamefont {Stein}}, \bibinfo
  {author} {\bibfnamefont {François}\ \bibnamefont {Hébert}}, \ and\ \bibinfo
  {author} {\bibfnamefont {Hao}\ \bibnamefont {Zhang}},\ }\bibfield  {title}
  {\enquote {\bibinfo {title} {{High-accuracy mass, spin, and recoil
  predictions of generic black-hole merger remnants}},}\ }\href {\doibase
  10.1103/PhysRevLett.122.011101} {\bibfield  {journal} {\bibinfo  {journal}
  {Phys. Rev. Lett.}\ }\textbf {\bibinfo {volume} {122}},\ \bibinfo {pages}
  {011101} (\bibinfo {year} {2019}{\natexlab{c}})},\ \Eprint
  {http://arxiv.org/abs/1809.09125} {arXiv:1809.09125 [gr-qc]} \BibitemShut
  {NoStop}%
\bibitem [{\citenamefont {Li}\ \emph {et~al.}(2022)\citenamefont {Li},
  \citenamefont {Sun}, \citenamefont {Lo}, \citenamefont {Payne},\ and\
  \citenamefont {Chen}}]{Li:2021wgz}%
  \BibitemOpen
  \bibfield  {author} {\bibinfo {author} {\bibfnamefont {Xiang}\ \bibnamefont
  {Li}}, \bibinfo {author} {\bibfnamefont {Ling}\ \bibnamefont {Sun}}, \bibinfo
  {author} {\bibfnamefont {Rico Ka~Lok}\ \bibnamefont {Lo}}, \bibinfo {author}
  {\bibfnamefont {Ethan}\ \bibnamefont {Payne}}, \ and\ \bibinfo {author}
  {\bibfnamefont {Yanbei}\ \bibnamefont {Chen}},\ }\bibfield  {title} {\enquote
  {\bibinfo {title} {{Angular emission patterns of remnant black holes}},}\
  }\href {\doibase 10.1103/PhysRevD.105.024016} {\bibfield  {journal} {\bibinfo
   {journal} {Phys. Rev. D}\ }\textbf {\bibinfo {volume} {105}},\ \bibinfo
  {pages} {024016} (\bibinfo {year} {2022})},\ \Eprint
  {http://arxiv.org/abs/2110.03116} {arXiv:2110.03116 [gr-qc]} \BibitemShut
  {NoStop}%
\bibitem [{\citenamefont {Hamilton}\ \emph {et~al.}(2023)\citenamefont
  {Hamilton}, \citenamefont {London},\ and\ \citenamefont
  {Hannam}}]{Hamilton:2023znn}%
  \BibitemOpen
  \bibfield  {author} {\bibinfo {author} {\bibfnamefont {Eleanor}\ \bibnamefont
  {Hamilton}}, \bibinfo {author} {\bibfnamefont {Lionel}\ \bibnamefont
  {London}}, \ and\ \bibinfo {author} {\bibfnamefont {Mark}\ \bibnamefont
  {Hannam}},\ }\bibfield  {title} {\enquote {\bibinfo {title} {{Ringdown
  frequencies in black holes formed from precessing black-hole binaries}},}\
  }\href {\doibase 10.1103/PhysRevD.107.104035} {\bibfield  {journal} {\bibinfo
   {journal} {Phys. Rev. D}\ }\textbf {\bibinfo {volume} {107}},\ \bibinfo
  {pages} {104035} (\bibinfo {year} {2023})},\ \Eprint
  {http://arxiv.org/abs/2301.06558} {arXiv:2301.06558 [gr-qc]} \BibitemShut
  {NoStop}%
\bibitem [{\citenamefont {Zhu}\ \emph {et~al.}(2025)\citenamefont {Zhu} \emph
  {et~al.}}]{Zhu:2023fnf}%
  \BibitemOpen
  \bibfield  {author} {\bibinfo {author} {\bibfnamefont {Hengrui}\ \bibnamefont
  {Zhu}} \emph {et~al.},\ }\bibfield  {title} {\enquote {\bibinfo {title}
  {{Black hole spectroscopy for precessing binary black hole coalescences}},}\
  }\href {\doibase 10.1103/PhysRevD.111.064052} {\bibfield  {journal} {\bibinfo
   {journal} {Phys. Rev. D}\ }\textbf {\bibinfo {volume} {111}},\ \bibinfo
  {pages} {064052} (\bibinfo {year} {2025})},\ \Eprint
  {http://arxiv.org/abs/2312.08588} {arXiv:2312.08588 [gr-qc]} \BibitemShut
  {NoStop}%
\bibitem [{\citenamefont {Field}\ \emph {et~al.}(2025)\citenamefont {Field},
  \citenamefont {Varma}, \citenamefont {Blackman}, \citenamefont {Gadre},
  \citenamefont {Galley}, \citenamefont {Islam}, \citenamefont {Mitman},
  \citenamefont {Pürrer}, \citenamefont {Ravichandran}, \citenamefont
  {Scheel}, \citenamefont {Stein},\ and\ \citenamefont {Yoo}}]{Field:2025isp}%
  \BibitemOpen
  \bibfield  {author} {\bibinfo {author} {\bibfnamefont {Scott~E.}\
  \bibnamefont {Field}}, \bibinfo {author} {\bibfnamefont {Vijay}\ \bibnamefont
  {Varma}}, \bibinfo {author} {\bibfnamefont {Jonathan}\ \bibnamefont
  {Blackman}}, \bibinfo {author} {\bibfnamefont {Bhooshan}\ \bibnamefont
  {Gadre}}, \bibinfo {author} {\bibfnamefont {Chad~R.}\ \bibnamefont {Galley}},
  \bibinfo {author} {\bibfnamefont {Tousif}\ \bibnamefont {Islam}}, \bibinfo
  {author} {\bibfnamefont {Keefe}\ \bibnamefont {Mitman}}, \bibinfo {author}
  {\bibfnamefont {Michael}\ \bibnamefont {Pürrer}}, \bibinfo {author}
  {\bibfnamefont {Adhrit}\ \bibnamefont {Ravichandran}}, \bibinfo {author}
  {\bibfnamefont {Mark~A.}\ \bibnamefont {Scheel}}, \bibinfo {author}
  {\bibfnamefont {Leo~C.}\ \bibnamefont {Stein}}, \ and\ \bibinfo {author}
  {\bibfnamefont {Jooheon}\ \bibnamefont {Yoo}},\ }\bibfield  {title} {\enquote
  {\bibinfo {title} {{GWSurrogate: A Python package for gravitational wave
  surrogate models}},}\ }\href {\doibase 10.21105/joss.07073} {\bibfield
  {journal} {\bibinfo  {journal} {J. Open Source Softw.}\ }\textbf {\bibinfo
  {volume} {10}},\ \bibinfo {pages} {7073} (\bibinfo {year} {2025})},\ \Eprint
  {http://arxiv.org/abs/2504.08839} {arXiv:2504.08839 [astro-ph.IM]}
  \BibitemShut {NoStop}%
\bibitem [{\citenamefont {Ravishankar}\ \emph {et~al.}(2026)\citenamefont
  {Ravishankar}, \citenamefont {Varma}, \citenamefont {Field} \emph
  {et~al.}}]{gwsurrogate_benchmark_version}%
  \BibitemOpen
  \bibfield  {author} {\bibinfo {author} {\bibfnamefont {Abhishek}\
  \bibnamefont {Ravishankar}}, \bibinfo {author} {\bibfnamefont {Vijay}\
  \bibnamefont {Varma}}, \bibinfo {author} {\bibfnamefont {Scott~E.}\
  \bibnamefont {Field}},  \emph {et~al.},\ }\href
  {https://github.com/Abhishek-Ravishankar/gwsurrogate/tree/5864af24b496bb55cd98d5dcf4d4295a5ecf4ca4}
  {\enquote {\bibinfo {title} {\texttt{gwsurrogate}: Benchmarked version},}\ }
  (\bibinfo {year} {2026}),\ \bibinfo {note} {code snapshot used for Fig.
  \ref{fig:Timing_Comparison}}\BibitemShut {NoStop}%
\bibitem [{\citenamefont {Field}\ \emph {et~al.}(2011)\citenamefont {Field},
  \citenamefont {Galley}, \citenamefont {Herrmann}, \citenamefont {Hesthaven},
  \citenamefont {Ochsner},\ and\ \citenamefont {Tiglio}}]{Field:2011mf}%
  \BibitemOpen
  \bibfield  {author} {\bibinfo {author} {\bibfnamefont {Scott~E.}\
  \bibnamefont {Field}}, \bibinfo {author} {\bibfnamefont {Chad~R.}\
  \bibnamefont {Galley}}, \bibinfo {author} {\bibfnamefont {Frank}\
  \bibnamefont {Herrmann}}, \bibinfo {author} {\bibfnamefont {Jan~S.}\
  \bibnamefont {Hesthaven}}, \bibinfo {author} {\bibfnamefont {Evan}\
  \bibnamefont {Ochsner}}, \ and\ \bibinfo {author} {\bibfnamefont {Manuel}\
  \bibnamefont {Tiglio}},\ }\bibfield  {title} {\enquote {\bibinfo {title}
  {{Reduced basis catalogs for gravitational wave templates}},}\ }\href
  {\doibase 10.1103/PhysRevLett.106.221102} {\bibfield  {journal} {\bibinfo
  {journal} {Phys. Rev. Lett.}\ }\textbf {\bibinfo {volume} {106}},\ \bibinfo
  {pages} {221102} (\bibinfo {year} {2011})},\ \Eprint
  {http://arxiv.org/abs/1101.3765} {arXiv:1101.3765 [gr-qc]} \BibitemShut
  {NoStop}%
\end{thebibliography}%
